\documentclass[twocolumn,floatfix,showpacs,showkeys,preprintnumbers,nofootinbib,superscriptaddress]{revtex4}
\usepackage[utf8]{inputenc}
\usepackage[sort&compress]{natbib}
\usepackage{ulem}
\usepackage{bm}
\usepackage{times}
\usepackage{amssymb,amsbsy,amsmath,amsfonts}
\usepackage{graphicx}
\usepackage{float}
\usepackage{color}
\usepackage{morefloats}
\usepackage{rotating}
\usepackage{srcltx}
\usepackage{slashed}
\usepackage{subfigure}
\usepackage{multirow}
\usepackage{verbatim}
\usepackage{hyperref}
\usepackage{tabularx}
\usepackage{adjustbox}

\usepackage{overpic}
\usepackage{makecell}

\begin{document}

\title{Production rates of hidden-charm  pentaquark molecules in $\Lambda_b$ decays}

\author{Ya-Wen Pan}
\affiliation{School of Physics, Beihang University, Beijing 102206, China}

\author{Ming-Zhu Liu}\email{zhengmz11@buaa.edu.cn}
\affiliation{ School of Nuclear Science and Technology, Lanzhou University, Lanzhou 730000, China}
\affiliation{School of Physics, Beihang University, Beijing 102206, China}

\author{Li-Sheng Geng}\email{lisheng.geng@buaa.edu.cn}
\affiliation{School of Physics, Beihang University, Beijing 102206, China}
\affiliation{Beijing Key Laboratory of Advanced Nuclear Materials and Physics, Beihang University, Beijing 102206, China}
\affiliation{Peng Huanwu Collaborative Center for Research and Education, Beihang University, Beijing 100191, China}
\affiliation{Southern Center for Nuclear-Science Theory (SCNT), Institute of Modern Physics, Chinese Academy of Sciences, Huizhou 516000, Guangdong Province, China}

\date{\today}
\begin{abstract}
The  partial decay widths  and production  mechanism  of  the three pentaquark states, $P_{\psi}^{N}(4312)$,{$P_{\psi}^{N}(4440)$}, and {$P_{\psi}^{N}(4457)$},  discovered by the LHCb Collaboration in 2019,  are still under debate.   In this work, we employ the contact-range effective field theory approach to construct the    $\bar{D}^{(*)}\Sigma_{c}^{(*)}$,  $\bar{D}^{*}\Lambda_c$, $\bar{D}\Lambda_c$,  $J/\psi p$, and $\eta_c p$ coupled-channel interactions to dynamically generate the multiplet of hidde-charm pentaquark molecules by reproducing  the masses and widths of  { $P_{\psi}^{N}(4312)$},  {$P_{\psi}^{N}(4440)$}, and  {$P_{\psi}^{N}(4457)$}.      
 Assuming that the pentaquark molecules are produced   in the $\Lambda_b$ decay via the  triangle diagrams, where   $\Lambda_{b}$ firstly decays into $D_{s}^{(\ast)}\Lambda_{c}$, then $D_{s}^{(\ast)}$ { scatters } into $\bar{D}^{(\ast)}K$, and finally the molecules are  dynamically generated by the $\bar{D}^{(\ast)}\Lambda_{c}$ interactions,  we calculate the branching fractions of the decays  $\Lambda_b \to  {P_{\psi}^{N}}K$ using  the effective Lagrangian approach.  With the partial decay widths of these pentaquark molecules, we further  estimate  the branching fraction of the decays   $ \Lambda_b \to (  {P_{\psi}^{N}} \to J/\psi p )K $ and  $ \Lambda_b \to (  {P_{\psi}^{N}}\to \bar{D}^* \Lambda_c )K $. Our results show that the  pentaquark states  { $P_{\psi}^{N}(4312)$},   {$P_{\psi}^{N}(4440)$}, and  {$P_{\psi}^{N}(4457)$} as hadronic molecules can be produced in the $\Lambda_b$ decay, and on the other hand their heavy quark spin symmetry  partners are invisible in the $J/\psi p$ invariant mass distribution because of  the small production rates.  Our studies show that is possible to observe some of the pentaquark states in the $\Lambda_b\to \bar{D}^*\Lambda_c K$ decays.
\end{abstract}


\maketitle

\section{Introduction}

In 2015, two pentaquark states   {$P_{\psi}^{N}(4380)$ } and  {$P_{\psi}^{N}(4450)$ }  were observed by the LHCb Collaboration  in the $J/\psi p$ invariant mass distributions  of  the $\Lambda_{b}\to J/\psi p K$ decay~\cite{Aaij:2015tga}.
Four years later, they updated the data sample and found  that the original  {$P_{\psi}^{N}(4450)$ }  state splits into two states,  {$P_{\psi}^{N}(4440)$} and  {$P_{\psi}^{N}(4457)$}, and a new state  { $P_{\psi}^{N}(4312)$} emerges below the $\bar{D}\Sigma_c$ threshold~\cite{Aaij:2019vzc}. Recently   the LHCb Collaboration found  the evidence of the hidden-charm pentaquark state  {$P_{\psi }^{N}(4337) $} in the $B_s$ meson decay~\cite{LHCb:2021chn}, as well as   the hidden-charm pentaquark states  with strangeness  {$P_{\psi s}^{\Lambda}(4459)$} in the  $\Xi_b$ decay~\cite{LHCb:2020jpq}, the existence of which need to be confirmed because at present the significance of the observation is only about 3$\sigma$. Very recently the LHCb Collaboration reported another   pentaquark  state   {$P_{\psi s}^{\Lambda}(4338)$} in  the $B$   decay with a high  significance~\cite{LHCb:2022ogu}.    In this work, we only focus on the three pentaquark states  { $P_{\psi}^{N}(4312)$},  {$P_{\psi}^{N}(4440)$}, and {$P_{\psi}^{N}(4457)$}, which have been extensively studied in a series of theoretical works. We note that although the $\bar{D}^{(*)}\Sigma_c$ molecular interpretations for these pentaquark states   are the most popular~~\cite{Chen:2019asm,He:2019ify,Chen:2019bip,Xiao:2019aya,Sakai:2019qph,Yamaguchi:2019seo,He:2019rva,Liu:2019zvb,Valderrama:2019chc,Meng:2019ilv,Du:2019pij,Ling:2021lmq,Dong:2021juy,Ozdem:2021ugy,Pan:2022xxz,Zhang:2023czx,Pan:2022whr,Liu:2023wfo}, there exist other explanations, e.g.,  hadro-charmonia~\cite{Eides:2019tgv}, compact pentaquark states~\cite{Ali:2019npk,Wang:2019got,Cheng:2019obk,Weng:2019ynv,Zhu:2019iwm,Pimikov:2019dyr,Ruangyoo:2021aoi}, 
virtual states~\cite{Fernandez-Ramirez:2019koa}, triangle singularities~\cite{Nakamura:2021qvy}, and cusp effects~\cite{Burns:2022uiv}.

 From the perspective of masses,  the  three pentaquark states can be nicely arranged into the $\bar{D}^{(*)}\Sigma_c^{(*)}$ multiplet. However, their  widths obtained  in the hadronic molecule picture    always deviate a bit from the experimental data. In Ref.~\cite{Xie:2022hhv}, we found that their partial decay widths into three-body final states $\bar{D}^{(\ast)}\Lambda_c\pi$ are only at the order of a few hundreds of keV, which indicates that the two-body decay modes are dominant.   The chiral unitary study found that the partial decay  widths of ${P_{\psi}^{N}} \to J/\psi p (\eta_c p)$ account for the largest portion of their total decay widths~\cite{Xiao:2019aya}, while the study based on the triangle diagrams  shows  that the three ${P_{\psi}^{N}}$ states mainly decay into  $\bar{D}^{(*)}\Lambda_c$~\cite{Lin:2017mtz}.   In Refs.~\cite{Yamaguchi:2019seo, He:2019rva, Burns:2021jlu}, the authors argued that the one-pion exchange is responsible for the  $\bar{D}^{(*)}\Sigma_c \to \bar{D}^{(*)}\Lambda_c$ interactions,  and therefore  dominantly  contributes to   the widths of the pentaquark states.   From these studies, we conclude that these  three     molecules  mainly  decay via  two modes: hidden-charm $J/\psi p (\eta_c p)$ and open-charm $\bar{D}^{(*)}\Lambda_c$.  Considering the upper limit of the branching fraction   $\mathcal{B}({P_{\psi}^{N}}\to J/\psi p)<2\% $ measured in the photoproduction processes~\cite{Meziani:2016lhg,GlueX:2019mkq}, the partial decays  ${P_{\psi}^{N}}\to \bar{D}^{(*)}\Lambda_c$ are expected to play   a dominant role. However, such small upper limits cannot be easily reconciled with the current LHCb data~\cite{Cao:2019kst}. In this work, we employ the  contact-range effective field theory(EFT) approach to revisit the  partial decay widths  of the hidden-charm  pentaquark molecules by studying their two- and three-body   decays.
  
 Up to now, the hidden-charm pentaquark states have only
 been observed in the exclusive $b$ decays  in proton-proton collisions.  
 The productions of pentaquark states in other processes have been proposed. In Refs.~\cite{Wang:2015jsa,HillerBlin:2016odx,Karliner:2015voa,Wang:2019krd,Wu:2019adv}, the authors  claimed that the hidden-charm pentaquark states can be produced in the $J/\psi$ photoproduction off proton. This process could  distinguish  whether these pentaquark states are  genuine states or anomalous triangle singularities.   Moreover, it is suggested that the hidden-charm pentaquark states can be  produced in the $e^+e^-$ collisions~\cite{Li:2017ghe} and antiproton-deuteron collisions~\cite{Voloshin:2019wxx}. Based on Monte Carlo simulations, the inclusive production rates of these  pentaquark states are estimated in proton-proton collisions~\cite{Chen:2021ifb,Ling:2021sld} and electron-proton collisions~\cite{Shi:2022ipx}, which are helpful for future experimental searches for the pentaquark states.   In this work, based on the LHCb data, we primarily  focus on the production mechanism of the pentaquark states  in the $\Lambda_b$ decays.

The production mechanism of the pentaquark states in the $\Lambda_b$ decays can be classified into two categories. In mechanism I,   the mother particle  $M$ weakly decays  into three particles  $A$, $B$ and $C$, and the hadronic molecule under study can be dynamically generated via the   rescattering of any  two particles of $A$, $B$ and $C$.  This mechanism has already  been applied to study the production rates  of $X(3872)$ as a  $\bar{D}D^*$  molecule via the  weak decays   $B \to \bar{D}D^{*} K$~\cite{Braaten:2004fk,Braaten:2004ai}. For the pentaquark states, it was proposed that the weak decays of   $\Lambda_b \to \bar{D}^{(*)}\Sigma_c K$ and  $\Lambda_b \to J/\psi p K$ can dynamically generate the hidden-charm pentaquark molecules  via  the  $\bar{D}^{(*)}\Sigma_c$ rescattering~\cite{Du:2019pij,Du:2021fmf}  and   $J/\psi p $ rescatterring~\cite{Roca:2015dva}, respectively, which  can well describe the experimental  invariant mass distribution of $J/\psi p$, while  their absolute  production rates are not quantitatively   estimated. In particular, as pointed out in Ref.~\cite{Burns:2022uiv}, the branching fractions    $Br(\Lambda_b \to \bar{D}^{(*)}\Sigma_c K)$ are so tiny that the pentaquark molecules are rather difficult to be produced via the weak decays $\Lambda_b \to \bar{D}^{(*)}\Sigma_c K$. Therefore, whether these  pentaquark molecules can be produced via Mechanism I remains unsettled.

In Mechanism II, the  mother particle $M$  weakly   decays into two states $A$ and $B$, then $A$ {\color{black}  scatters  ( or decays) }  into  $C$ and $D$, and finally the final-state interaction of $B$ and $C$ dynamically generates the molecules of interest~\cite{Wang:2013cya,Guo:2013zbw,Hsiao:2019ait,Liu:2022dmm,Wu:2023rrp}. A  typical example is that the $X(3872)$ as  a   $\bar{D}D^*$  molecule can  be generated through  the weak decays $B \to \bar{D}^{(*)} D_{s}^{(*)}$ following $D_{s}^{(*)}$ {\color{black} scattering} into $D^{(*)}K$~\cite{Wu:2023rrp}.    
In Ref.~\cite{Wu:2019rog}, Wu et al. proposed that  $\Lambda_b$ weakly decays into $\Sigma_c$ and ${D}_{s}^{(*)}$, then ${D}_{s}^{(*)}$ {\color{black} scatter}  into $\bar{D}^{(*)}$ and $K$, and the pentaquark molecules are finally generated via the  $\bar{D}^{(*)}\Sigma_c$ interactions.   We note that the  $\Lambda_b$ decaying into $\Sigma_c^{(*)}$ is highly suppressed due to the fact the   light quark pair transition  between a symmetric and antisymmetric spin-flavor configuration   is  forbidden~\cite{Falk:1992ws,Gutsche:2018utw},     which indicates that  the production of  pentaquark molecules is difficult (if not impossible) via the weak decays of $\Lambda_b \to {D}_s^{(*)}\Sigma_c$ \footnote{In Ref.~\cite{Wu:2019rog}, the $\Lambda_b \to \Sigma_c$ transition is assumed to be proportional to the  $\Lambda_b \to \Lambda_c$ transition, characterized by  an unknown parameter $R$.   By reproducing the experimental  production rates of the pentaquark molecules,    $R$ is found to be about 0.1. }.   In Ref.~\cite{Burns:2022uiv}, the authors select the Color favorable weak decays $\Lambda_b \to {D}_{s}^{(*)}\Lambda_c $ to produce the pentaquark molecules  as well as to analyse their mass distributions,  but   did not explicitly  calculate their productions rates.    Following  Refs.~\cite{Liu:2022dmm,Wu:2023rrp}, we take the effective Lagrangian approach    to calculate the production rates of the pentaquark molecules in $\Lambda_b$ decays with no free parameters, and try to answer the questions whether the three pentaquark states $P_{\psi}^{N}(4312)$, {$P_{\psi}^{N}(4440)$}, and {$P_{\psi}^{N}(4457)$}  as hadronic molecules    can be produced in the $\Lambda_b$ decays, as well as  why their  HQSS partners   have not been observed in the same decays.

This work is organized as follows. We  first calculate the two-body partial decay widths of  the pentaquark molecules obtained  by the contact range EFT, and  the amplitudes of   their  production mechanism   in $\Lambda_b$ decays via the triangle diagrams  using the effective Lagrangian approach in section \ref{theoretical}.  The results and discussions on the  widths of the pentaquark molecules  and the branching fractions of the decays ${P_{\psi}^{N}} \to J/\psi p$ and ${P_{\psi}^{N}} \to \bar{D}^{(*)}\Lambda_c$, as well as the branching fractions of the weak decays $\Lambda_b\to {P_{\psi}^{N}} K$,  $\Lambda_b\to ({P_{\psi}^{N}}\to J/\psi p ) K$, and $\Lambda_b\to ({P_{\psi}^{N}}\to \bar{D}^{*}\Lambda_c) K$  are provided in section \ref{results}, followed by a short summary in the last section.

\section{Theoretical framework}
\label{theoretical}

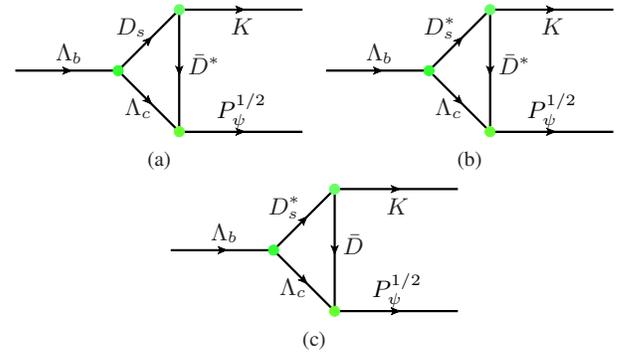
\begin{figure}[!h]
\begin{center}
\subfigure[]
{
\begin{minipage}[t]{0.45\linewidth}
\begin{center}
\begin{overpic}[scale=.48]{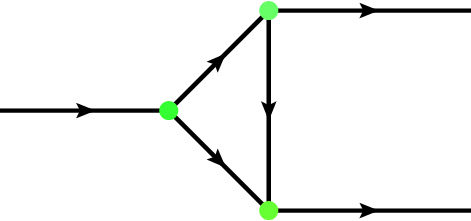}
	\put(70,7){$ \textcolor{black}{P_{\psi}^{  1/2}}$}
	\put(38,8){$\Lambda_{c}$}		
	\put(35,36){$D_{s}$}	
	\put(14,27){$\Lambda_{b}$ }
	\put(75,36){$K$} 
	\put(60,22){$\bar{D}^*$}
\end{overpic}
\end{center}
\end{minipage}
}
\subfigure[]
{
\begin{minipage}[t]{0.45\linewidth}
\begin{center}
\begin{overpic}[scale=.48]{triangle.eps}
	\put(70,7){$ \textcolor{black}{P_{\psi}^{  1/2}}$}
	\put(38,8){$\Lambda_{c}$}		
	\put(34,36){$D_{s}^*$}	
	\put(14,27){$\Lambda_{b}$ }
	\put(75,36){$K$} 
	\put(60,22){$\bar{D}^*$}
\end{overpic}
\end{center}
\end{minipage}
}
\subfigure[]
{
\begin{minipage}[t]{0.45\linewidth}
\begin{center}
\begin{overpic}[scale=.48]{triangle.eps}
	\put(70,7){$ \textcolor{black}{P_{\psi}^{  1/2}}$}
	\put(38,8){$\Lambda_{c}$}		
	\put(34,36){$D_{s}^*$}	
	\put(14,27){$\Lambda_{b}$ }
	\put(75,36){$K$} 
	\put(60,22){$\bar{D}$}
\end{overpic}
\end{center}
\end{minipage}
}
\caption{Triangle diagrams accounting for $\Lambda_b\to D_s^{(\ast)}\Lambda_{c}\to 	 \textcolor{black}{P_{\psi}^{  1/2}} K$ with $\textcolor{black}{P_{\psi}^{  1/2}}$ representing one of the pentaquark molecules of spin $1/2$   }
\label{triangle2}
\end{center}
\end{figure}

\begin{figure}[ttt]
\begin{center}
\subfigure[]
{
\begin{minipage}[t]{0.45\linewidth}
\begin{center}
\begin{overpic}[scale=.48]{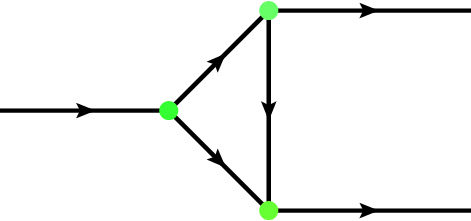}
	\put(70,7){$ \textcolor{black}{P_{\psi}^{  3/2}}$}
	\put(38,8){$\Lambda_{c}$}		
	\put(35,36){$D_{s}$}	
	\put(14,27){$\Lambda_{b}$ }
	\put(75,36){$K$} 
	\put(60,22){$\bar{D}^*$}
\end{overpic}
\end{center}
\end{minipage}
}
\subfigure[]
{
\begin{minipage}[t]{0.45\linewidth}
\begin{center}
\begin{overpic}[scale=.48]{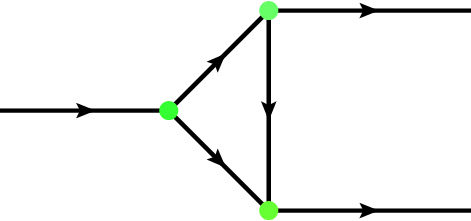}
	\put(70,7){$ \textcolor{black}{P_{\psi}^{  3/2}}$}
	\put(38,8){$\Lambda_{c}$}		
	\put(34,36){$D_{s}^*$}	
	\put(14,27){$\Lambda_{b}$ }
	\put(75,36){$K$} 
	\put(60,22){$\bar{D}^*$}
\end{overpic}
\end{center}
\end{minipage}
}
\caption{Triangle diagrams accounting for $\Lambda_b\to D_s^{(\ast)}\Lambda_{c}\to  \textcolor{black}{P_{\psi}^{  3/2}} K$ with $\textcolor{black}{P_{\psi}^{  3/2}} $ denoting one of  the pentaquark molecule of spin $3/2$.     }
\label{triangle3}
\end{center}
\end{figure}

In this work, we employ the triangle diagrams to describe the productions  of pentaquark molecules. We suppose that the colour favored weak decays $\Lambda_b\to  \Lambda_c D_{s}^{(*)-}$ are responsible for the short-range interactions because the branching fractions $\mathcal{B}(\Lambda_b\to  \Lambda_c D_{s}^{(*)-})$ are large  among the  nonleptonic decays of $\Lambda_b$.     Then the $D_{s}^{(\ast)-}$ mesons { scatter } into $\bar{D}^{(\ast)}$ and $K$ mesons, and the   pentaquark molecules with spin $1/2$ and $3/2$ are dynamically  generated via the $\bar{D}^{(\ast)} \Lambda_c$ interactions  as shown in Fig.~\ref{triangle2} and Fig.~\ref{triangle3}, respectively,     where $\textcolor{black}{P_{\psi}^{  1/2}}$ and $\textcolor{black}{P_{\psi}^{  3/2}}$ denote the  pentaquark molecules  of spin $1/2$ and $3/2$, respectively. As shown in a number of previous studies~\cite{Liu:2019tjn,Xiao:2019aya,Du:2019pij,Yamaguchi:2019seo,PavonValderrama:2019nbk,Lin:2019qiv,He:2019rva,Yalikun:2021bfm,Dong:2021juy,Zhang:2023czx} and also explicitly shown later,  there exists a complete  multiplet of hidden-charm pentaquark molecules dominantly generated by the $\bar{D}^{(*)}\Sigma_{c}^{(*)}$ interactions.    We denote the seven pentaquark  molecules  as $\textcolor{black}{P_{\psi 1}^{N}}$, $\textcolor{black}{P_{\psi 2}^{N}}$, ... ,$\textcolor{black}{P_{\psi 7}^{N}}$, following the order of  Scenario A of Table I in Ref.~\cite{Liu:2019tjn}.     {  It should be noted that such order specifies the spin of pentaquark molecules, i.e., $P_{\psi 3}^{N}$ and $P_{\psi 4}^{N}$ represent the pentaquark molecules of spin $1/2$ and $3/2$, respectively.     In this work, we study two scenarios A and B corresponding to different spin assignments of these pentaquark molecules.   In Scenario A,    $P_{\psi 3}^{N}$ and $P_{\psi 4}^{N}$ represent $P_{\psi}^{N}(4440)$ and $P_{\psi}^{N}(4457)$, while they represent  $P_{\psi}^{N}(4457)$  and $P_{\psi}^{N}(4440)$ in Scenario B.  $P_{\psi 1}^{N}$ represents  $P_{\psi}^{N}(4312)$ in both Scenario A and Scenario B.     }  
Considering only $S$-wave  $\bar{D}^{(*)}\Lambda_c$ interactions,  { the production of the pentaquark molecule of spin $5/2$ is not allowed by the mechanisms shown in either Fig.~\ref{triangle2} or Fig.~\ref{triangle3}, which indicate that $P_{\psi 7}^{N}$ can not be produced in our model.   }   Therefore, we only focus on the productions of  the remaining  six pentaquark molecules in the $\Lambda_b$ decays in this work.

\subsection {Effective Lagrangians}

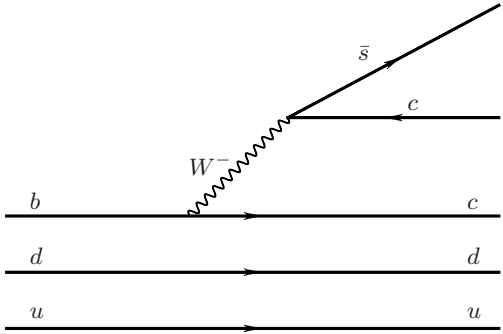
\begin{figure}[!h]
\begin{center}
\begin{overpic}[scale=.4]{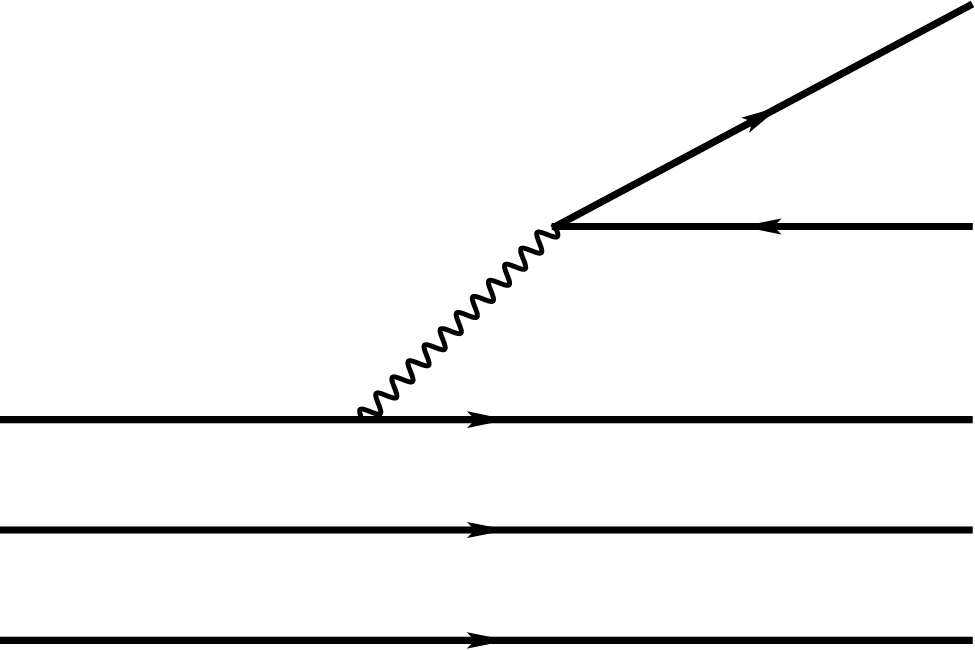}
		\put(5,25){$b$ }
		\put(5,3){$u$}	
		\put(93,3){$u$}	
				\put(5,14){$d$}	
		\put(93,14){$d$}
		\put(93,25){$c$}	\put(81,45){$c$} \put(71,55){$\bar{s}$}
    	\put(37,32){$W^-$}
\end{overpic}
  \caption{External $W$-emission  accounting for $\Lambda_{b}\to D_{s}^{(*)}\Lambda_{c}$ at quark level.}
  \label{quarklambdab}
\end{center}
\end{figure}

In this work,   we adopt the effective  Lagrangian approach to calculate the  triangle diagrams of Figs.~\ref{triangle2} and ~\ref{triangle3}. In the following, we spell out the relevant Lagrangians.

First, we focus on the weak decays of $\Lambda_b\to  \Lambda_c D_{s}^{(\ast)-}$.   
At quark level, the decays of $\Lambda_b\to  \Lambda_c D_{s}^{(*)-}$   can occur via the external $W$-emission mechanism  shown in Fig.~\ref{quarklambdab}, which is usually the largest in terms of the   topological classification of  weak decays~\cite{Chau:1982da,Chau:1987tk,Molina:2019udw}. 
  As shown in Ref.~\cite{Wu:2023rrp}, the color favored weak decays $B\to D_{s}^{(\ast)} {D}^{(\ast)}$  are significant   to produce the $\bar{D}^*D^{(*)}$ molecules in $B$ decays, which  share similar topologies to the weak decays $\Lambda_b\to  \Lambda_c D_{s}^{(*)-}$  at quark level.  

The  effective Hamiltonian  describing the weak decays of $\Lambda_b\to  \Lambda_c D_{s}^{(*)-}$   has the following form
\begin{equation}
\mathcal{H}_{eff}=\frac{G_F}{\sqrt{2}}V_{cb}V_{cs}[c_1(\mu)\mathcal{O}_{1}(\mu)+c_2(\mu)\mathcal{O}_{2}(\mu)]+h.c.
\end{equation}
where $G_F $ is the Fermi constant, $V_{bc}$ and $V_{cs}$ are the Cabibbo-Kobayashi-Maskawa~(CKM) matrix elements, $c_{1,2}(\mu)$ are the  Wilson coefficients, and $\mathcal{O}_{1}(\mu)$ and $\mathcal{O}_{2}(\mu)$ are the four-fermion operators of  $(s\bar{c})_{V-A}(c\bar{b})_{V-A}$ and $(\bar{c}c)_{V-A}(s\bar{b})_{V-A}$ with $(\bar{q}q)_{V-A}$ standing for $\bar{q}\gamma_\mu(1-\gamma_5)q$~\cite{Cheng:1996cs,Ali:1998eb,Li:2012cfa}. The Wilson coefficients $c_{1,2}(\mu)$ include the short-distance  quantum chromodynamics (QCD) dynamic scaling from $\mu=M_{W}$ to $\mu=m_{c}$.

In the naive factorisation approach~\cite{Bauer:1986bm}, the amplitudes of $\Lambda_b\to  \Lambda_c D_{s}^{(\ast)-}$  can be expressed  as the products of two current hadronic matrix elements
\begin{eqnarray}
&&\mathcal{A}\left(\Lambda_b\to  \Lambda_c D_{s}^{-}\right)\nonumber\\
&&=\frac{G_{F}}{\sqrt{2}} V_{cb}V_{cs} a_{1}\left\langle D_{s}^{-}|(s\bar{c})| 0\right\rangle\left\langle \Lambda_c|(c \bar{b})| \Lambda_b\right\rangle  \label{Ds-KK}  \\
&&\mathcal{A}\left(\Lambda_b\to  \Lambda_c D_{s}^{\ast-}\right)\nonumber\\
&&=\frac{G_{F}}{\sqrt{2}} V_{cb}V_{cs} a_{1}\left\langle D_{s}^{\ast-}|(s\bar{c})| 0\right\rangle \left\langle \Lambda_c|(c \bar{b})| \Lambda_b\right\rangle  \label{Ds-KK1}
\end{eqnarray}
where the effective Wilson coefficient $a_{1}$ is expressed as   $a_{1}=c_{1}(\mu)+c_{2}(\mu) / N_{c}$  with $N_c=3$ the number of colors~\cite{Bauer:1986bm,Li:2012cfa}.

 The  matrix elements between a pseudoscalar meson or vector meson and the vacuum have the following form:
\begin{eqnarray}
&&\left\langle D_{s}^{-}|(s \bar{c})| 0\right\rangle  =i\, f_{D_{s}^{-}} p^{\mu}_{D_{s}^{-}}\,, \\
&&\left\langle D_{s}^{\ast-} |(s\bar{c} )| 0\right\rangle = m_{D_{s}^{\ast-}}f_{D_{s}^{\ast-}}\epsilon_\mu^* \,.
\end{eqnarray}
where $f_{D_{s}^{-}}$ and $f_{D_{s}^{\ast-}}$  are the decay constants for $D_s^{-}$ and $D_s^{\ast-}$, respectively, and $\epsilon_\mu^*$ denotes the polarization vector of $D_s^{\ast-}$.

The $\Lambda_b\to\Lambda_c$ transition form factors  are parameterized as follows~\cite{Gutsche:2018utw} 
\begin{eqnarray}
&&\langle B(p^{\prime})|V_\mu-A_\mu|B(p)\rangle\\
&&=\bar{u}(p^{\prime})[f_{1}^V(q^2)\gamma_{\mu}-f_2^V(q^2)\frac{i\sigma_{\mu\nu}q^{\nu}}{
m}+f_3^V(q^2)\frac{q^{\mu}}{m} \nonumber \\ 
&& -(f_{1}^A(q^2)\gamma_{\mu}-f_2^A(q^2)\frac{i\sigma_{\mu\nu}q^{\nu}}{
m}+f_3^A(q^2)\frac{q^{\mu}}{m})\gamma^5
]u(p) \nonumber
\end{eqnarray}
where $\sigma^{\mu\nu}=\frac{i}{2}(\gamma^\mu\gamma^\nu-\gamma^\nu\gamma^\mu)$ and $q=p-p^{\prime}$.
As a result, the weak decays  $\Lambda_b\rightarrow\Lambda_c {D}_s^{(*)}$  can be characterised by the following Lagrangian~\cite{Cheng:1996cs}:
\begin{eqnarray}
\label{Eq:weakdecayV1}
\mathcal{L}_{\Lambda_b\Lambda_cD_s}&=&i\bar{\Lambda}_c(A+B \gamma_5)\Lambda_bD_s\,,\\
\mathcal{L}_{\Lambda_b\Lambda_cD_s^*}&=&\bar{\Lambda}_c(A_1\gamma_\mu\gamma_5+A_2\frac{p_{2\mu}}{m}\gamma_5+B_1\gamma_\mu +B_2\frac{p_{2\mu}}{m})\Lambda_bD_s^{*\mu}\,.\nonumber
\end{eqnarray}
where  $A_1$, $A_2$, $B_1$, $B_2$, $A$, and $B$  are:
\begin{equation}\label{Eq:weakdecayV2}
\begin{split}
    A&=\lambda f_{D_s}[(m-m_2)f_1^V+\frac{m_1^2}{m}f_3^V]\,, \\
    B&=\lambda f_{D_s}[(m+m_2)f_1^A-\frac{m_1^2}{m}f_3^A]\,, \\
    A_1&=-\lambda f_{D_s^*}m_1[f_1^A-f_2^A\frac{m-m_2}{m}]\,, \\
    B_1&=\lambda f_{D_s^*}m_1[f_1^V+f_2^V\frac{m+m_2}{m}]\,, \\
    A_2&=2\lambda f_{D_s^*}m_1f_2^A\,,\\
    B_2&=-2\lambda f_{D_s^*}m_1f_2^V \,.
\end{split}
\end{equation}
with $\lambda=\frac{G_F}{\sqrt{2}}V_{cb}V_{cs}a_1$ and $m, m_1, m_2$ referring  to the masses of $\Lambda_b$, ${D}_s^{(*)}$, and $\Lambda_c$, respectively. The form factors can be expressed in a double-pole  form: 
\begin{equation}
   f_i^{V/A}(q^2)=F_i^{V/A}(0)\frac{F_{0}}{1-a\, \varphi+ b\, \varphi^2}
\end{equation}
with $\varphi=q^2/m^2$. The values of $F_0$, $a$ and $b$  in the $\Lambda_b\rightarrow\Lambda_c$ transition  form factors are taken from  Ref.~\cite{Gutsche:2015mxa} and shown in Table~\ref{Tab:FormFactor}.

\begin{table}[ttt]
 \centering
 \caption{Values of  $F(0)$, $a$, $b$ in the $\Lambda_b\rightarrow\Lambda_c$ transition  form factors~\cite{Gutsche:2015mxa}. \label{Tab:FormFactor} }
 \begin{tabular}{ccccccc}
 \hline\hline
   & $F_1^V$~~~ & $F_2^V$~~~ & $F_3^V$~~~ & $F_1^A$~~~ &  $F_2^A$~~~ & $F_3^A$~~~\\
 \hline
 $F(0)$~~~ &  0.549~~~  & 0.110~~~ & $-0.023$~~~ & 0.542~~~ & 0.018~~~ & $-0.123$~~~\\
 $a$~~~ &  1.459~~~  & 1.680~~~ & 1.181~~~ & 1.443~~~ & 0.921~~~ & 1.714~~~\\
 $b$~~~ & 0.571~~~  & 0.794~~~ & 0.276~~~ &0.559~~~ & 0.255~~~ & 0.828~~~\\
\hline\hline
 \end{tabular}
 \end{table}

In this work, we take  $G_F = 1.166 \times 10^{-5}~{\rm GeV}^{-2}$, $V_{cb}=0.041$, $V_{cs}=0.987$, $f_{D_{s}^-} = 250$ MeV, and $f_{D_{s}^{\ast-}}=272$~MeV as in Refs.~\cite{ParticleDataGroup:2020ssz,Verma:2011yw,Gutsche:2018utw,FlavourLatticeAveragingGroup:2019iem}. We note that the value of $a_1$  as a function of  the energy scale $\mu$ differs from process to process~\cite{Chen:1999nxa,Cheng:2010ry}.  Therefore, we take  the branching fraction   $\mathcal{B}(\Lambda_b \to \Lambda_c D_s^{-})=(1.10\pm 0.10) \%$ to determine the effective  Wilson coefficient to be $a_1=0.883$.    We note that in Ref.~\cite{Wu:2023rrp}, the effective Wilson Coefficient $a_1$ are determined to be 0.79 and 0.81   by reproducing the branching fractions of the decays $B^+ \to \bar{D}^0 D_{s}^+ $ and $B^+ \to \bar{D}^0 D_{s}^{\ast+ }$, respectively. These values are consistent with the value obtained from the   weak decay $\Lambda_b \to \Lambda_c D_s^{-}$, showing that  the naive factorisation approach works well for the external $W$-emission mechanism.       Due to the non availability of   experimental data for the branching fraction   $\mathcal{B}(\Lambda_b \to \Lambda_c D_s^{*-})$, we assume that the effective  Wilson coefficient in Eq.(\ref{Ds-KK1}) is the same as that in Eq.(\ref{Ds-KK}).  With the so obtained effective  Wilson coefficient $a_1$ we  predict  the branching fraction  $\mathcal{B}(\Lambda_b \to \Lambda_c D_s^{*-})=(2.47\pm 0.26)\%$, consistent with the results of Refs.~\cite{Gutsche:2018utw,Chua:2019yqh}.  
As a matter of fact, the experimental  branching fraction  $\mathcal{B}(\Lambda_b \to \Lambda_c D_s^{-})$ helps reduce the uncertainty in the weak vertices. 

The Lagrangians describing the $D_{s}^{(*)}$ mesons { scattering}  into $\bar{D}^{(*)}$ and $K$ mesons  are:  
\begin{eqnarray}\label{efflag}
    \mathcal{L}_{KD_sD^*}&=&ig_{KD_sD^*}D^{*\mu}[\bar{D}_s\partial_{\mu}K-(\partial_\mu \bar{D}_s)K]+H.c. \nonumber\\
    \mathcal{L}_{KD_s^*D^*}&=&-g_{KD_s^*D^*} \epsilon^{\mu\nu\alpha\beta}[\partial_\mu\bar{D}_\nu^*\partial_\alpha D^*_{s\beta}\bar{K} \nonumber\\
    & &+\partial_\mu D_\nu^*\partial_\alpha\bar{D}^*_{s\beta}K] +H.c.
\end{eqnarray}
where  $g_{KD_sD^*}$ and $g_{KD_s^*D^*}$ are the kaon couplings to $D_{s}D^{\ast}$ and $D_{s}^{\ast}D^{*}$, respectively.  Unfortunately, there exists  no experimental data to determine the values of these couplings.     The coupling $g_{D_{s} D^{\ast} K}$ is estimated to be  $16.6$ and $10$ assuming  SU(3)-flavor symmetry~\cite{Xie:2022lyw} and SU(4)-flavor symmetry~\cite{Azevedo:2003qh}, respectively,  while the QCD sum rule yields $5$~\cite{Bracco:2006xf,Wang:2006ida}.  
 In view of this large variance, we  adopt the couplings estimated by SU(4) symmetry, which are in between those estimated  utilizing SU(3) symmetry and by the QCD sum rule, i.e.,  $g_{D_{s} D^{\ast} K}=g_{D_{s}^{\ast} D K}=10$ and  $g_{D_{s}^{\ast} D^{\ast} K}=7.0$~GeV$^{-1}$~\cite{Azevedo:2003qh}.

The effective Lagrangians describing the interactions between   pentaquark  molecules and their constituents  $\bar{D}^{(*)}\Lambda_{c}$ are written as 
\begin{eqnarray}\label{efflag}  &&\mathcal{L}_{\textcolor{black}{P_{\psi}^{  1/2}}\Lambda_c\bar{D}}=g_{\textcolor{black}{P_{\psi}^{ 1/2}}\Lambda_c\bar{D}}\textcolor{black}{P_{\psi}^{ 1/2}}\Lambda_c\bar{D}\,, \nonumber\\
&&\mathcal{L}_{\textcolor{black}{P_{\psi}^{ 1/2}}\Lambda_c\bar{D}^*}=g_{\textcolor{black}{P_{\psi}^{ 1/2}}\Lambda_c\bar{D}^*} \bar{\Lambda}_c\gamma_5(g_{\mu\nu}-\frac{p_{\mu}p_{\nu}}{m_{\textcolor{black}{P_{\psi}^{ 1/2}}}^2})\gamma^\nu
\textcolor{black}{P_{\psi}^{ 1/2}}D^{*\mu}\,, \nonumber\\
&&\mathcal{L}_{\textcolor{black}{P_{\psi}^{  3/2}}\Lambda_c\bar{D}^*}=g_{\textcolor{black}{P_{\psi}^{  3/2}}\Lambda_c\bar{D}^*} \bar{\Lambda}_c \textcolor{black}{P_{\psi\mu}^{  3/2}} D^{*\mu}\,.
\end{eqnarray}
where  $g_{\textcolor{black}{P_{\psi}^{  1/2}}\Lambda_c\bar{D}}$, $g_{\textcolor{black}{P_{\psi}^{  1/2}}\Lambda_c\bar{D}^*}$, and $g_{\textcolor{black}{P_{\psi}^{  3/2}}\Lambda_c\bar{D}^*}$ are the couplings of the $\textcolor{black}{P_{\psi}^{  1/2}}$ and $\textcolor{black}{P_{\psi}^{  3/2}}$ pentaquark  molecules  to their constituents  $\bar{D}^{(*)}\Lambda_c$. One should note that although these pentaquark  molecules  are dominantly generated by the $\bar{D}^{(*)}\Sigma_{c}^{(*)}$ interactions~\cite{Liu:2019tjn}, the $\bar{D}^{(*)}\Lambda_c$ and $J/\psi(\eta_c) p$ coupled channels also play a relevant role~\cite{Guo:2019kdc,Xiao:2019aya,Lin:2019qiv,Yamaguchi:2019seo,He:2019rva,Burns:2021jlu}. Below, we estimate the couplings of the pentaquark molecules  to their constituents $\bar{D}^{(*)}\Lambda_c$ and $J/\psi(\eta_c) p$  by the contact range EFT approach, which is widely applied to study the dynamical  generation of hadronic molecules~\cite{Ji:2022vdj,Pan:2022whr}.

\subsection{Contact-range EFT approach }

In this subsection, we introduce the contact-range EFT approach.   
The  scattering amplitude $T$ is responsible for the dynamical generation of the pentaquark molecules via  the   Lippmann-Schwinger equation
\begin{eqnarray}
T(\sqrt{s})=(1-VG(\sqrt{s}))^{-1}V,
\label{lsequation}
\end{eqnarray}
where $V$ is the coupled-channel potential determined by the contact-range EFT approach (see Appendix~\ref{appendix B}), and $G(\sqrt{s})$ is the two-body propagator. In this work, we consider the following coupled channels $\bar{D}^*\Sigma_c^*-\bar{D}^*\Sigma_c-\bar{D}\Sigma_c-\bar{D}^*\Lambda_c-\bar{D}\Lambda_c-J/\psi p-\eta_c p$ with $J^{P}=1/2^{-}$ and $\bar{D}^*\Sigma_c^*-\bar{D}^*\Sigma_c-\bar{D}^*\Lambda_c-J/\psi p$ with $J^{P}=3/2^{-}$. Since  the mass splitting between $\bar{D}^*\Sigma_c^*$ and $\eta_c p$ is about 600 MeV, we take a relativistic propagator:
\begin{eqnarray}
G(\sqrt{s})= 2m_{1}\int \frac{d^{3}q}{(2\pi)^{3}}\frac{\omega_{1}+\omega_{2}}{2~\omega_{1}\omega_{2}} \frac{F(q^2,k)}{(\sqrt{s})^2-(\omega_1+\omega_2)^2+i \varepsilon}
\label{loopfunction}
\end{eqnarray}
where   $\sqrt{s}$ is  the total energy in the center-of-mass (c.m.) frame of $m_{1}$ and $m_{2}$,   $\omega_{i}=\sqrt{m_{i}^2+q^2}$ is the energy of the particle   and   the c.m. momentum $k$ is 
\begin{eqnarray}
k(\sqrt{s})=\frac{\sqrt{{\sqrt{s}^2}-(m_{1}+m_{2})^2}\sqrt{\sqrt{s}^2-(m_{1}-m_{2})^2}}{2{\sqrt{s}}}.
\end{eqnarray}
A regulator of Gaussian form  $F(q^2,k)=e^{-2q^2/\Lambda^2}/e^{-2k^2/\Lambda^2}$ is used to regulate the loop function. We note that the  loop function can also be  regularized by other methods such as  the  momentum cut off scheme and  dimensional regularization scheme~\cite{Oset:1997it,Jido:2003cb,Wu:2010jy,Hyodo:2011ur,Debastiani:2017ewu}. 

The dynamically generated pentaquark molecules  correspond to poles on  the unphysical sheet, which is defined as~\cite{Oller:1997ti,Roca:2005nm},   
\begin{eqnarray}
G_{II}(\sqrt{s})=G_{I}(\sqrt{s}){+~i\frac{2m_1 }{4\pi} \frac{k(\sqrt{s})}{\sqrt{s}}},
\end{eqnarray}
where  $m_1$ stands for  the mass of the  baryon.

With the potentials obtained in Eq.~(\ref{contact11}) and Eq.~(\ref{contact22}) of the Appendix,  we  search for  poles in  the vicinity of the $\bar{D}^{(*)}\Sigma_{c}^{(*)}$  channels, and then determine the 
 couplings between the pentaquark molecules  and their constituents from  the residues of the corresponding poles, 
\begin{eqnarray}
g_{i}g_{j}=\lim_{{\sqrt{s}}\to {\sqrt{s_0}}}\left({\sqrt{s}}-{\sqrt{s_0}}\right)T_{ij}(\sqrt{s}),
\end{eqnarray}
where $g_{i}$ denotes the coupling of channel $i$ to the  dynamically generated molecules and ${\sqrt{s_0}}$ is the pole position.

Using  the couplings $g_i$ obtained above,   one can estimate the  partial decay  widths of the pentaquark molecules \cite{Yu:2018yxl} 
\begin{eqnarray}
 \Gamma_{i}= g_{i  }^2 \frac{1}{2\pi} \frac{m_{i}}{m_{\textcolor{black}{P_{\psi}^{ N}}}} p_{i} 
 \label{partialwiths}
\end{eqnarray}
where  $m_{i}$ stands for the mass of the baryon of channel $i$, $m_{\textcolor{black}{P_{\psi}^{ N}}}$ is the mass of the pentaquark molecule (the real part of the pole position), and $p_{i}$ is the momentum of the baryon (meson) of channel $i$ in the $\textcolor{black}{P_{\psi}^{ N}}$ rest frame.

\subsection {Decay Amplitudes}

With the above effective Lagragians, we obtain the following decay amplitudes for  $\Lambda_b \to \textcolor{black}{P_{\psi}^{ 1/2}} K$ of  Fig.~\ref{triangle2}  
\begin{widetext}

\begin{eqnarray}
\begin{split}
\mathcal{M}_{1}^{a}=& i^3 \int\frac{d^4 q}{(2\pi)^4}[g_{\textcolor{black}{P_{\psi}^{ 1/2}}\Lambda_c \bar{D}^\ast}\bar{u}(p_2)\gamma^\nu \gamma_5(g_{\mu\nu}-\frac{p_{2\mu}p_{2\nu}}{m^2_{\textcolor{black}{P_{\psi}^{ 1/2}}}})(q_2\!\!\!\!\!\slash+m_2)\,i(A+B\gamma_5)u(k_0)] \\
&[-g_{KD^\ast D_s}(q_1+p_1)_\alpha](-g^{\mu\alpha}+\frac{q^\mu q^\alpha}{m^2_E}) \frac{1}{q^2_1-m^2_1}\frac{1}{q^2_2-m^2_2}\frac{1}{q^2-m^2_E}\, ,\\
\mathcal{M}_{1}^{b} =& i^3\int\frac{d^4 q}{(2\pi)^4}[g_{\textcolor{black}{P_{\psi}^{ 1/2}} \Lambda_c \bar{D}^\ast}\bar{u}(p_2)\gamma^\nu \gamma_5(g_{\mu\nu}-\frac{p_{2\mu}p_{2\nu}}{m^2_{\textcolor{black}{P_{\psi}^{ 1/2}}}})(q_2\!\!\!\!\!\slash+m_2)\\
&[(A_1 \gamma_\alpha \gamma_5+A_2\frac{q_{2\alpha}}{m}\gamma_5+B_1 \gamma_\alpha+B_2\frac{q_{2\alpha}}{m})u(k_0)][-g_{KD^\ast D^\ast_s}\varepsilon_{\rho\lambda\eta\tau}q^\rho q^\eta_1]\\
&(-g^{\mu\lambda}+\frac{q^{\mu}q^{\lambda}}{m^2_E})(-g^{\alpha\tau}+\frac{q_1^{\alpha}q_1^{\tau}}{m^2_1}) \frac{1}{q^2_1-m^2_1}\frac{1}{q^2_2-m^2_2}\frac{1}{q^2-m^2_E}\, ,  \\
\mathcal{M}_{1}^{c} =& i^3\int\frac{d^4 q}{(2\pi)^4}[g_{\textcolor{black}{P_{\psi}^{ 1/2}}\Lambda_c \bar{D}}\bar{u}(p_2)(q_2\!\!\!\!\!\slash+m_2)(A_1 \gamma_\alpha \gamma_5+A_2\frac{q_{2\alpha}}{m}\gamma_5+B_1 \gamma_\alpha+B_2\frac{q_{2\alpha}}{m})u(k_0)]\\
&[-g_{KD D^\ast_s}(-q+p_1)_{\tau} ] (-g^{\alpha\tau}+\frac{q_1^{\alpha}q_1^{\tau}}{m^2_1}) \frac{1}{q^2_1-m^2_1}\frac{1}{q^2_2-m^2_2}\frac{1}{q^2-m^2_E} \, .
\end{split}
\end{eqnarray}
    
\end{widetext}
where $k_0$, $q_1$,  $q_2$, $q$,  $p_1$, and $p_2$ refer to the momenta of $\Lambda_b$, $D_{s}^{(\ast)}$, $\Lambda_c$, $\bar{D}^{(\ast)}$, $K$,  and $\textcolor{black}{P_{\psi}^{ 1/2}} $, respectively, and $\bar{u}(p_2)$ and $u(k_0)$ represent the spinors of $\textcolor{black}{P_{\psi}^{ 1/2}} $ and $\Lambda_b$.  Similarly, we write the decay amplitudes for  $\Lambda_b \to \textcolor{black}{P_{\psi}^{ 3/2}} K$ of  Fig.~\ref{triangle3}  as follows
\begin{widetext}
\begin{eqnarray}\label{diade}
\begin{split}
\mathcal{M}_{3}^{a}=&i^3 \int\frac{d^4 q}{(2\pi)^4}[g_{\textcolor{black}{P_{\psi}^{ 3/2}}\Lambda_c \bar{D}^\ast}\bar{u}_\mu(p_2)](q_2\!\!\!\!\!\slash+m_2)[i(A+B\gamma_5)u(k_0)]\\
&[-g_{KD^\ast D_s}(q_1+p_1)_\nu](-g^{\mu\nu}+\frac{q^{\mu}q^{\nu}}{m^2_E})\,\frac{1}{q^2_1-m^2_1}\frac{1}{q^2_2-m^2_2}\frac{1}{q^2-m^2_E}\,,\\
\mathcal{M}_{3}^{b}=&i^3\int\frac{d^4 q}{(2\pi)^4}[g_{\textcolor{black}{P_{\psi}^{ 3/2}}\Lambda_c \bar{D}^\ast}\bar{u}_{\mu}(p_2)(q_2\!\!\!\!\!\slash+m_2)(-i)(A_1 \gamma_\alpha \gamma_5+A_2\frac{q_{2\alpha}}{m}\gamma_5+B_1 \gamma_\alpha+B_2\frac{q_{2\alpha}}{m})u(k_0)]\\
&[-g_{KD^\ast D^\ast_s}\varepsilon_{\rho\lambda\eta\tau}q^\rho q^\eta_1](-g^{\mu\lambda}+\frac{q^{\mu}q^{\lambda}}{m^2_E})(-g^{\alpha\tau}+\frac{q_1^{\alpha}q_1^{\tau}}{m^2_1}) \frac{1}{q^2_1-m^2_1}\frac{1}{q^2_2-m^2_2}\frac{1}{q^2-m^2_E}\,. 
\end{split}
\end{eqnarray}
\end{widetext}

 With  the amplitudes  for the decays of  $\Lambda_b \to \textcolor{black}{P_{\psi}^{ 1/2}}  K$  and $\Lambda_b \to \textcolor{black}{P_{\psi}^{ 3/2}} K$  given above, 
 one can compute the corresponding partial decay widths 
 \begin{eqnarray}
\Gamma=\frac{1}{2J+1}\frac{1}{8\pi}\frac{|\vec{p}|}{m_{\Lambda_b}^2}{|\overline{M}|}^{2}
\end{eqnarray}
where $J$ is the total angular momentum of the initial $\Lambda_b$ baryon and $|\vec{p}|$ is the momentum of either final state in the rest frame of  the $\Lambda_b$ baryon. 

 Regarding the three-body decays of the pentaquark molecules, we have systematically investigated two decay modes:  tree diagrams     and triangle diagrams, and found  that  the
 former 
  can almost saturate their total three-body   decay widths~\cite{Xie:2022hhv}.  
In this work, with the new  couplings between the pentaquark molecules and  $\bar{D}^{(*)} \Sigma_{c}^{(*)}$,  we  update the widths of the three-body   decays 
$ \textcolor{black}{P_{\psi}^{ N}} \to \bar{D}^{(*)}\Lambda_c \pi$, where   these   hidden-charm pentaquark molecules  decay into  $\bar{D}^{(*)}\Lambda_c \pi$    via the   off-shell $\Sigma_{c}^{(*)}$ baryons decaying into  $\Lambda_c \pi$.  The details for the calculations  can be found in our previous work~\cite{Xie:2022hhv}.

\section{Results and discussions}
\label{results}

Because the three-body partial  decay widths  of the pentaquark states \textcolor{black}{ $P_{\psi}^{N}(4312)$}, \textcolor{black}{$P_{\psi}^{N}(4440)$}, and \textcolor{black}{$P_{\psi}^{N}(4457)$} as hadronic molecules are  less than $1$ MeV,  we can   neglect their  three-body decays and assume that  their two-body  decays saturate their total  widths. Therefore, we suppose that these three  pentaquark molecules are dynamically generated via the $\bar{D}^{(\ast)}\Sigma_{c}^{(\ast)}$, $\bar{D}^{(\ast)}\Lambda_{c}$, $\eta_{c}N$ and $J/\psi N$ coupled-channel potentials. In the heavy quark limit, the contact potentials of this   coupled-channel  system are parameterized by  seven   parameters as shown in  Eq.(\ref{contact11}) and Eq.(\ref{contact22}). In this work, we set the potential $V_{J/\psi(\eta_{c}) N \to J/\psi(\eta_{c}) N }=0$, resulting  in six  unknown parameters.   The unknown   couplings  of the $\bar{D}^{(*)}\Sigma_{c}^{(*)} \to \bar{D}^{(*)}\Sigma_{c}^{(*)}$  potentials are well described by the  light meson saturation approach~\cite{Liu:2019zvb},   which   is widely applied  to study heavy hadronic molecules~\cite{Peng:2020xrf,Dong:2021juy,Peng:2021hkr,Dong:2021bvy}. 
Therefore, we expect that the light meson saturation approach(see Appendix) is also valid  for the 
   $\bar{D}^{(*)}\Lambda_{c}\to \bar{D}^{(*)}\Lambda_{c}$ interaction. With the $\bar{D}^{(*)}\Sigma_{c}^{(*)} \to \bar{D}^{(*)}\Sigma_{c}^{(*)}$  potentials determined in Ref.~\cite{Liu:2019tjn}, we obtain the $\bar{D}^{(*)}\Lambda_{c}\to \bar{D}^{(*)}\Lambda_{c}$ potentials using the light meson saturation approach, and then  search for poles  near the $\bar{D}^{(*)}\Lambda_{c}$ threshold, but find none, indicating that there exists no genuine states generated by the $\bar{D}^{(*)}\Lambda_{c}$ interactions, consistent with Refs.~\cite{Yang:2011wz,Xiao:2019aya,He:2019rva}. Very recently, Duan et al. argued that there exist the enhancements at the $\bar{D}^{(*)}\Lambda_{c}$ thresholds induced by the triangle and box singularities~\cite{Duan:2023dky}.   Therefore, even taking into account the  $\bar{D}^{(*)}\Lambda_{c}$, $\eta_{c}N$, and $J/\psi N$    channels, the number of hidden-charm pentaquark molecules does not change. These channels  mainly affect the imaginary part of the pole positions, i.e., the widths of the pentaquark states.

\begin{table}[ttt]
\setlength{\tabcolsep}{4pt}
\centering
\caption{Masses and quantum numbers of  hadrons relevant to  this work~\cite{ParticleDataGroup:2020ssz}. \label{mass}}
\begin{tabular}{ccc|ccc}
  \hline\hline 
   Hadron & $I (J^P)$ & M (MeV) & Hadron & $I (J^P)$ & M (MeV)    \\
  \hline
        $p$ & $\frac{1}{2}(1/2^+)$ & $938.27$  & 
      $n$ & $\frac{1}{2}(1/2^+)$ & $939.57$  \\
        $\Sigma_{c}^{++}$ & $1(1/2^+)$ & $2453.97$ & 
      $\Sigma_{c}^{+}$ & $1(1/2^+)$ & $2452.65$ \\
  $\Sigma_{c}^{\ast ++}$ & $1(3/2^+)$ & $2518.41$ & 
      $\Sigma_{c}^{\ast+}$ & $1(3/2^+)$ & $2517.4$ \\
         $\Sigma_{c}^{0}$ & $1(1/2^+)$ & $2453.75$  & 
      $\Sigma_{c}^{\ast0}$ & $1(3/2^+)$ & $2518.48$   \\ 
      $\Lambda_{c}^{+}$ & $0(1/2^+)$ & $2286.46$   &  $\Lambda_b$ & $0(1/2^+)$ & $5619.60$  \\
        $\pi^{\pm}$ & $1(0^-)$ & $139.57$ & 
      $\pi^{0}$ & $1(0^-)$ & $134.98$ \\
     $K^{\pm}$ & $1/2(0^-)$ & $493.677$ & 
      $K^{0}$ & $1/2(0^-)$ & $497.611$ \\
   $\bar{D}^{0}$ & $\frac{1}{2}(0^-)$ & $1864.84$  &    $D^{-}$ & $\frac{1}{2}(0^-)$ & $1869.66$  \\
  $\bar{D}^{\ast0}$ & $\frac{1}{2}(1^-)$ & $2006.85$ &  $D^{\ast-}$ & $\frac{1}{2}(1^-)$ & $2010.26$   \\
    $D_s^{\pm}$ & $0(0^-)$ & $1968.35$& 
$D_s^{\ast\pm}$ & $0(1^-)$ & $2112.2$    \\
  $J/\psi$ & $0(1^-)$ & $3096.90$& 
  $\eta_{c}$ & $0(0^-)$ & $2983.90$    \\
 \hline \hline 
\end{tabular}
\label{tab:masses}
\end{table}

   \subsection{ Widths of  hidden-charm pentaquark molecules  }
 
   In Table~\ref{tab:masses},  we tabulate the masses and quantum numbers of relevant particles. 
    For the cutoff in the Gaussian regulator, we choose the value of  $\Lambda=1.5$ GeV~\cite{Xie:2022hhv}.    To quantify the agreement with the experimental data, we use the $\chi^2$ defined as 
 \begin{eqnarray}
 \chi^2=\sum_{i=1}^{3}\frac{(M_{exp}^{i}-M_{fit}^{i})^2}{{d_{M}^i}^{2}} +\sum_{i=1}^{3}\frac{(\Gamma_{exp}^{i}-\Gamma_{fit}^{i})^2}{{d_{\Gamma}^i}^2}
 \end{eqnarray}
 where $M_{exp}^{i}( \Gamma_{exp}^{i})$ and $M_{fit}^{i}( \Gamma_{fit}^{i})$ are the masses(widths)  measured by the LHCb Collaboration and  those obtained in the contact-range EFT approach, $d_{M}^{i}$ and $d_{\Gamma}^{i}$ are the uncertainties of experimental masses and widths, and  the superscripts with $i=1$, $i=2$ and $i=3$ represent \textcolor{black}{ $P_{\psi}^{N}(4312)$}, \textcolor{black}{$P_{\psi}^{N}(4440)$}, and \textcolor{black}{$P_{\psi}^{N}(4457)$}, respectively.     
 The masses and decay widths of the three   states are~\cite{Aaij:2019vzc} 
\begin{eqnarray}
 \nonumber
M_{\textcolor{black}{ P_{\psi}^{N}(4312)}}&=&4311.9\pm 0.7^{+6.8}_{-0.6} ~\mbox{MeV}\,, \\    \Gamma_{\textcolor{black}{ P_{\psi}^{N}(4312)}}&=&9.8 \pm 2.7^{+3.7}_{-4.5} ~\mbox{MeV}\,,\nonumber   \\
M_{\textcolor{black}{P_{\psi}^{N}(4440)}}&=&4440.3\pm 1.3^{+4.1}_{-4.7} ~\mbox{MeV}\,,  \nonumber   \\\Gamma_{\textcolor{black}{P_{\psi}^{N}(4440)}}&=&20.6 \pm 4.9^{+8.7}_{-10.1}~\mbox{MeV}\,,
\\ \nonumber
M_{\textcolor{black}{P_{\psi}^{N}(4457)}}&=&4457.3\pm 0.6^{+4.1}_{-1.7} ~\mbox{MeV}\,,     \\ \Gamma_{\textcolor{black}{P_{\psi}^{N}(4457)}}&=&6.4 \pm 2.0^{+5.7}_{-1.9} ~\mbox{MeV}\,.\nonumber
\end{eqnarray}

\begin{table}[ttt]
\centering
\caption{ Parameters  $C_a$, $C_b$, $C_a^{\prime}$,  $C_b^{\prime}$, $g_{1}$ and $g_2$ (in units of GeV$^{-1}$) determined by fitting to the masses and widths of \textcolor{black}{ $P_{\psi}^{N}(4312)$}, \textcolor{black}{$P_{\psi}^{N}(4440)$}, and \textcolor{black}{$P_{\psi}^{N}(4457)$} and the corresponding $\chi^2$ in Scenario A and Scenario B.      }
\begin{tabular}{c |c c c  c c  c c c}
  \hline \hline
    Scenario & $C_{a}$ & $C_{b}$ & $C_{a}'$ & $C_{b}'$ & $g_{1}$ & $g_{2}$ & $\chi^{2}$ \\ \hline
 A & -52.533 & 6.265 & -11.366 & 2.669 & 20.360 & -19.590 & 1.155  \\ 
    B      & -56.030 & -5.418 & -12.120 & -5.136 & 58.963 & 28.587 & 2.980 \\
  \hline \hline
\end{tabular}
\label{fittingcoupling}
\end{table}

Given the fact that the light meson saturation is valid for the   $\bar{D}^{(*)}\Lambda_{c}\to \bar{D}^{(*)}\Lambda_{c}$ and    $\bar{D}^{(*)}\Sigma_{c}^{(*)}\to \bar{D}^{(*)}\Sigma_{c}^{(*)}$ potentials\footnote{ As indicated in Ref~\cite{Liu:2019zvb}, the ratio of $C_a$ to $C_b$ estimated by the light meson saturation approach is consistent  with that obtained by the contact-range EFT approach.  }, 
 we adopt the ratio $C_{a}^{\prime}/C_a=0.216$ obtained  in the light meson saturation approach.  As a result,  there only remain five parameters.   
  With the above preparations  we determine  the values of the  parameters $C_a$, $C_b$,  $C_b^{\prime}$, $g_{1}$ and $g_2$ as well as the $\chi^{2}$ for Scenario A and Scenario B  and show them  in Table~\ref{fittingcoupling}.   
  In the following, we compare  the fitted  parameters ($C_a$, $C_b$, and $C_b^{\prime}$) with those obtained in the light meson saturation approach (see the Appendix for details).    With the light meson saturation, we obtain the ratio $C_{b}/C_a=0.12$, while the ratio determined in the  EFT approach (by fitting to the data)  is $C_{b}/C_a=-0.12$ and  $C_{b}/C_a=0.10$ for Scenario A and Scenario B, respectively.  It seems that the light meson saturation approach prefers Scenario B, consistent with the single-channel analysis~\cite{Liu:2019zvb}. Moreover, the light meson saturation approach yields    $C_{b}^{\prime}/C_{b}=0.61$, while the value determined in the EFT approach  is  $C_{b}^{\prime}/C_{b}=0.43$ for Scenario A but   $C_{b}^{\prime}/C_{b}=0.95$ for Scenario B. One can see that they are quite different for either   Scenario A or Scenario B.  From the perspective of light meson saturation, only the  $\rho$ meson exchange is considered to saturate the  $\bar{D}^{(\ast)}\Sigma_{c} \to \bar{D}^{(*)}\Lambda_{c}$ interactions. However, the   one-pion exchange   plays a significant  role in the  $\bar{D}^{(\ast)}\Sigma_{c} \to \bar{D}^{(*)}\Lambda_{c}$ potentials~\cite{Yamaguchi:2019seo,He:2019rva,Burns:2021jlu}.     Since the one-pion exchange is not considered, the $C_b^{\prime}$ obtained in the light meson saturation approach is not consistent with   the $C_b^{\prime}$ determined in the EFT approach for either  Scenario A  or Scenario B. 

 In Table~\ref{molecularcouplings}, we present the pole positions of the hidden-charm pentaquark molecules  and  the   couplings to their constituents.    From  the obtained pole positions of \textcolor{black}{ $P_{\psi}^{N}(4312)$}, \textcolor{black}{$P_{\psi}^{N}(4440)$}, and \textcolor{black}{$P_{\psi}^{N}(4457)$}, it is obvious that  scenario A, yielding results consistent with the experimental data,  is  better than Scenario B, which is quite different from the single-channel study~\cite{Liu:2019tjn}. Our study shows that  the coupled-channel effects can help distinguish the two possible scenarios.  In a similar approach but without the $\bar{D}^{(*)}\Lambda_c$ channels,  Scenario A  is still slightly  better than Scenario B~\cite{Xie:2022hhv}.  We note in passing that  the   chiral unitary model~\cite{Xiao:2019aya} also  prefers  scenario A.  We  further note that the coefficients in the contact-range potentials of Eq.(\ref{contact11}) and Eq.(\ref{contact22}) are derived assuming the HQSS, while the  HQSS  breaking is not taken into  account.  In Ref.~\cite{Yamaguchi:2019seo}, it was shown that the tensor term of the one-pion exchange  potentials plays a crucial role in describing the widths of the pentaquark molecules,  while  the $D$-wave potentials are neglected   in this work.        Therefore, we can not conclude which scenario is more favorable at this stage.

 Up to now, the spins of \textcolor{black}{$P_{\psi}^{N}(4440)$} and \textcolor{black}{$P_{\psi}^{N}(4457)$} are still undetermined experimentally, which motivated many  theoretical discussions on how to determine their spins~\cite{Pan:2019skd,Zhang:2023czx,Liu:2023wfo}. One crucial issue is that  the strength of $\bar{D}^*\Sigma_c \to \bar{D}^*\Sigma_c $  potentials of  $J^P=1/2^-$   and  $J^P=3/2^-$ are  undetermined.  We can see that the  $J^P=1/2^-$ $\bar{D}^*\Sigma_c \to \bar{D}^*\Sigma_c $  potential is stronger  than  the   $J^P=3/2^-$ $\bar{D}^*\Sigma_c \to \bar{D}^*\Sigma_c $ potential in Scenario A, while their order reverses in Scenario B. In Refs.~\cite{Burns:2021jlu,Burns:2022uiv}, Burns et al. proposed  another case, named as Scenario C, which  actually corresponds to a special case of  Scenario B, where the $J^P=1/2^-$  $\bar{D}^*\Sigma_c \to \bar{D}^*\Sigma_c $ potential is not strong enough to form a bound state, and therefore  \textcolor{black}{$P_{\psi}^{N}(4457)$} is interpreted  as a kinetic effect rather than a genuine state. From their  values of $C_a$ and $C_b$~\cite{Burns:2022uiv}, the ratio  $C_b/C_a$ is  determined to be around 0.5, which 
implies the emergence of  a large spin-spin interaction, inconsistent with the
 principle of EFTs. It is  no surprise  that such a  large spin-spin  interaction breaks the completeness of the  multiplet picture of hidden-charme pentaquark molecules~\cite{Liu:2019tjn,Xiao:2019aya,Du:2019pij,Yamaguchi:2019seo,PavonValderrama:2019nbk,Lin:2019qiv,He:2019rva,Yalikun:2021bfm,Dong:2021juy,Zhang:2023czx}.     Therefore,  we strongly recommend that lattice QCD simulations could study the potentials of  $J^P=1/2^-$ $\bar{D}^*\Sigma_c$ and  $J^P=3/2^-$ $\bar{D}^*\Sigma_c$ to  address this issue\footnote{ A recent lattice QCD  study shows that there exists a $J^{P}=1/2^-$$\bar{D}^*\Sigma_c$ bound state with a binding energy of  $6$~MeV~\cite{Xing:2022ijm}.  }.

\begin{table}[ttt]
\setlength{\tabcolsep}{4pt}
\centering
\caption{Two-body  decay widths, three-body decay widths, and total decay widths (in units of MeV) of hidden-charm pentaquark molecules   in Scenario A and Scenario B. \label{total widths}
}
\begin{tabular}{c c c c c c c}
  \hline \hline 
     Scenario   &    \multicolumn{6}{c}{A}  \\ \hline
     Molecule  &  $\textcolor{black}{P_{\psi1}^{ N}}$ & $\textcolor{black}{P_{\psi2}^{ N}}$ & $\textcolor{black}{P_{\psi3}^{ N}}$ & $\textcolor{black}{P_{\psi4}^{ N}}$ & $\textcolor{black}{P_{\psi5}^{ N}}$ & $\textcolor{black}{P_{\psi6}^{ N}}$ \\
     Two-body decay & 7.00   & 5.40 & 17.20 & 1.40 & 19.80 & 15.40 \\
     Three-body decay & 0.20  & 1.47 & 0.03 & 0.33 & 3.83 & 6.85  \\
     Total decay& 7.20 & 6.87 & 17.23 & 1.73 & 23.63 & 22.25 \\  \hline
    Scenario   &   \multicolumn{6}{c}{B}   \\ \hline
     Molecule & $\textcolor{black}{P_{\psi1}^{ N}}$ & $\textcolor{black}{P_{\psi2}^{ N}}$ & $\textcolor{black}{P_{\psi3}^{ N}}$ & $\textcolor{black}{P_{\psi4}^{ N}}$ & $\textcolor{black}{P_{\psi5}^{ N}}$ & $\textcolor{black}{P_{\psi6}^{ N}}$ \\
     Two-body decay &  8.00  & 12.40 & 2.20 & 9.00 & 15.00 & 7.40 \\
     Three-body decay & 0.16  & 1.00 & 0.01 & 2.44 & 13.58 & 9.58  \\
     Total decay & 8.16 & 13.40 & 2.21 & 11.44 & 28.58 & 16.98 \\  \hline\hline
\end{tabular}
\end{table}

 The  imaginary parts of the pole positions in Table~\ref{molecularcouplings}  specify the two-body partial  decay widths of the pentaquark molecules, which can also be calculated via the  triangle diagrams using the effective Lagrangian approach~\cite{Xiao:2019mvs,Lin:2019qiv}. From the results of Table~\ref{molecularcouplings},   we  can calculate the two-body decay widths of these pentaquark molecules and tabulate them  in Table~\ref{total widths}. Moreover, with the newly obtained   couplings $g_{\textcolor{black}{P_{\psi}^{ N}}\bar{D}^{(*)}\Sigma_{c}^{(*)}}$, we update their  three-body decay widths  as shown  in Table~\ref{total widths}. Comparing with the results in Ref.~\cite{Xie:2022hhv}, we  find that the new results vary a bit because the  pole positions affect the phase space of the three-body  decays and the values of the couplings  $g_{\textcolor{black}{P_{\psi}^{ N}}\bar{D}^{(*)}\Sigma_{c}^{(*)}}$. Assuming that the two-body   and three-body  decays   are dominant decay channels for the pentaquark molecules,    we can obtain their  total decay widths  by  summing the two decay  modes. Our results indicate that the widths of    $\textcolor{black}{P_{\psi 5}^{ N}}$ and $\textcolor{black}{P_{\psi 6}^{ N}}$ as the $\bar{D}^*\Sigma_c^*$ molecules are larger than those of  \textcolor{black}{ $P_{\psi}^{N}(4312)$}, \textcolor{black}{$P_{\psi}^{N}(4440)$}, and \textcolor{black}{$P_{\psi}^{N}(4457)$} reported by the LHCb Collaboration, and their
three-body  decay widths  account for a large proportion of their total widths. In addition, we can see  that the three-body decay widths of  \textcolor{black}{ $P_{\psi}^{N}(4312)$}, \textcolor{black}{$P_{\psi}^{N}(4440)$}, and \textcolor{black}{$P_{\psi}^{N}(4457)$}  account for only a small proportion of their total widths, which confirms   our assumption that their total widths are almost saturated  by the two-body decays.

 \begin{widetext}

 \begin{table}[ttt]
 \setlength{\tabcolsep}{6pt}
\centering
\caption{Pole positions(in units of MeV) of six hidden-charm pentaquark molecules and the  couplings to their constituents  in Scenario A and Scenario B.  }
\begin{tabular}{c c c c c c c}
  \hline\hline 
     Scenario   &    \multicolumn{6}{c}{A}  \\ \hline
     Name &  $\textcolor{black}{P_{\psi1}^{ N}}$ & $\textcolor{black}{P_{\psi2}^{ N}}$ & $\textcolor{black}{P_{\psi3}^{ N}}$ & $\textcolor{black}{P_{\psi4}^{ N}}$ & $\textcolor{black}{P_{\psi5}^{ N}}$ & $\textcolor{black}{P_{\psi6}^{ N}}$ \\
     Molecule  &  $\bar{D}\Sigma_c $  & $\bar{D}\Sigma^{*}_c$ & $\bar{D}^{\ast}\Sigma_c$ & $\bar{D}^{\ast}\Sigma_c$  & $\bar{D}^{\ast}\Sigma_c^{\ast}$  & $\bar{D}^{\ast}\Sigma_c^{\ast}$   \\
     $J^P$ &  $\frac{1}{2}^-$  &  $\frac{3}{2}^-$ & $\frac{1}{2}^-$ & $\frac{3}{2}^-$  & $\frac{1}{2}^-$ & $\frac{3}{2}^-$  \\
     Pole (MeV) &  4310.6+3.5$i$  &4372.8 +2.7$i$  & 4440.6+8.6$i$ & 4458.4+0.7$i$ & 4500.0+9.9$i$ &  4513.2+7.7$i$  \\
     $g_{\textcolor{black}{P_{\psi}^{ N}}\Sigma_{c}^{*}\bar{D}^{*}}$ & -  & - & - & - & 2.686 & 2.194 \\
     $g_{\textcolor{black}{P_{\psi}^{ N}}\Sigma_{c}\bar{D}^{*}}$ &  - & - & 2.554 & 1.082 & 0.141 & 0.218 \\
     $g_{\textcolor{black}{P_{\psi}^{ N}}\Sigma_{c}^{*}\bar{D}}$ &  - & 2.133 & - & 0.179 & - & 0.237 \\
     $g_{\textcolor{black}{P_{\psi}^{ N}}\Sigma_{c}\bar{D}}$ &  2.089 & - & 0.254 & - & 0.139 & - \\
     $g_{\textcolor{black}{P_{\psi}^{ N}}\Lambda_{c}\bar{D}^{*}}$ & 0.234 & 0.074 & 0.177 & 0.050 & 0.110 & 0.241 \\
     $g_{\textcolor{black}{P_{\psi}^{ N}}\Lambda_{c}\bar{D}}$ &  0.014 & - & 0.158 & - & 0.207 & - \\
     $g_{\textcolor{black}{P_{\psi}^{ N}}J/\psi N}$ &  0.251 & 0.454 & 0.584 & 0.103 & 0.434 & 0.532 \\
     $g_{\textcolor{black}{P_{\psi}^{ N}}\eta_{c} N}$ &  0.420 & - & 0.261 & - & 0.527 & - \\\hline
    Scenario   &   \multicolumn{6}{c}{B}   \\ \hline
    Name &  $\textcolor{black}{P_{\psi1}^{ N}}$ & $\textcolor{black}{P_{\psi2}^{ N}}$ & $\textcolor{black}{P_{\psi3}^{ N}}$ & $\textcolor{black}{P_{\psi4}^{ N}}$ & $\textcolor{black}{P_{\psi5}^{ N}}$ & $\textcolor{black}{P_{\psi6}^{ N}}$\\
     Molecule  &  $\bar{D}\Sigma_c$  & $\bar{D}\Sigma^{*}_c$ & $\bar{D}^{\ast}\Sigma_c$ & $\bar{D}^{\ast}\Sigma_c$  & $\bar{D}^{\ast}\Sigma_c^{\ast}$  & $\bar{D}^{\ast}\Sigma_c^{\ast}$   \\
     $J^P$ &  $\frac{1}{2}^-$  &  $\frac{3}{2}^-$ & $\frac{1}{2}^-$ & $\frac{3}{2}^-$  & $\frac{1}{2}^-$ & $\frac{3}{2}^-$  \\
     Pole (MeV)  &  4309.9+4$i$  & 4365.8+6.2$i$ &  4458.4+4.5$i$ & 4441.4+1.1$i$ & 4521.6+7.5$i$ &  4522.5+3.7$i$  \\
     $g_{\textcolor{black}{P_{\psi}^{ N}}\Sigma_{c}^{*}\bar{D}^{*}}$ & -  & - & - & - & 1.841 & 1.621 \\
     $g_{\textcolor{black}{P_{\psi}^{ N}}\Sigma_{c}\bar{D}^{*}}$ &  - & - & 1.679 & 2.462 & 0.107 & 0.143 \\
     $g_{\textcolor{black}{P_{\psi}^{ N}}\Sigma_{c}^{*}\bar{D}}$ &  - & 2.451 & - & 0.099 & - & 0.171 \\
     $g_{\textcolor{black}{P_{\psi}^{ N}}\Sigma_{c}\bar{D}}$ &  2.072 & - & 0.161 & - & 0.131 & - \\
     $g_{\textcolor{black}{P_{\psi}^{ N}}\Lambda_{c}\bar{D}^{*}}$ & 0.392 & 0.090 & 0.247 & 0.159 & 0.232 & 0.223 \\
     $g_{\textcolor{black}{P_{\psi}^{ N}}\Lambda_{c}\bar{D}}$ & 0.020 & - & 0.191 & - & 0.281 & - \\
     $g_{\textcolor{black}{P_{\psi}^{ N}}J/\psi N}$ &  0.263 & 0.704 & 0.277 & 0.168 & 0.314 & 0.312 \\
     $g_{\textcolor{black}{P_{\psi}^{ N}}\eta_{c} N}$ & 0.413 & - & 0.164 & - & 0.328 & - \\\hline\hline
\end{tabular}
\label{molecularcouplings}
\end{table}
    
\end{widetext}

   \subsection{ Production rates of hidden-charm pentaquark molecules  }

  From   the values of the couplings given in Table~\ref{molecularcouplings},  one can see that the $\bar{D}^{(*)}\Sigma_{c}^{(*)}$ channel plays a dominant role in generating these  pentaquark molecules.   Yet their productions  in the $\Lambda_b$ decay can not proceed via the  $\bar{D}^{(*)}\Sigma_{c}^{(*)}$ interactions as discussed above. It is important to investigate the productions of hidden-charm pentaquark molecules  in the $\Lambda_b$ decays  via the $\bar{D}^{(*)}\Lambda_c$ interactions although the  couplings of the pentaquark states to  $\bar{D}^{(*)}\Lambda_c$ are small.  
With  the couplings $g_{\textcolor{black}{P_{\psi}^{ N}}\bar{D}^{(*)}\Lambda_c}$  given in Table~\ref{molecularcouplings},   we  employ the effective Lagrangian approach to   calculate  the decays of  $\Lambda_b \to \textcolor{black}{P_{\psi}^{ N}} K$ illustrated in Fig.~\ref{triangle2} and Fig.~\ref{triangle3}.

In Table~\ref{branchingratiospv}, we present the branching fractions of $\Lambda_b \to \textcolor{black}{P_{\psi}^{ N}} K$ in Scenario A and Scenario B.  Our results show that the branching factions of the three pentaquark states discovered by the LHCb Collaboration:  $\mathcal{B}(\Lambda_b \to \textcolor{black}{ P_{\psi}^{N}(4312)}K)=35.18\times 10^{-6}$, $\mathcal{B}(\Lambda_b \to \textcolor{black}{P_{\psi}^{N}(4440)}K)=15.30\times 10^{-6}$, and  $\mathcal{B}(\Lambda_b \to \textcolor{black}{P_{\psi}^{N}(4457)}K)=0.48\times 10^{-6}$ in Scenario A and   $\mathcal{B}(\Lambda_b \to \textcolor{black}{ P_{\psi}^{N}(4312)}K)=98.88\times 10^{-6}$, $\mathcal{B}(\Lambda_b \to \textcolor{black}{P_{\psi}^{N}(4440)}K)=5.21\times 10^{-6}$, and  $\mathcal{B}(\Lambda_b \to \textcolor{black}{P_{\psi}^{N}(4457)}K)=27.23\times 10^{-6}$ in Scenario B.  From the  
order of magnitude of the obtained branching fractions, we can conclude that   the pentaquark molecules can be produced via the triangle  diagrams  shown  in Fig.~\ref{triangle2} and Fig.~\ref{triangle3}.  The  branching fractions  of the decay $ \Lambda_b \to \textcolor{black}{ P_{\psi}^{N}(4312)}K $  are larger than those of  $\Lambda_b \to \textcolor{black}{P_{\psi}^{N}(4440)}K$ and $\Lambda_b \to \textcolor{black}{P_{\psi}^{N}(4457)}K$ in both cases, and  the branching fractions involving  \textcolor{black}{$P_{\psi}^{N}(4440)$} and \textcolor{black}{$P_{\psi}^{N}(4457)$} for  $J^{P}=1/2^{-}$ 
are always larger than those for $J^{P}=3/2^{-}$. Such results  reflect that  the branching fractions of the decays $\Lambda_b \to \textcolor{black}{P_{\psi}^{ N}} K$ are related to the couplings $g_{\textcolor{black}{P_{\psi}^{ N}}\bar{D}^{(*)}\Lambda_c}$, especially the coupling $g_{\textcolor{black}{P_{\psi}^{ N}}\bar{D}^{*}\Lambda_c}$, which shows that  the $\bar{D}^*\Lambda_c$ interactions play an important role in producing the pentaquark molecules in the $\Lambda_b$ decays.  Similarly, we predict the branching fractions of $\Lambda_b$ decaying into  $\textcolor{black}{P_{\psi2}^{ N}}$,  $\textcolor{black}{P_{\psi5}^{ N}}$ and $\textcolor{black}{P_{\psi 6}^{ N}}$ plus a kaon  as shown in Table~\ref{branchingratiospv}, the order of magnitude of which are similar to those involving \textcolor{black}{$P_{\psi}^{N}(4440)$} and \textcolor{black}{$P_{\psi}^{N}(4457)$}.  

\begin{table}[ttt]
\setlength{\tabcolsep}{5pt}
\centering
\caption{Branching fractions ($10^{-6}$) of  $\Lambda_b$ decaying into a $K$ meson and a hidden-charm pentaquark  molecule   in Scenario A and Scenario B.  }
\begin{tabular}{c c c c c c c}
  \hline \hline 
     Scenario   &    \multicolumn{6}{c}{A}  \\ \hline
     Molecule  &$\textcolor{black}{P_{\psi1}^{ N}}$ & $\textcolor{black}{P_{\psi2}^{ N}}$ & $\textcolor{black}{P_{\psi3}^{ N}}$ & $\textcolor{black}{P_{\psi4}^{ N}}$ & $\textcolor{black}{P_{\psi5}^{ N}}$ & $\textcolor{black}{P_{\psi6}^{ N}}$  \\
     $\mathcal{B}(\Lambda_b \to \textcolor{black}{P_{\psi}^{ N}}K)$ &  35.18  & 1.49 &  15.30 & 0.48 & 6.37 &  9.01 \\\hline
    Scenario   &   \multicolumn{6}{c}{B}   \\ \hline
     Molecule  & $\textcolor{black}{P_{\psi1}^{ N}}$ & $\textcolor{black}{P_{\psi2}^{ N}}$ & $\textcolor{black}{P_{\psi3}^{ N}}$ & $\textcolor{black}{P_{\psi4}^{ N}}$ & $\textcolor{black}{P_{\psi5}^{ N}}$ & $\textcolor{black}{P_{\psi6}^{ N}}$  \\
     $\mathcal{B}(\Lambda_b \to \textcolor{black}{P_{\psi}^{ N}} K)$ & 98.88   & 2.27 & 27.23 & 5.21 & 21.69  & 7.43  \\
  \hline\hline 
\end{tabular}
\label{branchingratiospv}
\end{table}

Up to now, 
there exist no available experimental data for the branching fractions of the decays  $\Lambda_b \to \textcolor{black}{P_{\psi}^{ N}}K$.   
The LHCb Collaboration measured the relevant  ratios of branching fractions for  the three pentaquark states: $R_{\textcolor{black}{ P_{\psi}^{N}(4312)}}=(0.30\pm 0.07^{+0.34}_{-0.09})\%$, $R_{\textcolor{black}{ P_{\psi}^{N}(4440)}}=(1.11\pm 0.33^{+0.22}_{-0.11})\%$, and $R_{\textcolor{black}{ P_{\psi}^{N}(4457)}}=(0.53\pm 0.16^{+0.15}_{-0.13})\%$, where $R$ is defined as 
\begin{eqnarray}
R_{\textcolor{black}{P_{\psi}^{ N+}}}=\frac{\mathcal{B}(\Lambda_b^0\to \textcolor{black}{P_{\psi}^{ N+}} K^{-}) \cdot \mathcal{B}(\textcolor{black}{P_{\psi}^{ N +}} \to J/\psi p)}{\mathcal{B}(\Lambda_b^0\to J/\psi p K^{-}) } 
\end{eqnarray}
According to  RPP~\cite{ParticleDataGroup:2020ssz}, the branching fraction of $\Lambda_b^0\to J/\psi p K^{-}$ is  $\mathcal{B}(\Lambda_b^0\to J/\psi p K^{-})=3.2^{+0.6}_{-0.5}\times 10^{-4}$, and then   we obtain the product of the branching fractions of the decays $\Lambda_b \to \textcolor{black}{P_{\psi}^{ N}} K$ and $ \textcolor{black}{P_{\psi}^{ N}} \to J/\psi p$
\begin{widetext}
\begin{eqnarray}
\label{pcbranching}
\mathcal{B}(\Lambda_b^0\to \textcolor{black}{ P_{\psi}^{N}(4312)^{+}} K^{-}) \cdot \mathcal{B}(\textcolor{black}{ P_{\psi}^{N}(4312)^{+}}\to J/\psi p)&= 0.96^{+1.13}_{-0.39}\times 10^{-6}\,,  \nonumber\\ 
\mathcal{B}(\Lambda_b^0\to \textcolor{black}{ P_{\psi}^{N}(4440)^{+}} K^{-}) \cdot \mathcal{B}(\textcolor{black}{ P_{\psi}^{N}(4440)^{+}}\to J/\psi p)&= 3.55^{+1.43}_{-1.24}\times 10^{-6}\,,   \\ \nonumber
\mathcal{B}(\Lambda_b^0\to \textcolor{black}{ P_{\psi}^{N}(4457)^{+}} K^{-}) \cdot \mathcal{B}(\textcolor{black}{ P_{\psi}^{N}(4457)^{+}}\to J/\psi p)&= 1.70^{+0.77}_{-0.71}\times 10^{-6}\,.
\end{eqnarray}  
\end{widetext}
To compare with the experimental data, we have to obtain the branching fractions of $\textcolor{black}{P_{\psi}^{ N}} $ decaying into $J/\psi p$, which 
are not yet determined experimentally. We note that the GlueX and JLab Collaborations investigated the production rates of pentaquark states in the photoproduction process and only gave the upper limits of $\mathcal{B}(\textcolor{black}{P_{\psi}^{ N}} \to J/\psi p)<2.0\%$~\cite{Meziani:2016lhg,GlueX:2019mkq}, which indicates that the branching fractions of $\mathcal{B}(\Lambda_b^0\to \textcolor{black}{P_{\psi}^{ N+ }}  K^{-})$ are at the order of $10^{-4}$,  approaching to the values of $\mathcal{B}(\Lambda_b^0\to J/\psi p K^{-})$. Such large values highlight the inconsistency  between the  LHCb results and the  GlueX/JLab results. Therefore, more precise experimental data are needed to settle this issue. 

\begin{widetext}

\begin{table}[ttt]
\setlength{\tabcolsep}{6pt}
\centering
\caption{Two-body partial decay widths (in units of MeV) of hidden-charm pentaquark molecules as well as their branching fractions in Scenario A and Scenario B .  }
\begin{tabular}{cc c c c c c c}
  \hline\hline
     Scenario   &    \multicolumn{6}{c}{A}  \\ \hline
     Molecule  & $\textcolor{black}{P_{\psi1}^{ N}}$ & $\textcolor{black}{P_{\psi2}^{ N}}$ & $\textcolor{black}{P_{\psi3}^{ N}}$ & $\textcolor{black}{P_{\psi4}^{ N}}$ & $\textcolor{black}{P_{\psi5}^{ N}}$ & $\textcolor{black}{P_{\psi6}^{ N}}$ \\
     $\Gamma_{2}(\Sigma_c \bar{D}^{\ast})$ & -  & - & - & - & \makecell{0.50 \\(2.52 $\%$)} & \makecell{1.38 \\(8.87 $\%$)} \\
     $\Gamma_{3}(\Sigma^{*}_c \bar{D})$ &  - & - & - & \makecell{1.14\\(73.23 $\%$)} & - & \makecell{2.62 \\(16.87 $\%$)} \\
     $\Gamma_{4}(\Sigma_c \bar{D})$ &  - & - & \makecell{2.87 \\(16.59 $\%$)} & - & \makecell{1.04 \\(5.29 $\%$)} & - \\
     $\Gamma_{5}(\Lambda_c \bar{D}^{\ast})$ &  \makecell{0.83\\ (11.71 $\%$)} & \makecell{0.19\\ (3.48 $\%$)} & \makecell{1.44 \\(8.33 $\%$)} & \makecell{0.12 \\(7.82 $\%$)} & \makecell{0.65\\ (3.32 $\%$)} & \makecell{3.24\\ (20.81 $\%$)} \\
     $\Gamma_{6}(\Lambda_c \bar{D})$ &  \makecell{0.01\\ (0.14 $\%$)} & - & \makecell{1.60\\ (9.22 $\%$)} & - & \makecell{2.98\\ (15.14 $\%$)} & - \\
     $\Gamma_{7}(J/\psi N)$ &  \makecell{1.43\\ (20.22 $\%$)} & \makecell{5.16\\ (96.52 $\%$)} & \makecell{9.29\\ (53.64 $\%$)} & \makecell{0.29\\ (18.95 $\%$)} & \makecell{5.46\\ (27.73 $\%$)} & \makecell{8.31\\ (53.45 $\%$)} \\
     $\Gamma_{8}(\eta_{c} N)$ &  \makecell{4.81\\ (67.93 $\%$)} & - & \makecell{2.12 \\(12.23 $\%$)} & - & \makecell{9.06 \\(45.98 $\%$)} & - \\\hline
    Scenario   &   \multicolumn{6}{c}{B}   \\ \hline
     Molecule  &  $\textcolor{black}{P_{\psi1}^{ N}}$ & $\textcolor{black}{P_{\psi2}^{ N}}$ & $\textcolor{black}{P_{\psi3}^{ N}}$ & $\textcolor{black}{P_{\psi4}^{ N}}$ & $\textcolor{black}{P_{\psi5}^{ N}}$ & $\textcolor{black}{P_{\psi6}^{ N}}$ \\
     $\Gamma_{2}(\Sigma_c \bar{D}^{\ast})$ &  - & - & - & - & \makecell{0.36\\ (2.17 $\%$)} & \makecell{0.64\\ (8.30 $\%$)} \\
     $\Gamma_{3}(\Sigma^{*}_c \bar{D})$ &  - & - & - & \makecell{0.31 \\(13.63 $\%$)} & - & \makecell{1.41 \\(18.19 $\%$)} \\
     $\Gamma_{4}(\Sigma_c \bar{D})$ &  - & - & \makecell{1.23 \\(12.88 $\%$)} & - & \makecell{0.98\\ (5.92 $\%$)} & - \\
     $\Gamma_{5}(\Lambda_c \bar{D}^{\ast})$ &  \makecell{2.27 \\(26.71 $\%$)} & \makecell{0.26\\ (2.10 $\%$)} & \makecell{2.97 \\(30.92 $\%$)} & \makecell{1.17\\ (52.05 $\%$)} & \makecell{3.05 (18.48 $\%$)} & \makecell{2.82 \\(36.37 $\%$)} \\
     $\Gamma_{6}(\Lambda_c \bar{D})$ &  \makecell{0.02\\ (0.23 $\%$)} & - & \makecell{2.40 \\(25.03 $\%$)} & - & \makecell{5.65 \\(34.18 $\%$)} & - \\
     $\Gamma_{7}(J/\psi N)$ &  \makecell{1.57 \\(18.45 $\%$)} & \makecell{12.28 \\(97.90 $\%$)} & \makecell{2.13 \\(22.26 $\%$)} & \makecell{0.77 \\(34.33 $\%$)} & \makecell{2.92\\ (17.67 $\%$)} & \makecell{2.88\\ (37.14 $\%$)} \\
     $\Gamma_{8}(\eta_{c} N)$ & \makecell{ 4.65 \\(54.61 $\%$)} & - & \makecell{0.85 \\(8.86 $\%$)} & - & \makecell{3.57 \\(21.58 $\%$)} & - \\
  \hline\hline
\end{tabular}
\label{partialdecaywith}
\end{table}

\end{widetext}

Using Eq.(\ref{partialwiths}), we calculate  the two-body  partial decay widths of hidden-charm  pentaquark molecules,  and then estimate the branching fractions of the decays $ \textcolor{black}{P_{\psi}^{ N}}\to J/\psi p$. The results are shown in Table~\ref{partialdecaywith}, where the three-body decay widths of the pentaquark molecules are not included.    
One can see that the partial decay widths of  
 $ \textcolor{black}{P_{\psi}^{ N}} \to \bar{D}^{(*)}\Lambda_c$  are  less than those of   $ \textcolor{black}{P_{\psi}^{ N}} \to J/\psi p$ in Scenario A, but their order reverses  in Scenario B. In Ref.~\cite{Xiao:2020frg}, the  estimated  branching fractions of the decays   $\textcolor{black}{P_{\psi}^{ N}} \to \bar{D}^{(*)}\Lambda_c$  are much smaller than those of the decays   $ \textcolor{black}{P_{\psi}^{ N}}\to J/\psi p$,  consistent with Scenario A. In terms of the meson exchange theory, the   branching fractions of the decays   $\textcolor{black}{P_{\psi}^{ N}} \to \bar{D}^{(*)}\Lambda_c$  are  larger than those of the decays  $ \textcolor{black}{P_{\psi}^{ N}}\to J/\psi p$, where the heavy  meson  ($\bar{D}^{(*)}$) exchange and the light meson ($\pi(\rho)$) exchange are responsible  for the   $\bar{D}^{(*)}\Sigma_{c}^{(*)} \to J/\psi p$  and  $\bar{D}^{(*)}\Sigma_{c}^{(*)} \to \bar{D}^{(*)}\Lambda_c$ interactions, respectively~\cite{Lin:2019qiv}.   
 It is obvious that the transitions $\bar{D}^{(*)}\Sigma_{c}^{(*)} \to J/\psi p$ are heavily suppressed, resulting in the small partial decay widths of  $\textcolor{black}{P_{\psi}^{ N}}  \to J/\psi p$ in the same theoretical  framework~\cite{Lin:2019qiv}. 
 We note that the meson exchange theory has been tested for light mesons exchanges, but  remains to be verified for  heavy meson exchanges, especially when  both heavy and light mesons can be exchanged.  
 The meson exchange theory dictates that charmed mesons are responsible for the very short range interaction, but they can not adequately describe such  short-range interactions because one gluon exchange may play a role.  In Ref.~\cite{Yamaguchi:2019djj}, the authors found that the strength of the short range potential provided by the one gluon exchange is much stronger  than that provided by the heavy meson exchange. In the present work, the hidden-charm meson-baryon potentials are provided by the contact-range EFT constrained by HQSS with the low-energy constants determined by fitting to data, which are plausible but the underlying mechanism  needs to be clarified.

\begin{table}[ttt]
\setlength{\tabcolsep}{5pt}
\centering
\caption{Branching fractions ($10^{-6}$) of the decays   $\Lambda_b \to (\textcolor{black}{P_{\psi}^{ N}} \to J/\psi p) K$  in Scenario A and Scenario B.  }
\begin{tabular}{c c c c c c c}
  \hline \hline 
     Scenario   &    \multicolumn{6}{c}{A}  \\ \hline
     Molecule  & $\textcolor{black}{P_{\psi1}^{ N}}$ & $\textcolor{black}{P_{\psi2}^{ N}}$ & $\textcolor{black}{P_{\psi3}^{ N}}$ & $\textcolor{black}{P_{\psi4}^{ N}}$ & $\textcolor{black}{P_{\psi5}^{ N}}$ & $\textcolor{black}{P_{\psi6}^{ N}}$ \\
     Ours &  7.11 & 1.44&  8.21& 0.09& 1.77& 4.82\\
     ChUA~\cite{Xiao:2020frg} &  1.82  & 8.62 & 0.13 & 0.83 & 0.04 & 2.36 \\
     Exp & 0.96  & - & 3.55 & 1.70 & - & - \\\hline
    Scenario   &   \multicolumn{6}{c}{B}   \\ \hline
     Molecule  & $\textcolor{black}{P_{\psi1}^{ N}}$ & $\textcolor{black}{P_{\psi2}^{ N}}$ & $\textcolor{black}{P_{\psi3}^{ N}}$ & $\textcolor{black}{P_{\psi4}^{ N}}$ & $\textcolor{black}{P_{\psi5}^{ N}}$ & $\textcolor{black}{P_{\psi6}^{ N}}$  \\
     Ours & 18.24 & 2.22& 6.06& 1.79& 3.83& 2.76\\
     ChUA~\cite{Xiao:2020frg} &  -  & - & - &  - & - & - \\
     Exp &  0.96  & - &  1.70 & 3.55 & - & - \\ 
  \hline \hline 
\end{tabular}
\label{branchingratiospv1sd}
\end{table}

With the   obtained branching fractions $\mathcal{B}(\Lambda_b \to \textcolor{black}{P_{\psi}^{ N}}K)$ in Table~\ref{branchingratiospv} and    $\mathcal{B}(\textcolor{black}{P_{\psi}^{ N}} \to J/\psi p)$ in Table~\ref{partialdecaywith}, we further calculate the branching fractions  $\mathcal{B}[\Lambda_b \to (\textcolor{black}{P_{\psi}^{ N}}\to J/\psi p) K]$ for Scenario A and Scenario B  as shown  Table~\ref{branchingratiospv1sd}.  Our results show  that   the branching fractions for  \textcolor{black}{ $P_{\psi}^{N}(4312)$} and  \textcolor{black}{$P_{\psi}^{N}(4440)$} are of   the same order as their experimental counterparts,  but the branching fraction for \textcolor{black}{$P_{\psi}^{N}(4457)$} is  smaller   by one order of magnitude.  For  Scenario B, the branching fractions for  \textcolor{black}{$P_{\psi}^{N}(4440)$} and  \textcolor{black}{$P_{\psi}^{N}(4457)$} are  of the same order as their experimental counterparts, but the branching fraction for  \textcolor{black}{ $P_{\psi}^{N}(4312)$} is larger by one order of magnitude. We  can see that our model  can not simultaneously describe the branching fractions of these  three pentaquark states.  
In Ref.~\cite{Xiao:2020frg}, the ChUA estimated the  couplings  $g_{\textcolor{black}{P_{\psi}^{ N}}\bar{D}^{(*)}\Lambda_c}$ and the branching fractions  $\mathcal{B}(\textcolor{black}{P_{\psi}^{ N}} \to J/\psi p)$, which actually corresponds to Scenario A of our results. 
Using the values estimated by ChUA  we  recalculate the branching fractions  $\mathcal{B}[\Lambda_b \to (\textcolor{black}{P_{\psi}^{ N}}\to J/\psi p) K]$ as shown in Table~\ref{branchingratiospv1sd}. The branching fractions for   \textcolor{black}{ $P_{\psi}^{N}(4312)$} and  \textcolor{black}{$P_{\psi}^{N}(4457)$} are of the same order as their experimental counterparts, but the branching fraction for  \textcolor{black}{$P_{\psi}^{N}(4440)$} is smaller by one order of magnitude. Obviously, the branching fractions for \textcolor{black}{ $P_{\psi}^{N}(4312)$}, \textcolor{black}{$P_{\psi}^{N}(4440)$}, and \textcolor{black}{$P_{\psi}^{N}(4457)$}  in our model are related to the couplings of the pentaquark molecules to 
$\bar{D}^{(*)}\Lambda_c$ and $J/\psi p$. Nevertheless,   the production mechanism of these three pentaquark states via the triangle diagrams shown in  Fig.~\ref{triangle2} and Fig.~\ref{triangle3} is capable of qualitatively  reproducing the experimental data, which  further corroborates the hadronic molecular picture of these  pentaquark states.

In Table~\ref{branchingratiospv1sd},  we show  the branching fractions for  the  HQSS partners of \textcolor{black}{ $P_{\psi}^{N}(4312)$}, \textcolor{black}{$P_{\psi}^{N}(4440)$}, and \textcolor{black}{$P_{\psi}^{N}(4457)$},  where only the  two-body decay modes contribute to  the branching fractions of the decays $\textcolor{black}{P_{\psi}^{ N}}\to J/\psi p$. 
  As shown  in Table~\ref{total widths}, the three-body decay widths of $\textcolor{black}{P_{\psi 2}^{ N}}$,  $\textcolor{black}{P_{\psi 5}^{ N}}$, and $\textcolor{black}{P_{\psi 6}^{ N}}$ are  up to several MeV.  If  taking into account the three-body decay widths,  the   branching fractions of  $\textcolor{black}{P_{\psi 2}^{ N}}$,  $\textcolor{black}{P_{\psi 5}^{ N}}$, and $\textcolor{black}{P_{\psi 6}^{ N}}$ decaying into $J/\psi p$ become $\mathcal{B}(\textcolor{black}{P_{\psi 2}^{ N}} \to J/\psi p)=75\%$,   $\mathcal{B}(\textcolor{black}{P_{\psi 5}^{ N}} \to J/\psi p)=23\%$, and $\mathcal{B}(\textcolor{black}{P_{\psi 6}^{ N}} \to J/\psi p)=37\%$ in Scenario A and $\mathcal{B}(\textcolor{black}{P_{\psi 2}^{ N}} \to J/\psi p)=91\%$,   $\mathcal{B}(\textcolor{black}{P_{\psi 5}^{ N}} \to J/\psi p)=10\%$, and $\mathcal{B}( \textcolor{black}{P_{\psi 6}^{ N}} \to J/\psi p)=17\%$ in Scenario B.   
  As result,  the corresponding branching fraction of the decays $\Lambda_b \to (\textcolor{black}{P_{\psi }^{ N}}\to J/\psi p) K$ reduce to 
  $\mathcal{B}[\Lambda_b \to (\textcolor{black}{P_{\psi 2}^{ N}} \to J/\psi p) K]=1.11\times 10^{-6}$,  $\mathcal{B}[\Lambda_b \to (\textcolor{black}{P_{\psi 5}^{ N}} \to J/\psi p) K]=1.47\times 10^{-6}$ and $\mathcal{B}[\Lambda_b \to (\textcolor{black}{P_{\psi 6}^{ N}}\to J/\psi p) K]=3.37\times 10^{-6}$ in Scenario A  and $\mathcal{B}[\Lambda_b \to (\textcolor{black}{P_{\psi 2}^{ N}}\to J/\psi p) K]=2.08\times 10^{-6}$,  $\mathcal{B}[\Lambda_b \to (\textcolor{black}{P_{\psi 5}^{ N}}\to J/\psi p) K]=2.22\times 10^{-6}$ and $\mathcal{B}[\Lambda_b \to (\textcolor{black}{P_{\psi 6}^{ N}}\to J/\psi p) K]=1.26\times 10^{-6}$ in Scenario B.  We can see that  the branching fractions of the pentaquark states $\textcolor{black}{P_{\psi 2}^{ N}}$,  $\textcolor{black}{P_{\psi 5}^{ N}}$, and $\textcolor{black}{P_{\psi 6}^{ N}}$ as hadronic molecules are smaller than those of \textcolor{black}{ $P_{\psi}^{N}(4312)$} and sum of \textcolor{black}{$P_{\psi}^{N}(4440)$} and \textcolor{black}{$P_{\psi}^{N}(4457)$} in Scenario A and Scenario B, which is consistent with the fact that these three HQSS partners have not been seen in the LHCb data sample of  2019.

 These  hidden-charm  pentaquark  molecules  can be seen in the $J\psi p$ invariant mass distribution, and one can also expect to see them  in the  $\bar{D}^* \Lambda_c$ invariant mass distribution. Therefore, with the same approach  we calculate the branching fractions of the decays $\Lambda_b \to (\textcolor{black}{P_{\psi }^{ N}}\to \bar{D}^* \Lambda_c) K$ and the results are shown in Table~\ref{branchingratiospv1sd11}.   We can see that the branching fractions  of  the pentaquark molecules  in the decays   $\Lambda_b \to (\textcolor{black}{P_{\psi }^{ N}} \to J/\psi p) K$ and   $\Lambda_b \to (\textcolor{black}{P_{\psi }^{ N}} \to \bar{D}^* \Lambda_c) K$ are similar except for $\textcolor{black}{P_{\psi 2}^{ N}}$.  The branching fraction $\mathcal{B}[\Lambda_b \to (\textcolor{black}{P_{\psi 2}^{ N}}\to \bar{D}^* \Lambda_c) K]$ is smaller than the branching fraction   $\mathcal{B}[\Lambda_b \to (\textcolor{black}{P_{\psi 2}^{ N}} \to J/\psi p) K]$ by two order of magnitude. We encourage experimental searches for these pentaquark states in the $\bar{D}^*\Lambda_c$ invariant mass distributions of the $\Lambda_b$ decays.
 

\begin{table}[ttt]
\setlength{\tabcolsep}{5pt}
\centering
\caption{Branching fractions($10^{-6}$) of the decays  $\Lambda_b \to (\textcolor{black}{P_{\psi}^{ N}}  \to \bar{D}^*\Lambda_c) K$  in Scenario A and Scenario B.  }
\begin{tabular}{c c c c c c c}
  \hline \hline 
     Scenario   &    \multicolumn{6}{c}{A}  \\ \hline
     Molecule  & $\textcolor{black}{P_{\psi1}^{ N}}$ & $\textcolor{black}{P_{\psi2}^{ N}}$ & $\textcolor{black}{P_{\psi3}^{ N}}$ & $\textcolor{black}{P_{\psi4}^{ N}}$ & $\textcolor{black}{P_{\psi5}^{ N}}$ & $\textcolor{black}{P_{\psi6}^{ N}}$ \\
                    Ours &  4.12  & 0.05 & 1.27  & 0.00 & 0.21 & 1.87 \\    \hline
    Scenario   &   \multicolumn{6}{c}{B}   \\ \hline
     Molecule  & $\textcolor{black}{P_{\psi1}^{ N}}$ & $\textcolor{black}{P_{\psi2}^{ N}}$ & $\textcolor{black}{P_{\psi3}^{ N}}$ & $\textcolor{black}{P_{\psi4}^{ N}}$ & $\textcolor{black}{P_{\psi5}^{ N}}$ & $\textcolor{black}{P_{\psi6}^{ N}}$  \\
         Ours & 26.41   & 0.05 & 8.42  & 2.71 & 4.01 & 2.70 \\
  \hline \hline 
\end{tabular}
\label{branchingratiospv1sd11}
\end{table}

\section{Summary and outlook}
\label{sum}

The three pentaquark states \textcolor{black}{ $P_{\psi}^{N}(4312)$}, \textcolor{black}{$P_{\psi}^{N}(4440)$}, and \textcolor{black}{$P_{\psi}^{N}(4457)$}  can be nicely arranged into a complete  multiplet of $\bar{D}^{(*)}\Sigma_{c}^{(*)}$ hadronic molecules, while their   partial decay widths and production rates in the $\Lambda_b$ decay remain undetermined.     
In this work, we employed the contact-range effective field theory approach to dynamically generate the pentaquark molecules via the $\bar{D}^{(*)}\Sigma_{c}^{(*)}$,  $\bar{D}^{(*)}\Lambda_c$,   $J/\psi p$, and $\eta_c p$ coupled-channel interactions, where the six relevant unknown parameters   were determined by fitting to the experimental data. With the obtained pole positions, we estimated the couplings of the pentaquark molecules to their constituents  $J/\psi p$  and  $\bar{D}^*\Lambda_c$, and then calculated the productions rates of these   molecules in the $\Lambda_b$ decays via the triangle diagrams, where the $\Lambda_b$ baryon weakly decays into $\Lambda_c{D}_s^{(*)}$, then the ${D}_s^{(*)}$ mesons { scatter } into $\bar{D}^{(*)} K$, and finally  the pentaquark molecules are dynamically generated by the $\bar{D}^*\Lambda_c$ interactions. In this work,  with no extra parameters (except those contained in the contact range EFT approach and determined by their masses and widths) we took the effective Lagrangian approach to calculate the triangle diagrams and their production rates in the $\Lambda_b$ decays.   

Our results showed that the masses of the three pentaquark states are well described either in Scenario A or Scenario B, which confirmed our previous conclusion that  we can not determine the favorable scenario in terms of their masses alone. However,     we found that  Scenario A is more favored than Scenario B once their widths are taken into account.  Moreover, our results showed  that  their couplings to $\bar{D}^{(*)}\Lambda_c$  are smaller than those to  $J/\psi p$ in Scenario A,   but   larger in Scenario B. For the branching fractions of the decays $\Lambda_b \to \textcolor{black}{P_{\psi }^{ N}} K$,  that of the \textcolor{black}{ $P_{\psi}^{N}(4312)$} is the largest and those of \textcolor{black}{$P_{\psi}^{N}(4440)$} and \textcolor{black}{$P_{\psi}^{N}(4457)$} with $J=1/2$ are always larger than those with $J=3/2$   in both Scenario A and Scenario B. Moreover, we predicted the following branching fractions:  $\mathcal{B}[\Lambda_b \to \textcolor{black}{P_{\psi 2}^{ N}} K]=(1\sim 2)\times 10^{-6}$,    $\mathcal{B}[\Lambda_b \to \textcolor{black}{P_{\psi 5}^{ N}} K]=(6\sim 22)\times 10^{-6}$ and  $\mathcal{B}[\Lambda_b \to \textcolor{black}{P_{\psi 6}^{ N}} K]=(7\sim 9)\times 10^{-6}$, respectively. 
 
 With the couplings between the molecules and their constituents determined,  we estimated the  branching fractions   $\mathcal{B}(\textcolor{black}{P_{\psi }^{ N}} \to J/\psi p)$, and then obtained the   branching fraction  $\mathcal{B}[\Lambda_b \to (\textcolor{black}{P_{\psi }^{ N}}  \to J/\psi p) K]$. Our results showed that such branching fractions for \textcolor{black}{ $P_{\psi}^{N}(4312)$} and   \textcolor{black}{$P_{\psi}^{N}(4440)$} are consistent with the experimental data,  
while that for \textcolor{black}{$P_{\psi}^{N}(4457)$} is   larger than the experimental data    in Scenario A. For Scenario B, the branching fractions for  \textcolor{black}{$P_{\psi}^{N}(4440)$} and   \textcolor{black}{$P_{\psi}^{N}(4457)$} are consistent with the experimental data, while  that for \textcolor{black}{ $P_{\psi}^{N}(4312)$} is   larger than the experimental data. Given the complicated nature of these decays and the various physical processes involved, we deem  the agreements with the existing data acceptable. Therefore, we conclude that the three  pentaquark states as hidden-charm meson-baryon molecules  can be dynamically generated via the $\bar{D}^{(*)}\Lambda_c$ interactions in the $\Lambda_b$ decay, which further corroborated the molecular interpretations of the pentaquark states. Moreover, the branching fractions of the HQSS partners of \textcolor{black}{ $P_{\psi}^{N}(4312)$}, \textcolor{black}{$P_{\psi}^{N}(4440)$}, and \textcolor{black}{$P_{\psi}^{N}(4457)$}  are estimated to be the order of $10^{-6}$,  smaller than that of \textcolor{black}{ $P_{\psi}^{N}(4312)$} and the sum of those of \textcolor{black}{$P_{\psi}^{N}(4440)$} and \textcolor{black}{$P_{\psi}^{N}(4457)$}.   Therefore, we can attribute the non-observation of the other HQSS partners in the decay $\Lambda_b \to (\textcolor{black}{P_{\psi }^{ N}}  \to J/\psi p) K$ to their small production rates.  As a byproduct, we further predicted the production rates of the pentaquark molecules in the decays $\Lambda_b \to (\textcolor{black}{P_{\psi }^{ N}}  \to \bar{D}^* \Lambda_{c}) K$.

\appendix
\section{Contact-range potentials}
\label{appendix B}
To systematically generate the complete multiplet of hidden-charm pentaquark  molecules, we take into account  the  $\bar{D}^{(*)}\Lambda_c$ channels in the $\bar{D}^{(*)}\Sigma_c^{(*)}$ coupled-channel systems, where the  HQSS plays an important role. First, we express the spin wave function of the $\bar{D}^{(\ast)}\Sigma_{c}^{(\ast)}$ pairs  in terms of the spins of the heavy quarks $s_{1h}$ and $s_{2h}$ and those of the light quark(s) (often referred to as brown 
mucks~\cite{Isgur:1991xa,Flynn:1992fm}) $s_{1l}$ and $s_{2l}$, where 1 and 2 denote $\bar{D}^{(*)}$ and $\Sigma_c^{(*)}$, respectively, via the following spin coupling formula,
\begin{eqnarray}
&&|s_{1l}, s_{1h}, j_{1}; s_{2l}, s_{2h},j_{2}; J\rangle =
  \nonumber\\  &&
  \sqrt{(2j_{1}+1)(2j_{2}+1)(2s_{L}+1)(2s_{H}+1)}\left(\begin{matrix}
s_{1l} & s_{2l} & s_{L} \\
s_{1h} & s_{2h} & s_{H} \\
j_{1} & j_{2} & J%
\end{matrix}\right)\nonumber\\&&|s_{1l},
s_{2l}, s_{L}; s_{1h}, s_{2h},s_{H}; J\rangle.
\label{9j}
\end{eqnarray}
The total light quark spin $s_{L}$  and heavy  quark spin $s_{H}$  are given by   $s_L=s_{1l} \otimes s_{2l}$   and $s_H=s_{1H} \otimes s_{2H}$, respectively.   

More explicitly, for the  $\bar{D}^{(\ast)}\Sigma_{c}$ states, the decompositions read
\begin{widetext}
\begin{eqnarray}
&&|\Sigma_{c}\bar{D}(1/2^{-})\rangle= \frac{1}{2}0_{H}\otimes {1/2}_{L}+ \frac{1}{2\sqrt{3}}1_{H}\otimes {1/2}_{L}+
\sqrt{\frac{2}{3}}1_{H}\otimes {3/2}_{L}\,, \nonumber  \\ \nonumber
&&|\Sigma_{c}^{\ast}\bar{D}(3/2^{-})\rangle =
-\frac{1}{2}0_{H}\otimes {3/2}_{L}+ \frac{1}{\sqrt{3}}1_{H}\otimes
{1/2}_{L}+ \frac{\sqrt{\frac{5}{3}}}{2}1_{H}\otimes {3/2}_{L}\,,  \\ \nonumber
&&|\Sigma_{c}\bar{D}^{\ast}(1/2^{-})\rangle=
\frac{1}{2\sqrt{3}}0_{H}\otimes {1/2}_{L}+ \frac{5}{6}1_{H}\otimes
{1/2}_{L}-\frac{\sqrt{2}}{3}1_{H}\otimes {3/2}_{L}\,,  \\ 
&&|\Sigma_{c}\bar{D}^{\ast}(3/2^{-})\rangle =
\frac{1}{\sqrt{3}}0_{H}\otimes {3/2}_{L}- \frac{1}{3}1_{H}\otimes
{1/2}_{L}+ \frac{\sqrt{5}}{3}1_{H}\otimes {3/2}_{L}\,, \\ \nonumber
&&|\Sigma_{c}^{\ast}\bar{D}^{\ast}(1/2^{-})\rangle=
\sqrt{\frac{2}{3}}0_{H}\otimes {1/2}_{L}-\frac{\sqrt{2}}{3}1_{H}\otimes {1/2}_{L}-\frac{1}{3}1_{H}\otimes
{3/2}_{L}\,,                                 \\ \nonumber
&&|\Sigma_{c}^{\ast}\bar{D}^{\ast}(3/2^{-})\rangle =
\frac{\sqrt{\frac{5}{3}}}{2}0_{H}\otimes {3/2}_{L}+
\frac{\sqrt{5}}{3}1_{H}\otimes {1/2}_{L}- \frac{1}{6}1_{H}\otimes
{3/2}_{L}\,.
\end{eqnarray}
\end{widetext}

The total  light quark spin ${1/2}_{L}$ of the $\bar{D}^{(*)}\Sigma_c^{(*)}$ system   is given by   the  coupling of the light quark spins,  ${1/2}_{1l}\otimes{1}_{2l} $.   Since the light quark spin of $\Lambda_c$ is $0$,   the total light quark spin ${1/2}_{L}^{\prime}$   of the  $\bar{D}^{(*)}\Lambda_c$ system is given by   ${1/2}_{1l}\otimes{0}_{2l} $.   
The decompositions of the  $\bar{D}^{(\ast)}\Lambda_{c}$ states are written as 
\begin{eqnarray}
&&|\bar{D}\Lambda_{c}(J^P=1/2^{-})\rangle = -\frac{1}{2}{0}_{H}\otimes{1/2}_{L}^{\prime}+ \frac{\sqrt{3}}{2}{1}_{H}\otimes
{1/2}_{L}^{\prime}\,, \nonumber  \\ 
&&|\bar{D}^{*}\Lambda_{c}(J^P=1/2^{-})\rangle =
\frac{\sqrt{3}}{2}{0}_{H}\otimes
{1/2}_{L}^{\prime}+ \frac{1}{2}{1}_{H}\otimes
{1/2}_{L}^{\prime}\,,\nonumber  \\ 
&&|\bar{D}^{*}\Lambda_{c}(J^P=3/2^{-})\rangle ={1}_{H}\otimes
{1/2}_{L}^{\prime} \,.
\end{eqnarray}

In the heavy quark limit, the $\bar{D}^{(\ast)}\Sigma_{c}^{(\ast)}\to \bar{D}^{(\ast)}\Sigma_{c}^{(\ast)}$ interactions are independent of the spin of the heavy quark, and therefor the potentials can be parameterized by two coupling constants  describing the interactions between light quarks of spin 1/2 and 3/2, respectively, i.e.,  $F_{1/2}=\langle 1/2_{L} | V| 1/2_{L} \rangle$  and $F_{3/2}=\langle 3/2_{L} | V| 3/2_{L} \rangle$:
\begin{eqnarray}
V_{\Sigma_{c}\bar{D}}(1/2^{-}) &=& \frac{1}{3}F_{1/2L}+
\frac{2}{3}F_{3/2L}\,,   \\ \nonumber
V_{\Sigma_{c}^{\ast}\bar{D}}(3/2^{-}) &=&
\frac{1}{3}F_{1/2L}+
\frac{2}{3}F_{3/2L}\,, \\ \nonumber
V_{\Sigma_{c}\bar{D}^{\ast}}(1/2^{-}) &=&
\frac{7}{9}F_{1/2L}+
\frac{2}{9}F_{3/2L}\,, \\ \nonumber
V_{\Sigma_{c}\bar{D}^{\ast}}(3/2^{-}) &=&
\frac{1}{9}F_{1/2L}+
\frac{8}{9}F_{3/2L}\,,  \\ \nonumber
V_{\Sigma_{c}^{\ast}\bar{D}^{\ast}}(1/2^{-}) &=&\frac{8}{9}F_{1/2L}+
\frac{1}{9}F_{3/2L}\,,   \\ \nonumber
V_{\Sigma_{c}^{\ast}\bar{D}^{\ast}}(3/2^{-}) &=&\frac{5}{9}F_{1/2L}+
\frac{4}{9}F_{3/2L} \,.
\end{eqnarray}
which can be rewritten as a combination of $C_{a}$ and $C_{b}$, i.e., $F_{1/2} = C_a-2C_b$ and  $F_{3/2} = C_a+C_b$~\cite{Liu:2019tjn}.

In the heavy quark limit, the $\bar{D}^{(\ast)}\Lambda_c\to \bar{D}^{(\ast)}\Lambda_c$ interactions are parameterised by one  coupling constant, i.e.,  $F_{1/2L}^{\prime}=\langle 1/2_{L}^{\prime} | V| 1/2_{L}^{\prime} \rangle$: 
\begin{eqnarray}
V_{\bar{D}\Lambda_{c}}(1/2^{-})= V_{\bar{D}^{*}\Lambda_{c}}(1/2^{-})=V_{\bar{D}^{*}\Lambda_{c}}(3/2^{-})=F_{1/2L}^{\prime},
\end{eqnarray}
where the parameter $F_{1/2L}^{\prime}$ can be rewritten as $C_a^{\prime}$.   

The  potentials of $J/\psi N \to J/\psi N  $, $J/\psi N  \to \eta_{c}N$ and $\eta_{c} N  \to \eta_{c}N$ are suppressed  due to the Okubo-Zweig-Iizuka (OZI) rule, which is also supported by  lattice QCD simulations~\cite{Skerbis:2018lew}. In this work, we set $V_{J/\psi(\eta_{c}) N \to J/\psi(\eta_{c}) N }=0$. 

In the heavy quark limit, the $\bar{D}^{(\ast)}\Sigma_{c}^{(\ast)} \to J/\psi(\eta_{c}) N$ and  $\bar{D}^{(\ast)}\Sigma_{c}^{(\ast)} \to J/\psi \Delta$ potentials are allowed, while the  $\bar{D}^{(\ast)}\Sigma_{c}^{(\ast)} \to J/\psi \Delta$ potentials are suppressed due to  isospin symmetry breaking.
From  HQSS, the  $\bar{D}^{(\ast)}\Sigma_{c}^{(\ast)} \to J/\psi(\eta_{c}) N$ interactions  are only related to  the spin of the light quark $1/2$, denoted by one coupling:  $g_2=\langle \bar{D}^{(\ast)}\Sigma_{c}^{(\ast)} | 1_{H}\otimes 1/2_{L} \rangle= \langle \bar{D}^{(\ast)}\Sigma_{c}^{(\ast)} | 0_{H}\otimes 1/2_{L} \rangle $.  Similarly, we can express the $\bar{D}^{(*)}\Lambda_c \to J/\psi(\eta_{c}) N$ interactions  by another parameter: $g_1= \langle \bar{D}^{(\ast)}\Lambda_{c} | 1_{H}\otimes 1/2_{L} \rangle =\langle \bar{D}^{(\ast)}\Lambda_{c}| 0_{H}\otimes 1/2_{L} \rangle$. As for the $\bar{D}^{(\ast)}\Sigma_{c}^{(\ast)} \to \bar{D}^{(*)}\Lambda_c$ interactions, they are   dependent on  only one  coupling  constant in the heavy quark limit. Therefore,  we parameter the $\bar{D}^{(\ast)}\Sigma_{c}^{(\ast)} \to \bar{D}^{(*)}\Lambda_c$ potential by  one  coupling:   $C_b^{\prime}= \langle \bar{D}^{(\ast)}\Sigma_{c}^{(\ast)}  | \bar{D}^{(\ast)}\Lambda_{c} \rangle$.

In the heavy quark limit, the contact-range potentials of $\bar{D}^*\Sigma_{c}^*-\bar{D}^*\Sigma_{c}-\bar{D}\Sigma_{c}-\bar{D}^*\Lambda_{c}-\bar{D}\Lambda_{c}-J/\psi N-\eta_{c}N$ system with $J^{P}=1/2^-$ can be expressed as 
\begin{widetext}
\begin{equation}
    V^{J=1/2}   =   \\   \\ 
    \begin{pmatrix}
  C_a-\frac{5}{3}C_b  &  -\frac{\sqrt{2}}{3}C_b&-\sqrt{\frac{2}{3}}C_b&\sqrt{\frac{2}{3}}C_b^{\prime}&\sqrt{2}C_b^{\prime}&   -\frac{\sqrt{2}}{3}g_{2}  &\sqrt{\frac{2}{3}}g_{2} \\ -\frac{\sqrt{2}}{3}C_b &C_{a}-\frac{4}{3}C_b&\frac{2}{\sqrt{3}}C_b&-\frac{2}{\sqrt{3}}C_b^{\prime}&C_b^{\prime}&  \frac{5}{6}g_{2} & \frac{1}{2\sqrt{3}}g_{2}\\
  -\sqrt{\frac{2}{3}}C_b&    \frac{2}{\sqrt{3}}C_b&C_{a}&C_b^{\prime}& 0 &\frac{1}{2\sqrt{3}}g_{2}   &\frac{1}{2}g_{2} \\
\sqrt{\frac{2}{3}}C_b^{\prime}&  -\frac{2}{\sqrt{3}}C_b^{\prime}&C_b^{\prime}&  C_{a}^{\prime} & 0 & \frac{1}{2}g_{1}  & \frac{\sqrt{3}}{2}g_{1}\\
\sqrt{2}C_b^{\prime} &  C_b^{\prime}&0  &0 &C_a^{\prime}&\frac{\sqrt{3}}{2}g_{1}&-\frac{1}{2}g_{1} \\
-\frac{\sqrt{2}}{3}g_{2}  & \frac{5}{6}g_{2}&\frac{1}{2\sqrt{3}}g_{2}&\frac{1}{2}g_{1}&\frac{\sqrt{3}}{2}g_{1} &0&0  \\
\sqrt{\frac{2}{3}}g_{2} &\frac{1}{2\sqrt{3}}g_{2}&\frac{1}{2}g_{2}&\frac{\sqrt{3}}{2}g_{1}&-\frac{1}{2}g_{1}&0&0\end{pmatrix}
\label{contact11}
\end{equation}
\end{widetext}
and the contact potentials of $\bar{D}^*\Sigma_{c}^*-\bar{D}^*\Sigma_{c}-\bar{D}\Sigma_{c}^*-\bar{D}^*\Lambda_{c}- J/\psi N$ system with $J^{P}=3/2^-$
are written as  
\begin{equation}
    V^{J=3/2}=\begin{pmatrix}
        C_a-\frac{2}{3}C_{b}& -\frac{\sqrt{5}}{3}C_b & \sqrt{\frac{5}{3}}C_{b} & \sqrt{\frac{5}{3}}C_{b}^{\prime}&\frac{\sqrt{5}}{3} g_{2}\\
   -\frac{\sqrt{5}}{3}C_b & C_a+\frac{2}{3}C_{b}& \frac{1}{\sqrt{3}}C_{b} & \frac{1}{\sqrt{3}}C_{b}^{\prime}&-\frac{1}{3} g_{2}\\
  \sqrt{\frac{5}{3}}C_{b}&   \frac{1}{\sqrt{3}}C_{b}&C_{a}&-C_{b}^{\prime}&\frac{1}{\sqrt{3}} g_{2}\\
 \sqrt{\frac{5}{3}}C_{b}^{\prime} &   \frac{1}{\sqrt{3}}C_{b}^{\prime}&-C_{b}^{\prime} &  C_a^{\prime}&g_{1}\\
   \frac{\sqrt{5}}{3} g_{2} &  -\frac{1}{3}g_{2}&\frac{1}{\sqrt{3}} g_{2} & g_{1}&0\end{pmatrix}
   \label{contact22}
\end{equation}

\section{light meson saturation}

Following Ref.~\cite{Peng:2020xrf}, we expect the EFT couplings $C_a(C_a^\prime)$ and  $C_b(C_b^{\prime})$
 to be saturated by scalar and vector meson exchanges
\begin{eqnarray}
  C_a^{\rm sat(\prime)}(\Lambda \sim m_{\sigma}, m_V) &\propto& C_a^{S} + C_a^V \, , \\
  C_b^{\rm sat(\prime)}(\Lambda \sim m_{\sigma}, m_V) &\propto& C_b^V \, . 
\end{eqnarray}

The value of the saturated couplings is expected to be proportional to the light meson $M$
potential  $V_M(\vec{q})$ 
at $|\vec{q}| = 0$ once we have removed
the  Dirac-delta term~\cite{Peng:2020xrf}.
According to  the one-boson exchange(OBE) model,  the $\sigma$ and $\rho(\omega)$ exchanges are responsible for the  $\bar{D}^{(\ast)}\Sigma_{c}^{(\ast)} \to \bar{D}^{(\ast)}\Sigma_{c}^{(\ast)} $ potentials, while  $\sigma$,$\omega$ exchanges  and $\rho$ exchange  are  allowed for the $\bar{D}^{(*)}\Lambda_{c} \to \bar{D}^{(*)}\Lambda_{c}$ potentials and  $\bar{D}^{(\ast)}\Sigma_{c}^{(\ast)} \to \bar{D}^{(*)}\Lambda_{c}$ potentials  due to  isospin symmetry. 
This gives us~\cite{Liu:2019zvb}
\begin{eqnarray}
  C_a^{{\rm sat} (\sigma)}(\Lambda \sim m_{\sigma})
  &\propto& - \frac{g_{\sigma 1} g_{\sigma 2}}{m_{\sigma}^2} \, , \\
  C_a^{{\rm sat } (V)}(\Lambda \sim m_{\rho})
  &\propto&  \frac{g_{V 1} g_{V 2}}{m_{V}^2}\,(1 + \vec{\tau}_1 \cdot \vec{T}_2)
  \, , \\
  C_b^{{\rm sat} (V)}(\Lambda \sim m_{\rho})
  &\propto&  \frac{f_{V 1} f_{V 2}}{6 M^2}
  (1 + \vec{\tau}_1 \cdot \vec{T}_2)
  \, ,  \\ 
    C_a^{ {\rm sat} (\sigma) \, \prime}(\Lambda \sim m_{\sigma})
  &\propto& - \frac{g_{\sigma 1} g_{\sigma 3}}{m_{\sigma}^2} \, , \\
  C_a^{{\rm sat } (V)\, \prime}(\Lambda \sim m_{\omega})
  &\propto&  \frac{g_{V 1} g_{V 3}}{m_{V}^2}\,
  , \\
  C_b^{{\rm sat} (V) \prime}(\Lambda \sim m_{\rho})
  &\propto&  \frac{f_{V 1} f_{V 3}}{6 M^2}
  ( \vec{\tau}_1 \cdot \vec{t}_2) \,.
\end{eqnarray}
where $V = \rho, \omega$ and we have made the simplification
that $m_\rho = m_\omega = m_V$.
The proportionality constant is unknown and depends on the details of
the renormalization process. In this work, we assume that these proportionality constants are the same. The $g_{\sigma_{1}}$,  $g_{\sigma_{2}}$, and $g_{\sigma_{3}}$ denote the couplings of the $\bar{D}^{(*)}$ mesons, $\Sigma_{c}^{(*)}$ baryons, and $\Lambda_c$ baryon   to the sigma meson, and  $g_{v1}$, $g_{v2}$, and $g_{v3}$ ($f_{v1}$, $f_{v2}$, and $f_{v3}$ ) denote the electric-type (magnetic-type) couplings between the $\bar{D}^{(*)}$ mesons, $\Sigma_{c}^{(*)}$ baryons, and $\Lambda_c$ baryon and a light vector meson.  $M$ is a mass scale to render $f_{v}$ dimensionless.  Following Refs.~\cite{Liu:2011xc,Liu:2019zvb}, we tabulate the values of these couplings in Table~\ref{tab:couplingsDS}. 
The  $\vec{\tau}_1 \cdot \vec{T}_2$ and   $\vec{\tau}_1 \cdot \vec{t}_2$ are the isospin factors of $\bar{D}^{(\ast)}\Sigma_{c}^{(\ast)} \to \bar{D}^{(\ast)}\Sigma_{c}^{(\ast)} $ potentials and $\bar{D}^{(\ast)}\Sigma_{c}^{(\ast)} \to \bar{D}^{(*)}\Lambda_{c}$ potentials, which are  $\vec{\tau}_1 \cdot \vec{T}_2=-2$ and $\vec{\tau}_1 \cdot \vec{t}_2=-\sqrt{3}$ for the total isospin $I=1/2$.

\begin{table}[ttt]
\setlength{\tabcolsep}{6pt}
\centering
\caption{Couplings of the light mesons of the OBE model
  ( $\sigma$, $\rho$, $\omega$) to the heavy-meson
  and heavy-baryon.
  For the magnetic-type coupling of the $\rho$ and $\omega$ vector mesons
  we have used the decomposition
  $f_{V} = \kappa_{V}\,g_{V}$, with $V=\rho,\omega$.
  $M=940$ MeV refers to the mass scale involved
  in the magnetic-type couplings~\cite{Liu:2011xc,Liu:2019zvb}.
}
\begin{tabular}{cc|cc|cc}
  \hline \hline 
  Coupling  & $P$/$P^*$ &   Coupling  & $\Sigma_Q$/$\Sigma_Q^*$    &   Coupling  & $\Lambda_Q$ \\
  \hline
  $g_{\sigma 1}$ & 3.4 &  $g_{\sigma 2}$ & 6.8 &  $g_{\sigma 3}$ & 3.4  \\
  $g_{v 1}$ & 2.6  & $g_{v 2}$ & 5.8 & $g_{v 3}$ & 2.9\\
  $\kappa_{v 1}$ & 2.3 &$\kappa_{v 2}$ & 1.7 &$\kappa_{v 3}$ & 1.2 \\
  \hline \hline 
\end{tabular}
\label{tab:couplingsDS}
\end{table}

\acknowledgments

We are grateful to Eulogio Oset, Fu-Sheng Yu,  Chu-Wen Xiao,  Jun-Xu Lu, and Qi Wu for useful discussions.   
  This work is supported in part by the National Natural Science Foundation of China under Grants No.11975041 and No.11961141004. Ming-Zhu Liu acknowledges support from the National Natural Science Foundation of
China under Grant No.12105007.

\bibliography{biblio.bib}

\begin{thebibliography}{108}
\expandafter\ifx\csname natexlab\endcsname\relax\def\natexlab#1{#1}\fi
\expandafter\ifx\csname bibnamefont\endcsname\relax
  \def\bibnamefont#1{#1}\fi
\expandafter\ifx\csname bibfnamefont\endcsname\relax
  \def\bibfnamefont#1{#1}\fi
\expandafter\ifx\csname citenamefont\endcsname\relax
  \def\citenamefont#1{#1}\fi
\expandafter\ifx\csname url\endcsname\relax
  \def\url#1{\texttt{#1}}\fi
\expandafter\ifx\csname urlprefix\endcsname\relax\def\urlprefix{URL }\fi
\providecommand{\bibinfo}[2]{#2}
\providecommand{\eprint}[2][]{\url{#2}}

\bibitem[{\citenamefont{Aaij et~al.}(2015)}]{Aaij:2015tga}
\bibinfo{author}{\bibfnamefont{R.}~\bibnamefont{Aaij}} \bibnamefont{et~al.} (\bibinfo{collaboration}{LHCb}), \bibinfo{journal}{Phys. Rev. Lett.} \textbf{\bibinfo{volume}{115}}, \bibinfo{pages}{072001} (\bibinfo{year}{2015}), \eprint{1507.03414}.

\bibitem[{\citenamefont{Aaij et~al.}(2019)}]{Aaij:2019vzc}
\bibinfo{author}{\bibfnamefont{R.}~\bibnamefont{Aaij}} \bibnamefont{et~al.} (\bibinfo{collaboration}{LHCb}), \bibinfo{journal}{Phys. Rev. Lett.} \textbf{\bibinfo{volume}{122}}, \bibinfo{pages}{222001} (\bibinfo{year}{2019}), \eprint{1904.03947}.

\bibitem[{\citenamefont{Aaij et~al.}(2022)}]{LHCb:2021chn}
\bibinfo{author}{\bibfnamefont{R.}~\bibnamefont{Aaij}} \bibnamefont{et~al.} (\bibinfo{collaboration}{LHCb}), \bibinfo{journal}{Phys. Rev. Lett.} \textbf{\bibinfo{volume}{128}}, \bibinfo{pages}{062001} (\bibinfo{year}{2022}), \eprint{2108.04720}.

\bibitem[{\citenamefont{Aaij et~al.}(2021)}]{LHCb:2020jpq}
\bibinfo{author}{\bibfnamefont{R.}~\bibnamefont{Aaij}} \bibnamefont{et~al.} (\bibinfo{collaboration}{LHCb}), \bibinfo{journal}{Sci. Bull.} \textbf{\bibinfo{volume}{66}}, \bibinfo{pages}{1278} (\bibinfo{year}{2021}), \eprint{2012.10380}.

\bibitem[{\citenamefont{Aaij et~al.}(2023)}]{LHCb:2022ogu}
\bibinfo{author}{\bibfnamefont{R.}~\bibnamefont{Aaij}} \bibnamefont{et~al.} (\bibinfo{collaboration}{LHCb}), \bibinfo{journal}{Phys. Rev. Lett.} \textbf{\bibinfo{volume}{131}}, \bibinfo{pages}{031901} (\bibinfo{year}{2023}), \eprint{2210.10346}.

\bibitem[{\citenamefont{Chen et~al.}(2019{\natexlab{a}})\citenamefont{Chen, Sun, Liu, and Zhu}}]{Chen:2019asm}
\bibinfo{author}{\bibfnamefont{R.}~\bibnamefont{Chen}}, \bibinfo{author}{\bibfnamefont{Z.-F.} \bibnamefont{Sun}}, \bibinfo{author}{\bibfnamefont{X.}~\bibnamefont{Liu}}, \bibnamefont{and} \bibinfo{author}{\bibfnamefont{S.-L.} \bibnamefont{Zhu}}, \bibinfo{journal}{Phys. Rev.} \textbf{\bibinfo{volume}{D100}}, \bibinfo{pages}{011502} (\bibinfo{year}{2019}{\natexlab{a}}), \eprint{1903.11013}.

\bibitem[{\citenamefont{He}(2019)}]{He:2019ify}
\bibinfo{author}{\bibfnamefont{J.}~\bibnamefont{He}}, \bibinfo{journal}{Eur. Phys. J.} \textbf{\bibinfo{volume}{C79}}, \bibinfo{pages}{393} (\bibinfo{year}{2019}), \eprint{1903.11872}.

\bibitem[{\citenamefont{Chen et~al.}(2019{\natexlab{b}})\citenamefont{Chen, Chen, and Zhu}}]{Chen:2019bip}
\bibinfo{author}{\bibfnamefont{H.-X.} \bibnamefont{Chen}}, \bibinfo{author}{\bibfnamefont{W.}~\bibnamefont{Chen}}, \bibnamefont{and} \bibinfo{author}{\bibfnamefont{S.-L.} \bibnamefont{Zhu}}, \bibinfo{journal}{Phys. Rev.} \textbf{\bibinfo{volume}{D100}}, \bibinfo{pages}{051501} (\bibinfo{year}{2019}{\natexlab{b}}), \eprint{1903.11001}.

\bibitem[{\citenamefont{Xiao et~al.}(2019{\natexlab{a}})\citenamefont{Xiao, Nieves, and Oset}}]{Xiao:2019aya}
\bibinfo{author}{\bibfnamefont{C.~W.} \bibnamefont{Xiao}}, \bibinfo{author}{\bibfnamefont{J.}~\bibnamefont{Nieves}}, \bibnamefont{and} \bibinfo{author}{\bibfnamefont{E.}~\bibnamefont{Oset}}, \bibinfo{journal}{Phys. Rev.} \textbf{\bibinfo{volume}{D100}}, \bibinfo{pages}{014021} (\bibinfo{year}{2019}{\natexlab{a}}), \eprint{1904.01296}.

\bibitem[{\citenamefont{Sakai et~al.}(2019)\citenamefont{Sakai, Jing, and Guo}}]{Sakai:2019qph}
\bibinfo{author}{\bibfnamefont{S.}~\bibnamefont{Sakai}}, \bibinfo{author}{\bibfnamefont{H.-J.} \bibnamefont{Jing}}, \bibnamefont{and} \bibinfo{author}{\bibfnamefont{F.-K.} \bibnamefont{Guo}}, \bibinfo{journal}{Phys. Rev.} \textbf{\bibinfo{volume}{D100}}, \bibinfo{pages}{074007} (\bibinfo{year}{2019}), \eprint{1907.03414}.

\bibitem[{\citenamefont{Yamaguchi et~al.}(2020)\citenamefont{Yamaguchi, Garc\'\i{}a-Tecocoatzi, Giachino, Hosaka, Santopinto, Takeuchi, and Takizawa}}]{Yamaguchi:2019seo}
\bibinfo{author}{\bibfnamefont{Y.}~\bibnamefont{Yamaguchi}}, \bibinfo{author}{\bibfnamefont{H.}~\bibnamefont{Garc\'\i{}a-Tecocoatzi}}, \bibinfo{author}{\bibfnamefont{A.}~\bibnamefont{Giachino}}, \bibinfo{author}{\bibfnamefont{A.}~\bibnamefont{Hosaka}}, \bibinfo{author}{\bibfnamefont{E.}~\bibnamefont{Santopinto}}, \bibinfo{author}{\bibfnamefont{S.}~\bibnamefont{Takeuchi}}, \bibnamefont{and} \bibinfo{author}{\bibfnamefont{M.}~\bibnamefont{Takizawa}}, \bibinfo{journal}{Phys. Rev. D} \textbf{\bibinfo{volume}{101}}, \bibinfo{pages}{091502} (\bibinfo{year}{2020}), \eprint{1907.04684}.

\bibitem[{\citenamefont{He and Chen}(2019)}]{He:2019rva}
\bibinfo{author}{\bibfnamefont{J.}~\bibnamefont{He}} \bibnamefont{and} \bibinfo{author}{\bibfnamefont{D.-Y.} \bibnamefont{Chen}}, \bibinfo{journal}{Eur. Phys. J.} \textbf{\bibinfo{volume}{C79}}, \bibinfo{pages}{887} (\bibinfo{year}{2019}), \eprint{1909.05681}.

\bibitem[{\citenamefont{Liu et~al.}(2021)\citenamefont{Liu, Wu, S\'anchez~S\'anchez, Valderrama, Geng, and Xie}}]{Liu:2019zvb}
\bibinfo{author}{\bibfnamefont{M.-Z.} \bibnamefont{Liu}}, \bibinfo{author}{\bibfnamefont{T.-W.} \bibnamefont{Wu}}, \bibinfo{author}{\bibfnamefont{M.}~\bibnamefont{S\'anchez~S\'anchez}}, \bibinfo{author}{\bibfnamefont{M.~P.} \bibnamefont{Valderrama}}, \bibinfo{author}{\bibfnamefont{L.-S.} \bibnamefont{Geng}}, \bibnamefont{and} \bibinfo{author}{\bibfnamefont{J.-J.} \bibnamefont{Xie}}, \bibinfo{journal}{Phys. Rev. D} \textbf{\bibinfo{volume}{103}}, \bibinfo{pages}{054004} (\bibinfo{year}{2021}), \eprint{1907.06093}.

\bibitem[{\citenamefont{Pavon~Valderrama}(2019{\natexlab{a}})}]{Valderrama:2019chc}
\bibinfo{author}{\bibfnamefont{M.}~\bibnamefont{Pavon~Valderrama}}, \bibinfo{journal}{Phys. Rev.} \textbf{\bibinfo{volume}{D100}}, \bibinfo{pages}{094028} (\bibinfo{year}{2019}{\natexlab{a}}), \eprint{1907.05294}.

\bibitem[{\citenamefont{Meng et~al.}(2019)\citenamefont{Meng, Wang, Wang, and Zhu}}]{Meng:2019ilv}
\bibinfo{author}{\bibfnamefont{L.}~\bibnamefont{Meng}}, \bibinfo{author}{\bibfnamefont{B.}~\bibnamefont{Wang}}, \bibinfo{author}{\bibfnamefont{G.-J.} \bibnamefont{Wang}}, \bibnamefont{and} \bibinfo{author}{\bibfnamefont{S.-L.} \bibnamefont{Zhu}}, \bibinfo{journal}{Phys. Rev.} \textbf{\bibinfo{volume}{D100}}, \bibinfo{pages}{014031} (\bibinfo{year}{2019}), \eprint{1905.04113}.

\bibitem[{\citenamefont{Du et~al.}(2020)\citenamefont{Du, Baru, Guo, Hanhart, Meißner, Oller, and Wang}}]{Du:2019pij}
\bibinfo{author}{\bibfnamefont{M.-L.} \bibnamefont{Du}}, \bibinfo{author}{\bibfnamefont{V.}~\bibnamefont{Baru}}, \bibinfo{author}{\bibfnamefont{F.-K.} \bibnamefont{Guo}}, \bibinfo{author}{\bibfnamefont{C.}~\bibnamefont{Hanhart}}, \bibinfo{author}{\bibfnamefont{U.-G.} \bibnamefont{Meißner}}, \bibinfo{author}{\bibfnamefont{J.~A.} \bibnamefont{Oller}}, \bibnamefont{and} \bibinfo{author}{\bibfnamefont{Q.}~\bibnamefont{Wang}}, \bibinfo{journal}{Phys. Rev. Lett.} \textbf{\bibinfo{volume}{124}}, \bibinfo{pages}{072001} (\bibinfo{year}{2020}), \eprint{1910.11846}.

\bibitem[{\citenamefont{Ling et~al.}(2021{\natexlab{a}})\citenamefont{Ling, Lu, Liu, and Geng}}]{Ling:2021lmq}
\bibinfo{author}{\bibfnamefont{X.-Z.} \bibnamefont{Ling}}, \bibinfo{author}{\bibfnamefont{J.-X.} \bibnamefont{Lu}}, \bibinfo{author}{\bibfnamefont{M.-Z.} \bibnamefont{Liu}}, \bibnamefont{and} \bibinfo{author}{\bibfnamefont{L.-S.} \bibnamefont{Geng}}, \bibinfo{journal}{Phys. Rev. D} \textbf{\bibinfo{volume}{104}}, \bibinfo{pages}{074022} (\bibinfo{year}{2021}{\natexlab{a}}), \eprint{2106.12250}.

\bibitem[{\citenamefont{Dong et~al.}(2021{\natexlab{a}})\citenamefont{Dong, Guo, and Zou}}]{Dong:2021juy}
\bibinfo{author}{\bibfnamefont{X.-K.} \bibnamefont{Dong}}, \bibinfo{author}{\bibfnamefont{F.-K.} \bibnamefont{Guo}}, \bibnamefont{and} \bibinfo{author}{\bibfnamefont{B.-S.} \bibnamefont{Zou}}, \bibinfo{journal}{Progr. Phys.} \textbf{\bibinfo{volume}{41}}, \bibinfo{pages}{65} (\bibinfo{year}{2021}{\natexlab{a}}), \eprint{2101.01021}.

\bibitem[{\citenamefont{\"Ozdem}(2021)}]{Ozdem:2021ugy}
\bibinfo{author}{\bibfnamefont{U.}~\bibnamefont{\"Ozdem}}, \bibinfo{journal}{Eur. Phys. J. C} \textbf{\bibinfo{volume}{81}}, \bibinfo{pages}{277} (\bibinfo{year}{2021}), \eprint{2102.01996}.

\bibitem[{\citenamefont{Pan et~al.}(2022{\natexlab{a}})\citenamefont{Pan, Wu, Liu, and Geng}}]{Pan:2022xxz}
\bibinfo{author}{\bibfnamefont{Y.-W.} \bibnamefont{Pan}}, \bibinfo{author}{\bibfnamefont{T.-W.} \bibnamefont{Wu}}, \bibinfo{author}{\bibfnamefont{M.-Z.} \bibnamefont{Liu}}, \bibnamefont{and} \bibinfo{author}{\bibfnamefont{L.-S.} \bibnamefont{Geng}}, \bibinfo{journal}{Phys. Rev. D} \textbf{\bibinfo{volume}{105}}, \bibinfo{pages}{114048} (\bibinfo{year}{2022}{\natexlab{a}}), \eprint{2204.02295}.

\bibitem[{\citenamefont{Zhang et~al.}(2023)\citenamefont{Zhang, Liu, Hu, Wang, and Mei\ss{}ner}}]{Zhang:2023czx}
\bibinfo{author}{\bibfnamefont{Z.}~\bibnamefont{Zhang}}, \bibinfo{author}{\bibfnamefont{J.}~\bibnamefont{Liu}}, \bibinfo{author}{\bibfnamefont{J.}~\bibnamefont{Hu}}, \bibinfo{author}{\bibfnamefont{Q.}~\bibnamefont{Wang}}, \bibnamefont{and} \bibinfo{author}{\bibfnamefont{U.-G.} \bibnamefont{Mei\ss{}ner}}, \bibinfo{journal}{Sci. Bull.} \textbf{\bibinfo{volume}{68}}, \bibinfo{pages}{981} (\bibinfo{year}{2023}), \eprint{2301.05364}.

\bibitem[{\citenamefont{Pan et~al.}(2022{\natexlab{b}})\citenamefont{Pan, Wu, Liu, and Geng}}]{Pan:2022whr}
\bibinfo{author}{\bibfnamefont{Y.-W.} \bibnamefont{Pan}}, \bibinfo{author}{\bibfnamefont{T.-W.} \bibnamefont{Wu}}, \bibinfo{author}{\bibfnamefont{M.-Z.} \bibnamefont{Liu}}, \bibnamefont{and} \bibinfo{author}{\bibfnamefont{L.-S.} \bibnamefont{Geng}}, \bibinfo{journal}{Eur. Phys. J. C} \textbf{\bibinfo{volume}{82}}, \bibinfo{pages}{908} (\bibinfo{year}{2022}{\natexlab{b}}), \eprint{2208.05385}.

\bibitem[{\citenamefont{Liu et~al.}(2023)\citenamefont{Liu, Lu, Liu, and Geng}}]{Liu:2023wfo}
\bibinfo{author}{\bibfnamefont{Z.-W.} \bibnamefont{Liu}}, \bibinfo{author}{\bibfnamefont{J.-X.} \bibnamefont{Lu}}, \bibinfo{author}{\bibfnamefont{M.-Z.} \bibnamefont{Liu}}, \bibnamefont{and} \bibinfo{author}{\bibfnamefont{L.-S.} \bibnamefont{Geng}}, \bibinfo{journal}{Phys. Rev. D} \textbf{\bibinfo{volume}{108}}, \bibinfo{pages}{L031503} (\bibinfo{year}{2023}), \eprint{2305.19048}.

\bibitem[{\citenamefont{Eides et~al.}(2020)\citenamefont{Eides, Petrov, and Polyakov}}]{Eides:2019tgv}
\bibinfo{author}{\bibfnamefont{M.~I.} \bibnamefont{Eides}}, \bibinfo{author}{\bibfnamefont{V.~Y.} \bibnamefont{Petrov}}, \bibnamefont{and} \bibinfo{author}{\bibfnamefont{M.~V.} \bibnamefont{Polyakov}}, \bibinfo{journal}{Mod. Phys. Lett. A} \textbf{\bibinfo{volume}{35}}, \bibinfo{pages}{2050151} (\bibinfo{year}{2020}), \eprint{1904.11616}.

\bibitem[{\citenamefont{Ali and Parkhomenko}(2019)}]{Ali:2019npk}
\bibinfo{author}{\bibfnamefont{A.}~\bibnamefont{Ali}} \bibnamefont{and} \bibinfo{author}{\bibfnamefont{A.~Y.} \bibnamefont{Parkhomenko}}, \bibinfo{journal}{Phys. Lett. B} \textbf{\bibinfo{volume}{793}}, \bibinfo{pages}{365} (\bibinfo{year}{2019}), \eprint{1904.00446}.

\bibitem[{\citenamefont{Wang}(2020)}]{Wang:2019got}
\bibinfo{author}{\bibfnamefont{Z.-G.} \bibnamefont{Wang}}, \bibinfo{journal}{Int. J. Mod. Phys. A} \textbf{\bibinfo{volume}{35}}, \bibinfo{pages}{2050003} (\bibinfo{year}{2020}), \eprint{1905.02892}.

\bibitem[{\citenamefont{Cheng and Liu}(2019)}]{Cheng:2019obk}
\bibinfo{author}{\bibfnamefont{J.-B.} \bibnamefont{Cheng}} \bibnamefont{and} \bibinfo{author}{\bibfnamefont{Y.-R.} \bibnamefont{Liu}}, \bibinfo{journal}{Phys. Rev.} \textbf{\bibinfo{volume}{D100}}, \bibinfo{pages}{054002} (\bibinfo{year}{2019}), \eprint{1905.08605}.

\bibitem[{\citenamefont{Weng et~al.}(2019)\citenamefont{Weng, Chen, Deng, and Zhu}}]{Weng:2019ynv}
\bibinfo{author}{\bibfnamefont{X.-Z.} \bibnamefont{Weng}}, \bibinfo{author}{\bibfnamefont{X.-L.} \bibnamefont{Chen}}, \bibinfo{author}{\bibfnamefont{W.-Z.} \bibnamefont{Deng}}, \bibnamefont{and} \bibinfo{author}{\bibfnamefont{S.-L.} \bibnamefont{Zhu}}, \bibinfo{journal}{Phys. Rev. D} \textbf{\bibinfo{volume}{100}}, \bibinfo{pages}{016014} (\bibinfo{year}{2019}), \eprint{1904.09891}.

\bibitem[{\citenamefont{Zhu et~al.}(2019)\citenamefont{Zhu, Liu, Huang, and Qiao}}]{Zhu:2019iwm}
\bibinfo{author}{\bibfnamefont{R.}~\bibnamefont{Zhu}}, \bibinfo{author}{\bibfnamefont{X.}~\bibnamefont{Liu}}, \bibinfo{author}{\bibfnamefont{H.}~\bibnamefont{Huang}}, \bibnamefont{and} \bibinfo{author}{\bibfnamefont{C.-F.} \bibnamefont{Qiao}}, \bibinfo{journal}{Phys. Lett.} \textbf{\bibinfo{volume}{B797}}, \bibinfo{pages}{134869} (\bibinfo{year}{2019}), \eprint{1904.10285}.

\bibitem[{\citenamefont{Pimikov et~al.}(2020)\citenamefont{Pimikov, Lee, and Zhang}}]{Pimikov:2019dyr}
\bibinfo{author}{\bibfnamefont{A.}~\bibnamefont{Pimikov}}, \bibinfo{author}{\bibfnamefont{H.-J.} \bibnamefont{Lee}}, \bibnamefont{and} \bibinfo{author}{\bibfnamefont{P.}~\bibnamefont{Zhang}}, \bibinfo{journal}{Phys. Rev.} \textbf{\bibinfo{volume}{D101}}, \bibinfo{pages}{014002} (\bibinfo{year}{2020}), \eprint{1908.04459}.

\bibitem[{\citenamefont{Ruangyoo et~al.}(2022)\citenamefont{Ruangyoo, Phumphan, Chen, Limphirat, and Yan}}]{Ruangyoo:2021aoi}
\bibinfo{author}{\bibfnamefont{W.}~\bibnamefont{Ruangyoo}}, \bibinfo{author}{\bibfnamefont{K.}~\bibnamefont{Phumphan}}, \bibinfo{author}{\bibfnamefont{C.-C.} \bibnamefont{Chen}}, \bibinfo{author}{\bibfnamefont{A.}~\bibnamefont{Limphirat}}, \bibnamefont{and} \bibinfo{author}{\bibfnamefont{Y.}~\bibnamefont{Yan}}, \bibinfo{journal}{J. Phys. G} \textbf{\bibinfo{volume}{49}}, \bibinfo{pages}{075001} (\bibinfo{year}{2022}), \eprint{2105.14249}.

\bibitem[{\citenamefont{Fernández-Ramírez et~al.}(2019)\citenamefont{Fernández-Ramírez, Pilloni, Albaladejo, Jackura, Mathieu, Mikhasenko, Silva-Castro, and Szczepaniak}}]{Fernandez-Ramirez:2019koa}
\bibinfo{author}{\bibfnamefont{C.}~\bibnamefont{Fernández-Ramírez}}, \bibinfo{author}{\bibfnamefont{A.}~\bibnamefont{Pilloni}}, \bibinfo{author}{\bibfnamefont{M.}~\bibnamefont{Albaladejo}}, \bibinfo{author}{\bibfnamefont{A.}~\bibnamefont{Jackura}}, \bibinfo{author}{\bibfnamefont{V.}~\bibnamefont{Mathieu}}, \bibinfo{author}{\bibfnamefont{M.}~\bibnamefont{Mikhasenko}}, \bibinfo{author}{\bibfnamefont{J.~A.} \bibnamefont{Silva-Castro}}, \bibnamefont{and} \bibinfo{author}{\bibfnamefont{A.~P.} \bibnamefont{Szczepaniak}} (\bibinfo{collaboration}{JPAC}), \bibinfo{journal}{Phys. Rev. Lett.} \textbf{\bibinfo{volume}{123}}, \bibinfo{pages}{092001} (\bibinfo{year}{2019}), \eprint{1904.10021}.

\bibitem[{\citenamefont{Nakamura}(2021)}]{Nakamura:2021qvy}
\bibinfo{author}{\bibfnamefont{S.~X.} \bibnamefont{Nakamura}}, \bibinfo{journal}{Phys. Rev. D} \textbf{\bibinfo{volume}{103}}, \bibinfo{pages}{111503} (\bibinfo{year}{2021}), \eprint{2103.06817}.

\bibitem[{\citenamefont{Burns and Swanson}(2022{\natexlab{a}})}]{Burns:2022uiv}
\bibinfo{author}{\bibfnamefont{T.~J.} \bibnamefont{Burns}} \bibnamefont{and} \bibinfo{author}{\bibfnamefont{E.~S.} \bibnamefont{Swanson}}, \bibinfo{journal}{Phys. Rev. D} \textbf{\bibinfo{volume}{106}}, \bibinfo{pages}{054029} (\bibinfo{year}{2022}{\natexlab{a}}), \eprint{2207.00511}.

\bibitem[{\citenamefont{Xie et~al.}(2022)\citenamefont{Xie, Ling, Liu, and Geng}}]{Xie:2022hhv}
\bibinfo{author}{\bibfnamefont{J.-M.} \bibnamefont{Xie}}, \bibinfo{author}{\bibfnamefont{X.-Z.} \bibnamefont{Ling}}, \bibinfo{author}{\bibfnamefont{M.-Z.} \bibnamefont{Liu}}, \bibnamefont{and} \bibinfo{author}{\bibfnamefont{L.-S.} \bibnamefont{Geng}}, \bibinfo{journal}{Eur. Phys. J. C} \textbf{\bibinfo{volume}{82}}, \bibinfo{pages}{1061} (\bibinfo{year}{2022}), \eprint{2204.12356}.

\bibitem[{\citenamefont{Lin et~al.}(2017)\citenamefont{Lin, Shen, Guo, and Zou}}]{Lin:2017mtz}
\bibinfo{author}{\bibfnamefont{Y.-H.} \bibnamefont{Lin}}, \bibinfo{author}{\bibfnamefont{C.-W.} \bibnamefont{Shen}}, \bibinfo{author}{\bibfnamefont{F.-K.} \bibnamefont{Guo}}, \bibnamefont{and} \bibinfo{author}{\bibfnamefont{B.-S.} \bibnamefont{Zou}}, \bibinfo{journal}{Phys. Rev. D} \textbf{\bibinfo{volume}{95}}, \bibinfo{pages}{114017} (\bibinfo{year}{2017}), \eprint{1703.01045}.

\bibitem[{\citenamefont{Burns and Swanson}(2022{\natexlab{b}})}]{Burns:2021jlu}
\bibinfo{author}{\bibfnamefont{T.~J.} \bibnamefont{Burns}} \bibnamefont{and} \bibinfo{author}{\bibfnamefont{E.~S.} \bibnamefont{Swanson}}, \bibinfo{journal}{Eur. Phys. J. A} \textbf{\bibinfo{volume}{58}}, \bibinfo{pages}{68} (\bibinfo{year}{2022}{\natexlab{b}}), \eprint{2112.11527}.

\bibitem[{\citenamefont{Meziani et~al.}(2016)}]{Meziani:2016lhg}
\bibinfo{author}{\bibfnamefont{Z.~E.} \bibnamefont{Meziani}} \bibnamefont{et~al.} (\bibinfo{year}{2016}), \eprint{1609.00676}.

\bibitem[{\citenamefont{Ali et~al.}(2019)}]{GlueX:2019mkq}
\bibinfo{author}{\bibfnamefont{A.}~\bibnamefont{Ali}} \bibnamefont{et~al.} (\bibinfo{collaboration}{GlueX}), \bibinfo{journal}{Phys. Rev. Lett.} \textbf{\bibinfo{volume}{123}}, \bibinfo{pages}{072001} (\bibinfo{year}{2019}), \eprint{1905.10811}.

\bibitem[{\citenamefont{Cao and Dai}(2019)}]{Cao:2019kst}
\bibinfo{author}{\bibfnamefont{X.}~\bibnamefont{Cao}} \bibnamefont{and} \bibinfo{author}{\bibfnamefont{J.-p.} \bibnamefont{Dai}}, \bibinfo{journal}{Phys. Rev. D} \textbf{\bibinfo{volume}{100}}, \bibinfo{pages}{054033} (\bibinfo{year}{2019}), \eprint{1904.06015}.

\bibitem[{\citenamefont{Wang et~al.}(2015)\citenamefont{Wang, Liu, and Zhao}}]{Wang:2015jsa}
\bibinfo{author}{\bibfnamefont{Q.}~\bibnamefont{Wang}}, \bibinfo{author}{\bibfnamefont{X.-H.} \bibnamefont{Liu}}, \bibnamefont{and} \bibinfo{author}{\bibfnamefont{Q.}~\bibnamefont{Zhao}}, \bibinfo{journal}{Phys. Rev. D} \textbf{\bibinfo{volume}{92}}, \bibinfo{pages}{034022} (\bibinfo{year}{2015}), \eprint{1508.00339}.

\bibitem[{\citenamefont{Hiller~Blin et~al.}(2016)\citenamefont{Hiller~Blin, Fern\'andez-Ram\'\i{}rez, Jackura, Mathieu, Mokeev, Pilloni, and Szczepaniak}}]{HillerBlin:2016odx}
\bibinfo{author}{\bibfnamefont{A.~N.} \bibnamefont{Hiller~Blin}}, \bibinfo{author}{\bibfnamefont{C.}~\bibnamefont{Fern\'andez-Ram\'\i{}rez}}, \bibinfo{author}{\bibfnamefont{A.}~\bibnamefont{Jackura}}, \bibinfo{author}{\bibfnamefont{V.}~\bibnamefont{Mathieu}}, \bibinfo{author}{\bibfnamefont{V.~I.} \bibnamefont{Mokeev}}, \bibinfo{author}{\bibfnamefont{A.}~\bibnamefont{Pilloni}}, \bibnamefont{and} \bibinfo{author}{\bibfnamefont{A.~P.} \bibnamefont{Szczepaniak}}, \bibinfo{journal}{Phys. Rev. D} \textbf{\bibinfo{volume}{94}}, \bibinfo{pages}{034002} (\bibinfo{year}{2016}), \eprint{1606.08912}.

\bibitem[{\citenamefont{Karliner and Rosner}(2016)}]{Karliner:2015voa}
\bibinfo{author}{\bibfnamefont{M.}~\bibnamefont{Karliner}} \bibnamefont{and} \bibinfo{author}{\bibfnamefont{J.~L.} \bibnamefont{Rosner}}, \bibinfo{journal}{Phys. Lett. B} \textbf{\bibinfo{volume}{752}}, \bibinfo{pages}{329} (\bibinfo{year}{2016}), \eprint{1508.01496}.

\bibitem[{\citenamefont{Wang et~al.}(2019)\citenamefont{Wang, Chen, and He}}]{Wang:2019krd}
\bibinfo{author}{\bibfnamefont{X.-Y.} \bibnamefont{Wang}}, \bibinfo{author}{\bibfnamefont{X.-R.} \bibnamefont{Chen}}, \bibnamefont{and} \bibinfo{author}{\bibfnamefont{J.}~\bibnamefont{He}}, \bibinfo{journal}{Phys. Rev.} \textbf{\bibinfo{volume}{D99}}, \bibinfo{pages}{114007} (\bibinfo{year}{2019}), \eprint{1904.11706}.

\bibitem[{\citenamefont{Wu et~al.}(2019)\citenamefont{Wu, Lee, and Zou}}]{Wu:2019adv}
\bibinfo{author}{\bibfnamefont{J.-J.} \bibnamefont{Wu}}, \bibinfo{author}{\bibfnamefont{T.~S.~H.} \bibnamefont{Lee}}, \bibnamefont{and} \bibinfo{author}{\bibfnamefont{B.-S.} \bibnamefont{Zou}}, \bibinfo{journal}{Phys. Rev. C} \textbf{\bibinfo{volume}{100}}, \bibinfo{pages}{035206} (\bibinfo{year}{2019}), \eprint{1906.05375}.

\bibitem[{\citenamefont{Li et~al.}(2018)\citenamefont{Li, Liu, Liu, Si, and Zhang}}]{Li:2017ghe}
\bibinfo{author}{\bibfnamefont{S.-Y.} \bibnamefont{Li}}, \bibinfo{author}{\bibfnamefont{Y.-R.} \bibnamefont{Liu}}, \bibinfo{author}{\bibfnamefont{Y.-N.} \bibnamefont{Liu}}, \bibinfo{author}{\bibfnamefont{Z.-G.} \bibnamefont{Si}}, \bibnamefont{and} \bibinfo{author}{\bibfnamefont{X.-F.} \bibnamefont{Zhang}}, \bibinfo{journal}{Commun. Theor. Phys.} \textbf{\bibinfo{volume}{69}}, \bibinfo{pages}{291} (\bibinfo{year}{2018}), \eprint{1706.04765}.

\bibitem[{\citenamefont{Voloshin}(2019)}]{Voloshin:2019wxx}
\bibinfo{author}{\bibfnamefont{M.~B.} \bibnamefont{Voloshin}}, \bibinfo{journal}{Phys. Rev. D} \textbf{\bibinfo{volume}{99}}, \bibinfo{pages}{093003} (\bibinfo{year}{2019}), \eprint{1903.04422}.

\bibitem[{\citenamefont{Chen et~al.}(2022)\citenamefont{Chen, Xie, Xu, Zhang, Zhou, She, and Chen}}]{Chen:2021ifb}
\bibinfo{author}{\bibfnamefont{C.-h.} \bibnamefont{Chen}}, \bibinfo{author}{\bibfnamefont{Y.-L.} \bibnamefont{Xie}}, \bibinfo{author}{\bibfnamefont{H.-g.} \bibnamefont{Xu}}, \bibinfo{author}{\bibfnamefont{Z.}~\bibnamefont{Zhang}}, \bibinfo{author}{\bibfnamefont{D.-M.} \bibnamefont{Zhou}}, \bibinfo{author}{\bibfnamefont{Z.-L.} \bibnamefont{She}}, \bibnamefont{and} \bibinfo{author}{\bibfnamefont{G.}~\bibnamefont{Chen}}, \bibinfo{journal}{Phys. Rev. D} \textbf{\bibinfo{volume}{105}}, \bibinfo{pages}{054013} (\bibinfo{year}{2022}), \eprint{2111.03241}.

\bibitem[{\citenamefont{Ling et~al.}(2021{\natexlab{b}})\citenamefont{Ling, Dai, Du, and Wang}}]{Ling:2021sld}
\bibinfo{author}{\bibfnamefont{P.}~\bibnamefont{Ling}}, \bibinfo{author}{\bibfnamefont{X.-H.} \bibnamefont{Dai}}, \bibinfo{author}{\bibfnamefont{M.-L.} \bibnamefont{Du}}, \bibnamefont{and} \bibinfo{author}{\bibfnamefont{Q.}~\bibnamefont{Wang}}, \bibinfo{journal}{Eur. Phys. J. C} \textbf{\bibinfo{volume}{81}}, \bibinfo{pages}{819} (\bibinfo{year}{2021}{\natexlab{b}}), \eprint{2104.11133}.

\bibitem[{\citenamefont{Shi et~al.}(2022)\citenamefont{Shi, Guo, and Yang}}]{Shi:2022ipx}
\bibinfo{author}{\bibfnamefont{P.-P.} \bibnamefont{Shi}}, \bibinfo{author}{\bibfnamefont{F.-K.} \bibnamefont{Guo}}, \bibnamefont{and} \bibinfo{author}{\bibfnamefont{Z.}~\bibnamefont{Yang}}, \bibinfo{journal}{Phys. Rev. D} \textbf{\bibinfo{volume}{106}}, \bibinfo{pages}{114026} (\bibinfo{year}{2022}), \eprint{2208.02639}.

\bibitem[{\citenamefont{Braaten et~al.}(2004)\citenamefont{Braaten, Kusunoki, and Nussinov}}]{Braaten:2004fk}
\bibinfo{author}{\bibfnamefont{E.}~\bibnamefont{Braaten}}, \bibinfo{author}{\bibfnamefont{M.}~\bibnamefont{Kusunoki}}, \bibnamefont{and} \bibinfo{author}{\bibfnamefont{S.}~\bibnamefont{Nussinov}}, \bibinfo{journal}{Phys. Rev. Lett.} \textbf{\bibinfo{volume}{93}}, \bibinfo{pages}{162001} (\bibinfo{year}{2004}), \eprint{hep-ph/0404161}.

\bibitem[{\citenamefont{Braaten and Kusunoki}(2005)}]{Braaten:2004ai}
\bibinfo{author}{\bibfnamefont{E.}~\bibnamefont{Braaten}} \bibnamefont{and} \bibinfo{author}{\bibfnamefont{M.}~\bibnamefont{Kusunoki}}, \bibinfo{journal}{Phys. Rev. D} \textbf{\bibinfo{volume}{71}}, \bibinfo{pages}{074005} (\bibinfo{year}{2005}), \eprint{hep-ph/0412268}.

\bibitem[{\citenamefont{Du et~al.}(2021)\citenamefont{Du, Baru, Guo, Hanhart, Mei\ss{}ner, Oller, and Wang}}]{Du:2021fmf}
\bibinfo{author}{\bibfnamefont{M.-L.} \bibnamefont{Du}}, \bibinfo{author}{\bibfnamefont{V.}~\bibnamefont{Baru}}, \bibinfo{author}{\bibfnamefont{F.-K.} \bibnamefont{Guo}}, \bibinfo{author}{\bibfnamefont{C.}~\bibnamefont{Hanhart}}, \bibinfo{author}{\bibfnamefont{U.-G.} \bibnamefont{Mei\ss{}ner}}, \bibinfo{author}{\bibfnamefont{J.~A.} \bibnamefont{Oller}}, \bibnamefont{and} \bibinfo{author}{\bibfnamefont{Q.}~\bibnamefont{Wang}}, \bibinfo{journal}{JHEP} \textbf{\bibinfo{volume}{08}}, \bibinfo{pages}{157} (\bibinfo{year}{2021}), \eprint{2102.07159}.

\bibitem[{\citenamefont{Roca et~al.}(2015)\citenamefont{Roca, Nieves, and Oset}}]{Roca:2015dva}
\bibinfo{author}{\bibfnamefont{L.}~\bibnamefont{Roca}}, \bibinfo{author}{\bibfnamefont{J.}~\bibnamefont{Nieves}}, \bibnamefont{and} \bibinfo{author}{\bibfnamefont{E.}~\bibnamefont{Oset}}, \bibinfo{journal}{Phys. Rev. D} \textbf{\bibinfo{volume}{92}}, \bibinfo{pages}{094003} (\bibinfo{year}{2015}), \eprint{1507.04249}.

\bibitem[{\citenamefont{Wang et~al.}(2013)\citenamefont{Wang, Hanhart, and Zhao}}]{Wang:2013cya}
\bibinfo{author}{\bibfnamefont{Q.}~\bibnamefont{Wang}}, \bibinfo{author}{\bibfnamefont{C.}~\bibnamefont{Hanhart}}, \bibnamefont{and} \bibinfo{author}{\bibfnamefont{Q.}~\bibnamefont{Zhao}}, \bibinfo{journal}{Phys. Rev. Lett.} \textbf{\bibinfo{volume}{111}}, \bibinfo{pages}{132003} (\bibinfo{year}{2013}), \eprint{1303.6355}.

\bibitem[{\citenamefont{Guo et~al.}(2013)\citenamefont{Guo, Hanhart, Mei\ss{}ner, Wang, and Zhao}}]{Guo:2013zbw}
\bibinfo{author}{\bibfnamefont{F.-K.} \bibnamefont{Guo}}, \bibinfo{author}{\bibfnamefont{C.}~\bibnamefont{Hanhart}}, \bibinfo{author}{\bibfnamefont{U.-G.} \bibnamefont{Mei\ss{}ner}}, \bibinfo{author}{\bibfnamefont{Q.}~\bibnamefont{Wang}}, \bibnamefont{and} \bibinfo{author}{\bibfnamefont{Q.}~\bibnamefont{Zhao}}, \bibinfo{journal}{Phys. Lett. B} \textbf{\bibinfo{volume}{725}}, \bibinfo{pages}{127} (\bibinfo{year}{2013}), \eprint{1306.3096}.

\bibitem[{\citenamefont{Hsiao et~al.}(2020)\citenamefont{Hsiao, Yu, and Ke}}]{Hsiao:2019ait}
\bibinfo{author}{\bibfnamefont{Y.-K.} \bibnamefont{Hsiao}}, \bibinfo{author}{\bibfnamefont{Y.}~\bibnamefont{Yu}}, \bibnamefont{and} \bibinfo{author}{\bibfnamefont{B.-C.} \bibnamefont{Ke}}, \bibinfo{journal}{Eur. Phys. J. C} \textbf{\bibinfo{volume}{80}}, \bibinfo{pages}{895} (\bibinfo{year}{2020}), \eprint{1909.07327}.

\bibitem[{\citenamefont{Liu et~al.}(2022)\citenamefont{Liu, Ling, Geng, En-Wang, and Xie}}]{Liu:2022dmm}
\bibinfo{author}{\bibfnamefont{M.-Z.} \bibnamefont{Liu}}, \bibinfo{author}{\bibfnamefont{X.-Z.} \bibnamefont{Ling}}, \bibinfo{author}{\bibfnamefont{L.-S.} \bibnamefont{Geng}}, \bibinfo{author}{\bibnamefont{En-Wang}}, \bibnamefont{and} \bibinfo{author}{\bibfnamefont{J.-J.} \bibnamefont{Xie}}, \bibinfo{journal}{Phys. Rev. D} \textbf{\bibinfo{volume}{106}}, \bibinfo{pages}{114011} (\bibinfo{year}{2022}), \eprint{2209.01103}.

\bibitem[{\citenamefont{Wu et~al.}(2023)\citenamefont{Wu, Liu, and Geng}}]{Wu:2023rrp}
\bibinfo{author}{\bibfnamefont{Q.}~\bibnamefont{Wu}}, \bibinfo{author}{\bibfnamefont{M.-Z.} \bibnamefont{Liu}}, \bibnamefont{and} \bibinfo{author}{\bibfnamefont{L.-S.} \bibnamefont{Geng}} (\bibinfo{year}{2023}), \eprint{2304.05269}.

\bibitem[{\citenamefont{Wu and Chen}(2019)}]{Wu:2019rog}
\bibinfo{author}{\bibfnamefont{Q.}~\bibnamefont{Wu}} \bibnamefont{and} \bibinfo{author}{\bibfnamefont{D.-Y.} \bibnamefont{Chen}}, \bibinfo{journal}{Phys. Rev.} \textbf{\bibinfo{volume}{D100}}, \bibinfo{pages}{114002} (\bibinfo{year}{2019}), \eprint{1906.02480}.

\bibitem[{\citenamefont{Falk and Neubert}(1993)}]{Falk:1992ws}
\bibinfo{author}{\bibfnamefont{A.~F.} \bibnamefont{Falk}} \bibnamefont{and} \bibinfo{author}{\bibfnamefont{M.}~\bibnamefont{Neubert}}, \bibinfo{journal}{Phys. Rev. D} \textbf{\bibinfo{volume}{47}}, \bibinfo{pages}{2982} (\bibinfo{year}{1993}), \eprint{hep-ph/9209269}.

\bibitem[{\citenamefont{Gutsche et~al.}(2018)\citenamefont{Gutsche, Ivanov, K\"orner, and Lyubovitskij}}]{Gutsche:2018utw}
\bibinfo{author}{\bibfnamefont{T.}~\bibnamefont{Gutsche}}, \bibinfo{author}{\bibfnamefont{M.~A.} \bibnamefont{Ivanov}}, \bibinfo{author}{\bibfnamefont{J.~G.} \bibnamefont{K\"orner}}, \bibnamefont{and} \bibinfo{author}{\bibfnamefont{V.~E.} \bibnamefont{Lyubovitskij}}, \bibinfo{journal}{Phys. Rev. D} \textbf{\bibinfo{volume}{98}}, \bibinfo{pages}{074011} (\bibinfo{year}{2018}), \eprint{1806.11549}.

\bibitem[{\citenamefont{Liu et~al.}(2019)\citenamefont{Liu, Pan, Peng, Sánchez~Sánchez, Geng, Hosaka, and Pavon~Valderrama}}]{Liu:2019tjn}
\bibinfo{author}{\bibfnamefont{M.-Z.} \bibnamefont{Liu}}, \bibinfo{author}{\bibfnamefont{Y.-W.} \bibnamefont{Pan}}, \bibinfo{author}{\bibfnamefont{F.-Z.} \bibnamefont{Peng}}, \bibinfo{author}{\bibfnamefont{M.}~\bibnamefont{Sánchez~Sánchez}}, \bibinfo{author}{\bibfnamefont{L.-S.} \bibnamefont{Geng}}, \bibinfo{author}{\bibfnamefont{A.}~\bibnamefont{Hosaka}}, \bibnamefont{and} \bibinfo{author}{\bibfnamefont{M.}~\bibnamefont{Pavon~Valderrama}}, \bibinfo{journal}{Phys. Rev. Lett.} \textbf{\bibinfo{volume}{122}}, \bibinfo{pages}{242001} (\bibinfo{year}{2019}), \eprint{1903.11560}.

\bibitem[{\citenamefont{Pavon~Valderrama}(2019{\natexlab{b}})}]{PavonValderrama:2019nbk}
\bibinfo{author}{\bibfnamefont{M.}~\bibnamefont{Pavon~Valderrama}}, \bibinfo{journal}{Phys. Rev. D} \textbf{\bibinfo{volume}{100}}, \bibinfo{pages}{094028} (\bibinfo{year}{2019}{\natexlab{b}}), \eprint{1907.05294}.

\bibitem[{\citenamefont{Lin and Zou}(2019)}]{Lin:2019qiv}
\bibinfo{author}{\bibfnamefont{Y.-H.} \bibnamefont{Lin}} \bibnamefont{and} \bibinfo{author}{\bibfnamefont{B.-S.} \bibnamefont{Zou}}, \bibinfo{journal}{Phys. Rev.} \textbf{\bibinfo{volume}{D100}}, \bibinfo{pages}{056005} (\bibinfo{year}{2019}), \eprint{1908.05309}.

\bibitem[{\citenamefont{Yalikun et~al.}(2021)\citenamefont{Yalikun, Lin, Guo, Kamiya, and Zou}}]{Yalikun:2021bfm}
\bibinfo{author}{\bibfnamefont{N.}~\bibnamefont{Yalikun}}, \bibinfo{author}{\bibfnamefont{Y.-H.} \bibnamefont{Lin}}, \bibinfo{author}{\bibfnamefont{F.-K.} \bibnamefont{Guo}}, \bibinfo{author}{\bibfnamefont{Y.}~\bibnamefont{Kamiya}}, \bibnamefont{and} \bibinfo{author}{\bibfnamefont{B.-S.} \bibnamefont{Zou}}, \bibinfo{journal}{Phys. Rev. D} \textbf{\bibinfo{volume}{104}}, \bibinfo{pages}{094039} (\bibinfo{year}{2021}), \eprint{2109.03504}.

\bibitem[{\citenamefont{Chau}(1983)}]{Chau:1982da}
\bibinfo{author}{\bibfnamefont{L.-L.} \bibnamefont{Chau}}, \bibinfo{journal}{Phys. Rept.} \textbf{\bibinfo{volume}{95}}, \bibinfo{pages}{1} (\bibinfo{year}{1983}).

\bibitem[{\citenamefont{Chau and Cheng}(1987)}]{Chau:1987tk}
\bibinfo{author}{\bibfnamefont{L.-L.} \bibnamefont{Chau}} \bibnamefont{and} \bibinfo{author}{\bibfnamefont{H.-Y.} \bibnamefont{Cheng}}, \bibinfo{journal}{Phys. Rev. D} \textbf{\bibinfo{volume}{36}}, \bibinfo{pages}{137} (\bibinfo{year}{1987}), \bibinfo{note}{[Addendum: Phys.Rev.D 39, 2788--2791 (1989)]}.

\bibitem[{\citenamefont{Molina et~al.}(2020)\citenamefont{Molina, Xie, Liang, Geng, and Oset}}]{Molina:2019udw}
\bibinfo{author}{\bibfnamefont{R.}~\bibnamefont{Molina}}, \bibinfo{author}{\bibfnamefont{J.-J.} \bibnamefont{Xie}}, \bibinfo{author}{\bibfnamefont{W.-H.} \bibnamefont{Liang}}, \bibinfo{author}{\bibfnamefont{L.-S.} \bibnamefont{Geng}}, \bibnamefont{and} \bibinfo{author}{\bibfnamefont{E.}~\bibnamefont{Oset}}, \bibinfo{journal}{Phys. Lett. B} \textbf{\bibinfo{volume}{803}}, \bibinfo{pages}{135279} (\bibinfo{year}{2020}), \eprint{1908.11557}.

\bibitem[{\citenamefont{Cheng}(1997)}]{Cheng:1996cs}
\bibinfo{author}{\bibfnamefont{H.-Y.} \bibnamefont{Cheng}}, \bibinfo{journal}{Phys. Rev. D} \textbf{\bibinfo{volume}{56}}, \bibinfo{pages}{2799} (\bibinfo{year}{1997}), \bibinfo{note}{[Erratum: Phys.Rev.D 99, 079901 (2019)]}, \eprint{hep-ph/9612223}.

\bibitem[{\citenamefont{Ali et~al.}(1998)\citenamefont{Ali, Kramer, and Lu}}]{Ali:1998eb}
\bibinfo{author}{\bibfnamefont{A.}~\bibnamefont{Ali}}, \bibinfo{author}{\bibfnamefont{G.}~\bibnamefont{Kramer}}, \bibnamefont{and} \bibinfo{author}{\bibfnamefont{C.-D.} \bibnamefont{Lu}}, \bibinfo{journal}{Phys. Rev. D} \textbf{\bibinfo{volume}{58}}, \bibinfo{pages}{094009} (\bibinfo{year}{1998}), \eprint{hep-ph/9804363}.

\bibitem[{\citenamefont{Li et~al.}(2012)\citenamefont{Li, Lu, and Yu}}]{Li:2012cfa}
\bibinfo{author}{\bibfnamefont{H.-n.} \bibnamefont{Li}}, \bibinfo{author}{\bibfnamefont{C.-D.} \bibnamefont{Lu}}, \bibnamefont{and} \bibinfo{author}{\bibfnamefont{F.-S.} \bibnamefont{Yu}}, \bibinfo{journal}{Phys. Rev. D} \textbf{\bibinfo{volume}{86}}, \bibinfo{pages}{036012} (\bibinfo{year}{2012}), \eprint{1203.3120}.

\bibitem[{\citenamefont{Bauer et~al.}(1987)\citenamefont{Bauer, Stech, and Wirbel}}]{Bauer:1986bm}
\bibinfo{author}{\bibfnamefont{M.}~\bibnamefont{Bauer}}, \bibinfo{author}{\bibfnamefont{B.}~\bibnamefont{Stech}}, \bibnamefont{and} \bibinfo{author}{\bibfnamefont{M.}~\bibnamefont{Wirbel}}, \bibinfo{journal}{Z. Phys. C} \textbf{\bibinfo{volume}{34}}, \bibinfo{pages}{103} (\bibinfo{year}{1987}).

\bibitem[{\citenamefont{Gutsche et~al.}(2015)\citenamefont{Gutsche, Ivanov, K\"orner, Lyubovitskij, Santorelli, and Habyl}}]{Gutsche:2015mxa}
\bibinfo{author}{\bibfnamefont{T.}~\bibnamefont{Gutsche}}, \bibinfo{author}{\bibfnamefont{M.~A.} \bibnamefont{Ivanov}}, \bibinfo{author}{\bibfnamefont{J.~G.} \bibnamefont{K\"orner}}, \bibinfo{author}{\bibfnamefont{V.~E.} \bibnamefont{Lyubovitskij}}, \bibinfo{author}{\bibfnamefont{P.}~\bibnamefont{Santorelli}}, \bibnamefont{and} \bibinfo{author}{\bibfnamefont{N.}~\bibnamefont{Habyl}}, \bibinfo{journal}{Phys. Rev. D} \textbf{\bibinfo{volume}{91}}, \bibinfo{pages}{074001} (\bibinfo{year}{2015}), \bibinfo{note}{[Erratum: Phys.Rev.D 91, 119907 (2015)]}, \eprint{1502.04864}.

\bibitem[{\citenamefont{Zyla et~al.}(2020)}]{ParticleDataGroup:2020ssz}
\bibinfo{author}{\bibfnamefont{P.~A.} \bibnamefont{Zyla}} \bibnamefont{et~al.} (\bibinfo{collaboration}{Particle Data Group}), \bibinfo{journal}{PTEP} \textbf{\bibinfo{volume}{2020}}, \bibinfo{pages}{083C01} (\bibinfo{year}{2020}).

\bibitem[{\citenamefont{Verma}(2012)}]{Verma:2011yw}
\bibinfo{author}{\bibfnamefont{R.~C.} \bibnamefont{Verma}}, \bibinfo{journal}{J. Phys. G} \textbf{\bibinfo{volume}{39}}, \bibinfo{pages}{025005} (\bibinfo{year}{2012}), \eprint{1103.2973}.

\bibitem[{\citenamefont{Aoki et~al.}(2020)}]{FlavourLatticeAveragingGroup:2019iem}
\bibinfo{author}{\bibfnamefont{S.}~\bibnamefont{Aoki}} \bibnamefont{et~al.} (\bibinfo{collaboration}{Flavour Lattice Averaging Group}), \bibinfo{journal}{Eur. Phys. J. C} \textbf{\bibinfo{volume}{80}}, \bibinfo{pages}{113} (\bibinfo{year}{2020}), \eprint{1902.08191}.

\bibitem[{\citenamefont{Chen et~al.}(1999)\citenamefont{Chen, Cheng, Tseng, and Yang}}]{Chen:1999nxa}
\bibinfo{author}{\bibfnamefont{Y.-H.} \bibnamefont{Chen}}, \bibinfo{author}{\bibfnamefont{H.-Y.} \bibnamefont{Cheng}}, \bibinfo{author}{\bibfnamefont{B.}~\bibnamefont{Tseng}}, \bibnamefont{and} \bibinfo{author}{\bibfnamefont{K.-C.} \bibnamefont{Yang}}, \bibinfo{journal}{Phys. Rev. D} \textbf{\bibinfo{volume}{60}}, \bibinfo{pages}{094014} (\bibinfo{year}{1999}), \eprint{hep-ph/9903453}.

\bibitem[{\citenamefont{Cheng and Chiang}(2010)}]{Cheng:2010ry}
\bibinfo{author}{\bibfnamefont{H.-Y.} \bibnamefont{Cheng}} \bibnamefont{and} \bibinfo{author}{\bibfnamefont{C.-W.} \bibnamefont{Chiang}}, \bibinfo{journal}{Phys. Rev. D} \textbf{\bibinfo{volume}{81}}, \bibinfo{pages}{074021} (\bibinfo{year}{2010}), \eprint{1001.0987}.

\bibitem[{\citenamefont{Chua}(2019)}]{Chua:2019yqh}
\bibinfo{author}{\bibfnamefont{C.-K.} \bibnamefont{Chua}}, \bibinfo{journal}{Phys. Rev. D} \textbf{\bibinfo{volume}{100}}, \bibinfo{pages}{034025} (\bibinfo{year}{2019}), \eprint{1905.00153}.

\bibitem[{\citenamefont{Xie et~al.}(2023)\citenamefont{Xie, Liu, and Geng}}]{Xie:2022lyw}
\bibinfo{author}{\bibfnamefont{J.-M.} \bibnamefont{Xie}}, \bibinfo{author}{\bibfnamefont{M.-Z.} \bibnamefont{Liu}}, \bibnamefont{and} \bibinfo{author}{\bibfnamefont{L.-S.} \bibnamefont{Geng}}, \bibinfo{journal}{Phys. Rev. D} \textbf{\bibinfo{volume}{107}}, \bibinfo{pages}{016003} (\bibinfo{year}{2023}), \eprint{2207.12178}.

\bibitem[{\citenamefont{Azevedo and Nielsen}(2004)}]{Azevedo:2003qh}
\bibinfo{author}{\bibfnamefont{R.~S.} \bibnamefont{Azevedo}} \bibnamefont{and} \bibinfo{author}{\bibfnamefont{M.}~\bibnamefont{Nielsen}}, \bibinfo{journal}{Phys. Rev. C} \textbf{\bibinfo{volume}{69}}, \bibinfo{pages}{035201} (\bibinfo{year}{2004}), \eprint{nucl-th/0310061}.

\bibitem[{\citenamefont{Bracco et~al.}(2006)\citenamefont{Bracco, Cerqueira, Chiapparini, Lozea, and Nielsen}}]{Bracco:2006xf}
\bibinfo{author}{\bibfnamefont{M.~E.} \bibnamefont{Bracco}}, \bibinfo{author}{\bibfnamefont{A.}~\bibnamefont{Cerqueira}, \bibfnamefont{Jr.}}, \bibinfo{author}{\bibfnamefont{M.}~\bibnamefont{Chiapparini}}, \bibinfo{author}{\bibfnamefont{A.}~\bibnamefont{Lozea}}, \bibnamefont{and} \bibinfo{author}{\bibfnamefont{M.}~\bibnamefont{Nielsen}}, \bibinfo{journal}{Phys. Lett. B} \textbf{\bibinfo{volume}{641}}, \bibinfo{pages}{286} (\bibinfo{year}{2006}), \eprint{hep-ph/0604167}.

\bibitem[{\citenamefont{Wang and Wan}(2006)}]{Wang:2006ida}
\bibinfo{author}{\bibfnamefont{Z.~G.} \bibnamefont{Wang}} \bibnamefont{and} \bibinfo{author}{\bibfnamefont{S.~L.} \bibnamefont{Wan}}, \bibinfo{journal}{Phys. Rev. D} \textbf{\bibinfo{volume}{74}}, \bibinfo{pages}{014017} (\bibinfo{year}{2006}), \eprint{hep-ph/0606002}.

\bibitem[{\citenamefont{Guo and Oller}(2019)}]{Guo:2019kdc}
\bibinfo{author}{\bibfnamefont{Z.-H.} \bibnamefont{Guo}} \bibnamefont{and} \bibinfo{author}{\bibfnamefont{J.}~\bibnamefont{Oller}}, \bibinfo{journal}{Phys. Lett. B} \textbf{\bibinfo{volume}{793}}, \bibinfo{pages}{144} (\bibinfo{year}{2019}), \eprint{1904.00851}.

\bibitem[{\citenamefont{Ji et~al.}(2023)\citenamefont{Ji, Dong, Albaladejo, Du, Guo, Nieves, and Zou}}]{Ji:2022vdj}
\bibinfo{author}{\bibfnamefont{T.}~\bibnamefont{Ji}}, \bibinfo{author}{\bibfnamefont{X.-K.} \bibnamefont{Dong}}, \bibinfo{author}{\bibfnamefont{M.}~\bibnamefont{Albaladejo}}, \bibinfo{author}{\bibfnamefont{M.-L.} \bibnamefont{Du}}, \bibinfo{author}{\bibfnamefont{F.-K.} \bibnamefont{Guo}}, \bibinfo{author}{\bibfnamefont{J.}~\bibnamefont{Nieves}}, \bibnamefont{and} \bibinfo{author}{\bibfnamefont{B.-S.} \bibnamefont{Zou}}, \bibinfo{journal}{Sci. Bull.} \textbf{\bibinfo{volume}{68}}, \bibinfo{pages}{688} (\bibinfo{year}{2023}), \eprint{2212.00631}.

\bibitem[{\citenamefont{Oset and Ramos}(1998)}]{Oset:1997it}
\bibinfo{author}{\bibfnamefont{E.}~\bibnamefont{Oset}} \bibnamefont{and} \bibinfo{author}{\bibfnamefont{A.}~\bibnamefont{Ramos}}, \bibinfo{journal}{Nucl. Phys.} \textbf{\bibinfo{volume}{A635}}, \bibinfo{pages}{99} (\bibinfo{year}{1998}), \eprint{nucl-th/9711022}.

\bibitem[{\citenamefont{Jido et~al.}(2003)\citenamefont{Jido, Oller, Oset, Ramos, and Meissner}}]{Jido:2003cb}
\bibinfo{author}{\bibfnamefont{D.}~\bibnamefont{Jido}}, \bibinfo{author}{\bibfnamefont{J.~A.} \bibnamefont{Oller}}, \bibinfo{author}{\bibfnamefont{E.}~\bibnamefont{Oset}}, \bibinfo{author}{\bibfnamefont{A.}~\bibnamefont{Ramos}}, \bibnamefont{and} \bibinfo{author}{\bibfnamefont{U.~G.} \bibnamefont{Meissner}}, \bibinfo{journal}{Nucl. Phys.} \textbf{\bibinfo{volume}{A725}}, \bibinfo{pages}{181} (\bibinfo{year}{2003}), \eprint{nucl-th/0303062}.

\bibitem[{\citenamefont{Wu et~al.}(2010)\citenamefont{Wu, Molina, Oset, and Zou}}]{Wu:2010jy}
\bibinfo{author}{\bibfnamefont{J.-J.} \bibnamefont{Wu}}, \bibinfo{author}{\bibfnamefont{R.}~\bibnamefont{Molina}}, \bibinfo{author}{\bibfnamefont{E.}~\bibnamefont{Oset}}, \bibnamefont{and} \bibinfo{author}{\bibfnamefont{B.~S.} \bibnamefont{Zou}}, \bibinfo{journal}{Phys. Rev. Lett.} \textbf{\bibinfo{volume}{105}}, \bibinfo{pages}{232001} (\bibinfo{year}{2010}), \eprint{1007.0573}.

\bibitem[{\citenamefont{Hyodo and Jido}(2012)}]{Hyodo:2011ur}
\bibinfo{author}{\bibfnamefont{T.}~\bibnamefont{Hyodo}} \bibnamefont{and} \bibinfo{author}{\bibfnamefont{D.}~\bibnamefont{Jido}}, \bibinfo{journal}{Prog. Part. Nucl. Phys.} \textbf{\bibinfo{volume}{67}}, \bibinfo{pages}{55} (\bibinfo{year}{2012}), \eprint{1104.4474}.

\bibitem[{\citenamefont{Debastiani et~al.}(2018)\citenamefont{Debastiani, Dias, Liang, and Oset}}]{Debastiani:2017ewu}
\bibinfo{author}{\bibfnamefont{V.~R.} \bibnamefont{Debastiani}}, \bibinfo{author}{\bibfnamefont{J.~M.} \bibnamefont{Dias}}, \bibinfo{author}{\bibfnamefont{W.~H.} \bibnamefont{Liang}}, \bibnamefont{and} \bibinfo{author}{\bibfnamefont{E.}~\bibnamefont{Oset}}, \bibinfo{journal}{Phys. Rev. D} \textbf{\bibinfo{volume}{97}}, \bibinfo{pages}{094035} (\bibinfo{year}{2018}), \eprint{1710.04231}.

\bibitem[{\citenamefont{Oller and Oset}(1997)}]{Oller:1997ti}
\bibinfo{author}{\bibfnamefont{J.~A.} \bibnamefont{Oller}} \bibnamefont{and} \bibinfo{author}{\bibfnamefont{E.}~\bibnamefont{Oset}}, \bibinfo{journal}{Nucl. Phys.} \textbf{\bibinfo{volume}{A620}}, \bibinfo{pages}{438} (\bibinfo{year}{1997}), \bibinfo{note}{[Erratum: Nucl. Phys.A652,407(1999)]}, \eprint{hep-ph/9702314}.

\bibitem[{\citenamefont{Roca et~al.}(2005)\citenamefont{Roca, Oset, and Singh}}]{Roca:2005nm}
\bibinfo{author}{\bibfnamefont{L.}~\bibnamefont{Roca}}, \bibinfo{author}{\bibfnamefont{E.}~\bibnamefont{Oset}}, \bibnamefont{and} \bibinfo{author}{\bibfnamefont{J.}~\bibnamefont{Singh}}, \bibinfo{journal}{Phys. Rev. D} \textbf{\bibinfo{volume}{72}}, \bibinfo{pages}{014002} (\bibinfo{year}{2005}), \eprint{hep-ph/0503273}.

\bibitem[{\citenamefont{Yu et~al.}(2019)\citenamefont{Yu, Pavao, Debastiani, and Oset}}]{Yu:2018yxl}
\bibinfo{author}{\bibfnamefont{Q.~X.} \bibnamefont{Yu}}, \bibinfo{author}{\bibfnamefont{R.}~\bibnamefont{Pavao}}, \bibinfo{author}{\bibfnamefont{V.~R.} \bibnamefont{Debastiani}}, \bibnamefont{and} \bibinfo{author}{\bibfnamefont{E.}~\bibnamefont{Oset}}, \bibinfo{journal}{Eur. Phys. J. C} \textbf{\bibinfo{volume}{79}}, \bibinfo{pages}{167} (\bibinfo{year}{2019}), \eprint{1811.11738}.

\bibitem[{\citenamefont{Peng et~al.}(2020)\citenamefont{Peng, Liu, S\'anchez~S\'anchez, and Pavon~Valderrama}}]{Peng:2020xrf}
\bibinfo{author}{\bibfnamefont{F.-Z.} \bibnamefont{Peng}}, \bibinfo{author}{\bibfnamefont{M.-Z.} \bibnamefont{Liu}}, \bibinfo{author}{\bibfnamefont{M.}~\bibnamefont{S\'anchez~S\'anchez}}, \bibnamefont{and} \bibinfo{author}{\bibfnamefont{M.}~\bibnamefont{Pavon~Valderrama}}, \bibinfo{journal}{Phys. Rev. D} \textbf{\bibinfo{volume}{102}}, \bibinfo{pages}{114020} (\bibinfo{year}{2020}), \eprint{2004.05658}.

\bibitem[{\citenamefont{Peng et~al.}(2022)\citenamefont{Peng, S\'anchez~S\'anchez, Yan, and Pavon~Valderrama}}]{Peng:2021hkr}
\bibinfo{author}{\bibfnamefont{F.-Z.} \bibnamefont{Peng}}, \bibinfo{author}{\bibfnamefont{M.}~\bibnamefont{S\'anchez~S\'anchez}}, \bibinfo{author}{\bibfnamefont{M.-J.} \bibnamefont{Yan}}, \bibnamefont{and} \bibinfo{author}{\bibfnamefont{M.}~\bibnamefont{Pavon~Valderrama}}, \bibinfo{journal}{Phys. Rev. D} \textbf{\bibinfo{volume}{105}}, \bibinfo{pages}{034028} (\bibinfo{year}{2022}), \eprint{2101.07213}.

\bibitem[{\citenamefont{Dong et~al.}(2021{\natexlab{b}})\citenamefont{Dong, Guo, and Zou}}]{Dong:2021bvy}
\bibinfo{author}{\bibfnamefont{X.-K.} \bibnamefont{Dong}}, \bibinfo{author}{\bibfnamefont{F.-K.} \bibnamefont{Guo}}, \bibnamefont{and} \bibinfo{author}{\bibfnamefont{B.-S.} \bibnamefont{Zou}}, \bibinfo{journal}{Commun. Theor. Phys.} \textbf{\bibinfo{volume}{73}}, \bibinfo{pages}{125201} (\bibinfo{year}{2021}{\natexlab{b}}), \eprint{2108.02673}.

\bibitem[{\citenamefont{Yang et~al.}(2012)\citenamefont{Yang, Sun, He, Liu, and Zhu}}]{Yang:2011wz}
\bibinfo{author}{\bibfnamefont{Z.-C.} \bibnamefont{Yang}}, \bibinfo{author}{\bibfnamefont{Z.-F.} \bibnamefont{Sun}}, \bibinfo{author}{\bibfnamefont{J.}~\bibnamefont{He}}, \bibinfo{author}{\bibfnamefont{X.}~\bibnamefont{Liu}}, \bibnamefont{and} \bibinfo{author}{\bibfnamefont{S.-L.} \bibnamefont{Zhu}}, \bibinfo{journal}{Chin. Phys. C} \textbf{\bibinfo{volume}{36}}, \bibinfo{pages}{6} (\bibinfo{year}{2012}), \eprint{1105.2901}.

\bibitem[{\citenamefont{Duan et~al.}(2023)\citenamefont{Duan, Qiu, Ling, and Zhao}}]{Duan:2023dky}
\bibinfo{author}{\bibfnamefont{M.-X.} \bibnamefont{Duan}}, \bibinfo{author}{\bibfnamefont{L.}~\bibnamefont{Qiu}}, \bibinfo{author}{\bibfnamefont{X.-Z.} \bibnamefont{Ling}}, \bibnamefont{and} \bibinfo{author}{\bibfnamefont{Q.}~\bibnamefont{Zhao}} (\bibinfo{year}{2023}), \eprint{2303.13329}.

\bibitem[{\citenamefont{Pan et~al.}(2020)\citenamefont{Pan, Liu, Peng, S\'anchez~S\'anchez, Geng, and Pavon~Valderrama}}]{Pan:2019skd}
\bibinfo{author}{\bibfnamefont{Y.-W.} \bibnamefont{Pan}}, \bibinfo{author}{\bibfnamefont{M.-Z.} \bibnamefont{Liu}}, \bibinfo{author}{\bibfnamefont{F.-Z.} \bibnamefont{Peng}}, \bibinfo{author}{\bibfnamefont{M.}~\bibnamefont{S\'anchez~S\'anchez}}, \bibinfo{author}{\bibfnamefont{L.-S.} \bibnamefont{Geng}}, \bibnamefont{and} \bibinfo{author}{\bibfnamefont{M.}~\bibnamefont{Pavon~Valderrama}}, \bibinfo{journal}{Phys. Rev. D} \textbf{\bibinfo{volume}{102}}, \bibinfo{pages}{011504} (\bibinfo{year}{2020}), \eprint{1907.11220}.

\bibitem[{\citenamefont{Xing et~al.}(2022)\citenamefont{Xing, Liang, Liu, Sun, and Yang}}]{Xing:2022ijm}
\bibinfo{author}{\bibfnamefont{H.}~\bibnamefont{Xing}}, \bibinfo{author}{\bibfnamefont{J.}~\bibnamefont{Liang}}, \bibinfo{author}{\bibfnamefont{L.}~\bibnamefont{Liu}}, \bibinfo{author}{\bibfnamefont{P.}~\bibnamefont{Sun}}, \bibnamefont{and} \bibinfo{author}{\bibfnamefont{Y.-B.} \bibnamefont{Yang}} (\bibinfo{year}{2022}), \eprint{2210.08555}.

\bibitem[{\citenamefont{Xiao et~al.}(2019{\natexlab{b}})\citenamefont{Xiao, Huang, Dong, Geng, and Chen}}]{Xiao:2019mvs}
\bibinfo{author}{\bibfnamefont{C.-J.} \bibnamefont{Xiao}}, \bibinfo{author}{\bibfnamefont{Y.}~\bibnamefont{Huang}}, \bibinfo{author}{\bibfnamefont{Y.-B.} \bibnamefont{Dong}}, \bibinfo{author}{\bibfnamefont{L.-S.} \bibnamefont{Geng}}, \bibnamefont{and} \bibinfo{author}{\bibfnamefont{D.-Y.} \bibnamefont{Chen}}, \bibinfo{journal}{Phys. Rev. D} \textbf{\bibinfo{volume}{100}}, \bibinfo{pages}{014022} (\bibinfo{year}{2019}{\natexlab{b}}), \eprint{1904.00872}.

\bibitem[{\citenamefont{Xiao et~al.}(2020)\citenamefont{Xiao, Lu, Wu, and Geng}}]{Xiao:2020frg}
\bibinfo{author}{\bibfnamefont{C.~W.} \bibnamefont{Xiao}}, \bibinfo{author}{\bibfnamefont{J.~X.} \bibnamefont{Lu}}, \bibinfo{author}{\bibfnamefont{J.~J.} \bibnamefont{Wu}}, \bibnamefont{and} \bibinfo{author}{\bibfnamefont{L.~S.} \bibnamefont{Geng}}, \bibinfo{journal}{Phys. Rev. D} \textbf{\bibinfo{volume}{102}}, \bibinfo{pages}{056018} (\bibinfo{year}{2020}), \eprint{2007.12106}.

\bibitem[{\citenamefont{Yamaguchi et~al.}(2019)\citenamefont{Yamaguchi, Abe, Fukukawa, and Hosaka}}]{Yamaguchi:2019djj}
\bibinfo{author}{\bibfnamefont{Y.}~\bibnamefont{Yamaguchi}}, \bibinfo{author}{\bibfnamefont{Y.}~\bibnamefont{Abe}}, \bibinfo{author}{\bibfnamefont{K.}~\bibnamefont{Fukukawa}}, \bibnamefont{and} \bibinfo{author}{\bibfnamefont{A.}~\bibnamefont{Hosaka}}, \bibinfo{journal}{EPJ Web Conf.} \textbf{\bibinfo{volume}{204}}, \bibinfo{pages}{01007} (\bibinfo{year}{2019}).

\bibitem[{\citenamefont{Isgur and Wise}(1992)}]{Isgur:1991xa}
\bibinfo{author}{\bibfnamefont{N.}~\bibnamefont{Isgur}} \bibnamefont{and} \bibinfo{author}{\bibfnamefont{M.~B.} \bibnamefont{Wise}}, \bibinfo{journal}{Adv. Ser. Direct. High Energy Phys.} \textbf{\bibinfo{volume}{10}}, \bibinfo{pages}{234} (\bibinfo{year}{1992}).

\bibitem[{\citenamefont{Flynn and Isgur}(1992)}]{Flynn:1992fm}
\bibinfo{author}{\bibfnamefont{J.~M.} \bibnamefont{Flynn}} \bibnamefont{and} \bibinfo{author}{\bibfnamefont{N.}~\bibnamefont{Isgur}}, \bibinfo{journal}{J. Phys. G} \textbf{\bibinfo{volume}{18}}, \bibinfo{pages}{1627} (\bibinfo{year}{1992}), \eprint{hep-ph/9207223}.

\bibitem[{\citenamefont{Skerbis and Prelovsek}(2019)}]{Skerbis:2018lew}
\bibinfo{author}{\bibfnamefont{U.}~\bibnamefont{Skerbis}} \bibnamefont{and} \bibinfo{author}{\bibfnamefont{S.}~\bibnamefont{Prelovsek}}, \bibinfo{journal}{Phys. Rev.} \textbf{\bibinfo{volume}{D99}}, \bibinfo{pages}{094505} (\bibinfo{year}{2019}), \eprint{1811.02285}.

\bibitem[{\citenamefont{Liu and Oka}(2012)}]{Liu:2011xc}
\bibinfo{author}{\bibfnamefont{Y.-R.} \bibnamefont{Liu}} \bibnamefont{and} \bibinfo{author}{\bibfnamefont{M.}~\bibnamefont{Oka}}, \bibinfo{journal}{Phys. Rev. D} \textbf{\bibinfo{volume}{85}}, \bibinfo{pages}{014015} (\bibinfo{year}{2012}), \eprint{1103.4624}.

\end{thebibliography}

\end{document}